\documentclass[fleqn,usenatbib]{mnras}
\usepackage[T1]{fontenc}

\usepackage{natbib}

\usepackage{graphicx}
\usepackage{amsmath}
\usepackage{amssymb}
\usepackage[version=4]{mhchem}
\usepackage{newtxtext,newtxmath}

\title[HCO$^{+}$ evolution]{The evolution of HCO$^{+}$ in molecular clouds using a novel chemical post-processing algorithm}

\author[M. Panessa et al.]{
M. Panessa,$^{1}$\thanks{panessa@ph1.uni-koeln.de} D. Seifried,$^{1}$ S. Walch,$^{1}$ B. Gaches,$^{1,2}$ A.~T.~Barnes,$^3$ F. Bigiel,$^3$ L. Neumann$^3$
 \\
  $^1$ Universit\"at zu K\"oln, I. Physikalisches Institut, Z\"ulpicher Str. 77, 50937 K\"oln, Germany\\
  $^2$ Department of Space, Earth \& Environment, Chalmers University of Technology, SE-412 96 Gothenburg, Sweden\\
  $^3$ Argelander-Institut f\"{u}r Astronomie, Universit\"{a}t Bonn, Auf dem H\"{u}gel 71, 53121, Bonn, Germany\\
  }

\date{Accepted XXX. Received YYY; in original form ZZZ}

\pubyear{2023}

\begin{document}
\label{firstpage}
\pagerange{\pageref{firstpage}--\pageref{lastpage}}

\maketitle

\begin{abstract}
{Modeling the chemistry of molecular clouds is critical to accurately simulating their evolution. To reduce computational cost, 3D simulations generally restrict their chemistry to species with strong heating and cooling effects. Time-dependent information about the evolution of other species is therefore often neglected. We address this gap by post-processing tracer particles in the SILCC-Zoom molecular cloud simulations. Using a chemical network of 39 species and 301 reactions (including freeze-out of CO and H\textsubscript{2}O), and a novel algorithm to reconstruct a density grid from sparse tracer particle data, we produce time-dependent density distributions for various species. We focus upon the evolution of HCO\textsuperscript{+}, which is a critical formation reactant of CO but is not typically modeled on-the-fly. We find that $\sim90\%$ of the HCO\textsuperscript{+} content of the cold molecular gas forms in situ around $n_\rmn{\ce{HCO+}}\simeq10^3$--$10^4$~cm$^{-3}$, over a time-scale of approximately 1~Myr. The remaining $\sim10\%$ forms at high extinction sites, with minimal turbulent mixing out into the less dense gas. We further show that the dominant HCO\textsuperscript{+} formation pathway is dependent on the visual extinction, with the reaction \mbox{H\textsubscript{3}\textsuperscript{+} + CO} contributing 90\% of the total HCO\textsuperscript{+} production above $A_\rmn{V,3D}=3$. We produce the very first maps of the HCO\textsuperscript{+} column density, $N$(HCO\textsuperscript{+}), and show that it reaches values as high as $10^{15}$~cm$^{-2}$. We find that 50\% of the HCO$^+$ mass is located within $A_\rmn{V}\sim10$--30, in a density range of $10^{3.5}$--$10^{4.5}$~cm$^{-3}$. Our maps of $N$(HCO\textsuperscript{+}) are shown to be in good agreement with recent observations of the W49A star-forming region.}
\end{abstract}


\section{Introduction}
\label{section-intro}

Astrophysical simulations have benefited enormously from modern advances in available computing power. Recent studies have shown that self-consistently tracking the chemical makeup of a molecular cloud is indispensable to understanding the cloud’s evolution \citep{walch_silcc_2015,girichidis_silcc_2016}. In particular, the abundances of common species like atomic carbon and carbon monoxide (CO) influence the cooling and heating rates via their line emission, in turn impacting the bulk dynamics of the cloud \citep{van_dishoeck_photodissociation_1988,wolre_dark_2010,glover_relationship_2011,Bisbas2021}.

Unfortunately, modeling the time-dependent chemistry of molecular clouds is both computationally expensive and theoretically challenging due to the sparsity and stiffness of the associated rate equations \citep{grassi_krome_2014}. For instance, \cite{seifried_modelling_2016} find that in a highly idealized filament simulation, the on-the-fly implementation of a chemical network of 37 species which self-consistently solves the CO abundance lengthens the computing time by up to a factor of seven, compared to implementing no network at all. As such, the gas dynamics in molecular cloud simulations are typically coupled to extremely simple networks, or the chemistry is not even computed on-the-fly \citep{liq-DMGsims-2018,Gong2018,gongm-2020,keating2020}. These minimal networks are restricted to the species which most strongly impact the cloud’s thermal state, with particular emphasis on modeling the abundance of CO through a limited ecosystem of reaction rates \citep[see e.g.][]{nelson_dynamics_1997,nelson_stability_1999,glover_approximations_2012,walch_silcc_2015,seifried_modelling_2017,mackey_non-equilibrium_2019,hu_metallicity_2021}.

While these networks suffice to model the bulk evolutionary dynamics of cold gas, they sacrifice the ability to study species which are dynamically less important but whose abundances and evolution could nevertheless supply further information about the molecular cloud. An example of a scientific question which restricted chemical networks cannot satisfactorily answer is the best way to trace `CO-dark' molecular gas. The greater photodissociation energy of H\textsubscript{2} relative to CO causes the formation of an extended envelope of molecular hydrogen outside the denser regions in which CO can survive \citep{van_dishoeck_photodissociation_1988,wolre_dark_2010,valdivia_h2_2016,gaches_model_2018}. This envelope of H\textsubscript{2}, which cannot be traced by CO emission, can represent several tens of percent of the cloud’s molecular hydrogen by mass \citep{wolre_dark_2010,smith_co-dark_2014,seifried_silcc-zoom_2020}. Simulations using restricted chemical networks can model the CO-dark molecular gas, but cannot suggest alternative tracers for the molecular hydrogen due to the paucity of other species included in the network. 

Because time-dependent networks are so costly to run on-the-fly, chemical post-processing is the chief way to investigate astrochemical problems. Typically, post-processing is performed by evolving a network to equilibrium given a set of fixed environmental parameters. But deferring analysis until equilibrium precludes any understanding of how the dynamical evolution of the cloud environment affects the chemistry. 

Several recent works have performed time-dependent chemical post-processing of astrophysical simulations, specifically to correct for the shortcomings of the equilibrium approach. For instance, \cite{ebagezio2022} compare time-dependent chemical data from SILCC-Zoom simulations of molecular clouds to the chemical outcome if the clouds are evolved to equilibrium. They find that evolving until equilibrium overestimates the total mass of H$_2$ and CO by up to 110 and 30 percent, respectively. The earlier in a cloud's dynamical lifetime that its chemistry is evolved to equilibrium, the less accurate are the final abundances. The distribution of species has also been shown to differ between equilibrium and non-equilibrium results by \cite{hu_metallicity_2021}. They post-process a simulation using time-dependent chemistry, and find that the transitions between the ionized, atomic, and molecular gas phases are more shallow and gradual with time-dependent processing than when the chemistry is evolved to equilibrium. \cite{ferrada-chamorro_chemical_2021}, meanwhile, post-process the chemistry of a 3D-MHD simulation of a collapsing pre-stellar core, coupled to a chemical network. They account for the dynamical evolution of the simulation by post-processing abundances associated with passive tracer particles, which are free to advect with local density gradients in the gas. However, these simulations concentrate on the dense, well-shielded gas with $\langle n \rangle \simeq 10^4$~cm$^{-3}$, rather than the full range of densities found in molecular cloud environments.

To accurately model the time-dependent chemistry of the multi-phase ISM on the scale of tens of parsecs, a simulation must be coupled to at least a simple network modeling the production of CO, such as discussed in \cite{nelson_dynamics_1997}, \cite{nelson_stability_1999}, \cite{glover_simulating_2007}, \cite{glover_modelling_2010}, and \cite{glover_relationship_2011}. Modeling a robust chemical network in the ISM requires an algorithm which accounts for the bulk motion of the gas, in a simulation which itself was already coupled to a simpler time-dependent network. Works which have made steps in this direction include \cite{gnedin_modeling_2009,clark_how_2012,richings_effects_2016,seifried_modelling_2016,valdivia_h2_2016,seifried_modelling_2017,capelo_effect_2018,lupi_natural_2018,lupi_c_2020} and \cite{hu_metallicity_2021}.

We present here a novel, time-dependent chemical post-processing scheme, intended for the analysis of 3D magnetohydrodynamic (MHD) simulations. Although we apply the method here specifically to molecular clouds with masses of approximately $10^5~\textrm{M}_\odot$ over a timespan around 4~Myr, it could be further applied to any astrophysical simulation of arbitrary domain size or simulation timespan which includes passive tracer particles. We showcase the technique using a chemical network of 39 species and 301 reactions which is based on the network in \cite{grassi_detailed_2017}, but the chemical network could also be of arbitrary size, provided all the included species are modeled comprehensively. Our method uses \textsc{Krome} \citep{grassi_krome_2014} to return the time-dependent density of every species in the chemical network, down to the scale of the individual gas elements associated with the simulation’s tracer particles. We additionally propose an algorithm for recovering the density distribution over the entire domain for any species modeled in the post-processing network. 

We validate each step of our methodology by investigating the non-equilibrium evolution of the formyl cation \ce{HCO+}, a critical formation reactant of CO \citep{van_dishoeck_photodissociation_1988,nikolic_hco_2007,gerin_co_2021}. These species generally share a density regime, with HCO\textsuperscript{+} most effectively tracing slightly denser gas than CO does \citep[see e.g.][]{teague_chemistry_2015,barnes_lego_2020,yang_search_2021,Jacob-Hygal2022}. Because the presence of HCO\textsuperscript{+} can regulate the production of CO, a deeper understanding of HCO\textsuperscript{+} evolution is critical to refining models of CO. However, HCO\textsuperscript{+} is not generally modeled on-the-fly in simulations. Therefore it is an excellent choice to showcase the scientific value of non-equilibrium chemical post-processing.

This paper is organized as follows. In Section~\ref{section-simulations}, we summarize the initial conditions of the reference simulations, as well as the mechanics of their passive tracer particles and the limited chemical network which was coupled to the simulations. In Section~\ref{section-postprocessing}, we discuss our more robust chemical network and the post-processing methodology. We then analyse the time-dependent evolution of the post-processed tracer abundances in Section~\ref{section-tracerhistory}, with the motivating example of the evolution of the dense-gas tracer HCO\textsuperscript{+}. The algorithm for constructing a three-dimensional density grid from a snapshot of tracer abundances is explained in Section~\ref{section-regridding}, and we present HCO\textsuperscript{+} column density maps and compare them to observations. Some caveats and opportunities for future improvements are outlined in Section~\ref{section:discussion}. Finally, in Section~\ref{section-conclusion} we summarize our work and briefly discuss potential future applications of our methodology.

\section{The Simulations}
\label{section-simulations}

In this work, we apply our chemical post-processing methodology to four SILCC-Zoom simulations, part of the SILCC collaboration \citep{walch_silcc_2015,girichidis_silcc_2016,gatto_silcc_2017}. The two hydrodynamic (HD) simulations used here were first introduced in \cite{seifried_silcc-zoom_2017}, and later the MHD simulations in \cite{seifried_silcc-zoom_2020}, with modifications to their original form described by \cite{seifried_accuracy_2021}. We will summarize here the details of these simulations most salient to the post-processing.

\subsection{The reference simulations}
\label{subsection-reference-simulation}

The SILCC simulations model a part of a galactic disk with solar neighbourhood conditions using a stratified box centred on the galactic midplane. This box measures $500~\textrm{pc}\, \times \, 500~\textrm{pc}\, \times \, \pm 5~\textrm{kpc}$, with a starting resolution of 3.9~pc. Particular subvolumes are selected for their propensity to form molecular clouds. Once these regions reach a density of a few $10$~cm$^{-3}$, they are then re-simulated with a higher resolution. These higher-resolution subvolumes (hereafter `zoom-in regions') are simulated in tandem with the surrounding multi-phase ISM. 

The SILCC-Zoom simulations are run using the adaptive mesh refinement (AMR) code FLASH v. 4.3 \citep{fryxell_flash_2000,dubey_challenges_2008}. The zoom-in regions measure approximately 100~pc in each dimension, located within the broader SILCC domain. The full domain evolves for a startup time $t_{0}=11.9$~Myr for the HD clouds and $t_0=16.0$~Myr for the MHD clouds, after which the zoom-in process begins. The resolution inside the zoom-in region progressively increases to a maximal refinement of 0.06~pc over a total time of 1.65~Myr, to suppress the development of spurious turbulent grid artefacts. Outside the zoom-in regions, the broader simulation continues at the initial resolution. In this paper, whenever we refer to an elapsed time value for a particular snapshot of a molecular cloud simulation, we expressly mean the time $t_{\textrm{evol}}=t-t_{0}$, or the time since the beginning of the simulation’s zoom-in refinement.

The global parameters of the SILCC simulations are set to solar neighbourhood fiducial values as follows. At the galactic midplane, the bulk density is $\rho_{0} = 9 \times 10^{-24}$~g~cm$^{-3}$, in a Gaussian profile with a scale height of 30~pc transverse to the \textit{x-y} plane. The gas surface density is set to $\Sigma_{\textrm{gas}} = 10\, \textrm{M}_{\odot}\,\textrm{pc}^{-2}$. Supernovae throughout the simulation volume are triggered in the `mixed driving' configuration described in \cite{walch_silcc_2015} and \cite{girichidis_silcc_2016} until the zoom-in refinement begins at $t_{0}$, at which point the supernova driving is deactivated altogether. 

The cosmic ray ionization rate (CRIR) for molecular hydrogen is set to a constant value of $\zeta=6\times 10^{-17}$~s\textsuperscript{-1}, and the strength of the interstellar radiation field (ISRF) is set to the value of the Draine field \citep{draine_photoelectric_1978}, which is $G_{0}=1.7$ in Habing units \citep{habing_interstellar_1968}. Attenuation of the ISRF is calculated using the \textsc{TreeRay/OpticalDepth} module \citep{clark_how_2012,wunsch_tree-based_2018} with respect to the column densities of H\textsubscript{2}, CO, and the dust distribution. This routine assigns each cell a three-dimensional visual extinction $A_\rmn{V,3D}$ by measuring the total gas column density $N(\rmn{H_{tot}})$ along $n_\rmn{pix}$ equally-weighted sight lines \citep{gorski_2011}, and then calculating: 

\begin{equation}
    A_\rmn{V,3D} = \frac{-1}{\gamma}\textrm{ln} \left( \frac{1}{n_\rmn{pix}} \sum_{i=1}^{n_\rmn{pix}}\textrm{exp}\left(-\gamma A_\rmn{V,i}\right) \right),
\end{equation}
where each of the $n_\rmn{pix}$ unidirectional extinction magnitudes are given by $A_\rmn{V,i} = (5.348 \times 10^{-22})\times N (\rmn{H_{tot,i}})~\textrm{cm}^{2}$ \citep{draine_structure_1996}, $\gamma = 2.5$ \citep[][see also \citealt{glover_approximations_2012}]{bergin_molecular_2004}, and $n_\rmn{pix}=48$ in the simulations used in this paper. Attenuation due to dust at a cell is then given by $\textrm{exp}(-\gamma A_\rmn{V,3D})$. The self-shielding of H\textsubscript{2} and CO are analogously computed from those species’ respective column densities.

The simulations are coupled on-the-fly to the chemical network first presented in \cite{nelson_stability_1999} \citep[see also][]{glover_simulating_2007,glover_simulating_2007-1,glover_approximations_2012,gong_simple_2017}, which has been updated in accordance with \cite{mackey_non-equilibrium_2019}. This network (hereafter `NL99') contains only a few hydrogen and carbon species, and consolidates all metals (principally Si and Si\textsuperscript{+}, by abundance) into the neutral and ionized placeholders M and M\textsuperscript{+}. The full list of species in this network is provided in Appendix~\ref{appendix:networks_nl99}. The NL99 network’s primary objective is to calculate the abundances of CO, C\textsuperscript{+}, and O so that their heating and cooling contributions can inform the dynamical evolution of the gas. 

We applied our post-processing scheme to four SILCC-Zoom simulations, two each including and not including magnetic fields, which originated in separate magnetized and unmagnetized SILCC runs. The two purely hydrodynamic (HD) clouds were first explored in \cite{seifried_silcc-zoom_2017}, and updated in \cite{seifried_accuracy_2021} with a greater maximum refinement and the application of the chemical network derived from  \cite{mackey_non-equilibrium_2019} instead of a precursor network \citep{nelson_dynamics_1997}. We denote these two simulations \mbox{MC1-HD} and \mbox{MC2-HD}. The two magnetohydrodynamic (MHD) clouds, which have also been updated in \cite{seifried_accuracy_2021}, originated in \cite{seifried_silcc-zoom_2020}; we denote them \mbox{MC1-MHD} and \mbox{MC2-MHD}. Despite the similarities in their names, these are four separate simulations, evolved independently from the beginning of their runs. In the MHD simulations, the magnetic field was unidirectional and initialized at $B_{x} = B_{x,0} \sqrt{\rho(z)/\rho_{0}}$, where the midplane magnetic field $B_{x,0} = 3\, \mu \textrm{G}$, following observations \citep{beck_magnetic_2013}.

\subsection{Tracer particles}
\label{subsection-tracers}

The FLASH code simulates astrophysical domains using a volume-filling Eulerian grid. However, the code can also inject tracer particles into the simulation volume. These tracers are passive and massless, with no dynamical impact upon the gas. At each time step, they are advected with the local density flow according to the velocity field of the gas at their particular location. 

When the zoom-in refinement begins, we initialize the tracers in a uniform lattice with a spacing in each dimension of 1~pc. This tracer density comprehensively recovers the local chemical abundances from the simulation grid. In Appendix~\ref{appendix:tenpercent}, we show that even if we reduce the tracer count by a factor of ten, the tracers still accurately capture the chemical state of a simulated cloud in most gas density regimes, indicating that the tracer count resulting from a 1~pc lattice is sufficient. The lattice covers the entire zoom-in region and an additional zone extending 10~pc out from the zoom-in region in the $x$, $y$, and $z$ directions. The total number of tracers in the simulations is about $9\times10^5$ in both \mbox{MC1-HD} and \mbox{MC2-HD}, $2\times10^6$ in \mbox{MC1-MHD}, and $1.6\times10^6$ in \mbox{MC2-MHD}. The MHD zoom-in regions are slightly wider in each dimension than the HD regions (and therefore contain more tracers) because of their more diffuse distribution.

Every 3.3~kyr, a snapshot is taken of every tracer particle. This interval is a compromise between the available disk space and the necessity to capture the timescale on which relevant changes in MCs occur. The snapshot records each tracer's local values of the gas density, gas and dust temperature, self-shielding factors of H\textsubscript{2} and CO as well as $A_\rmn{V,3D}$, and the mass fractions of H, H\textsubscript{2}, C, CO, and C\textsuperscript{+}. These local readings proportionally represent those reported in a cell-sized box projected around the tracer particle, a system called Cloud-in-Cell interpolation. Because we simultaneously model the zoom-in region with the surrounding multi-phase ISM, some of the tracer particles move in or out of the zoom-in region over the simulation lifetime. Since the surrounding ISM is not well resolved, we restrict our analysis to particles which lie inside the zoom-in region at particular timesteps. 

Post-processing the abundances reported by the tracer particles unlocks the time-dependent chemistry of individual gas complexes in the molecular cloud simulations \citep{genel_following_2013,ferrada-chamorro_chemical_2021}. Prior works have shown the significant impact of turbulent mixing on the abundance distribution of H\textsubscript{2} \citep{glover_modelling_2010,valdivia_h2_2016,seifried_silcc-zoom_2017,ebagezio2022}. If we post-processed solely the simulation grids, we would be restricted to snapshots which do not preserve multi-timestep dynamical information, limiting us to equilibrium chemistry. Post-processing the chemical abundance evolution reported by the tracer particles gives a fuller picture of the simulated cloud's non-equilibrium chemical evolution than post-processing the AMR grids alone.

\section{Chemical Post-processing}
\label{section-postprocessing}

Our post-processing calculations are handled by the chemistry and microphysics package \textsc{Krome} \citep{grassi_krome_2014}. Rather than post-processing the full AMR grids, we apply the solver instead to the history files produced over the lifetime of each tracer particle. Because the tracers record the local values of the temperature, density, and radiation shielding, these time-dependent quantities can be used in \textsc{Krome} to solve the network of reaction rates for each tracer individually. The chemical abundances recorded by the tracer particles at $t_\rmn{evol}=0$ are used to initialize our post-processed chemistry network, but thereafter only the density, temperature, and shielding factors are used as inputs for each subsequent post-processing step.

\subsection{The chemical network}
\label{subsection-chemicalnetwork}

For post-processing the abundances reported by the tracers, we use a chemical network containing 301 reactions, 37 gas-phase species, and 2 species frozen on to dust grains. The network is derived from the \texttt{react\_COthin} network included with the \textsc{Krome} package and described in \cite{grassi_detailed_2017}. The network solves for not only the simplest hydrogen and carbon species which were present in the simulation's on-the-fly network (see Section \ref{subsection-reference-simulation}), but also for HCO\textsuperscript{+}, OH, CH, H\textsubscript{2}O, the cosmic ray tracer H\textsubscript{3}\textsuperscript{+}, neutral and ionized silicon, and more. The full list of 39 species is included in Appendix~\ref{appendix:networks_pp}. Our principal modification to the network comes in the addition of two species to represent CO and H\textsubscript{2}O which have frozen on to grains, as well as reactions to model their adsorption and desorption rates (see Appendix~\ref{appendix:networks_freeze} for details).

\begin{figure}
    \centering
    \includegraphics[width=\linewidth]{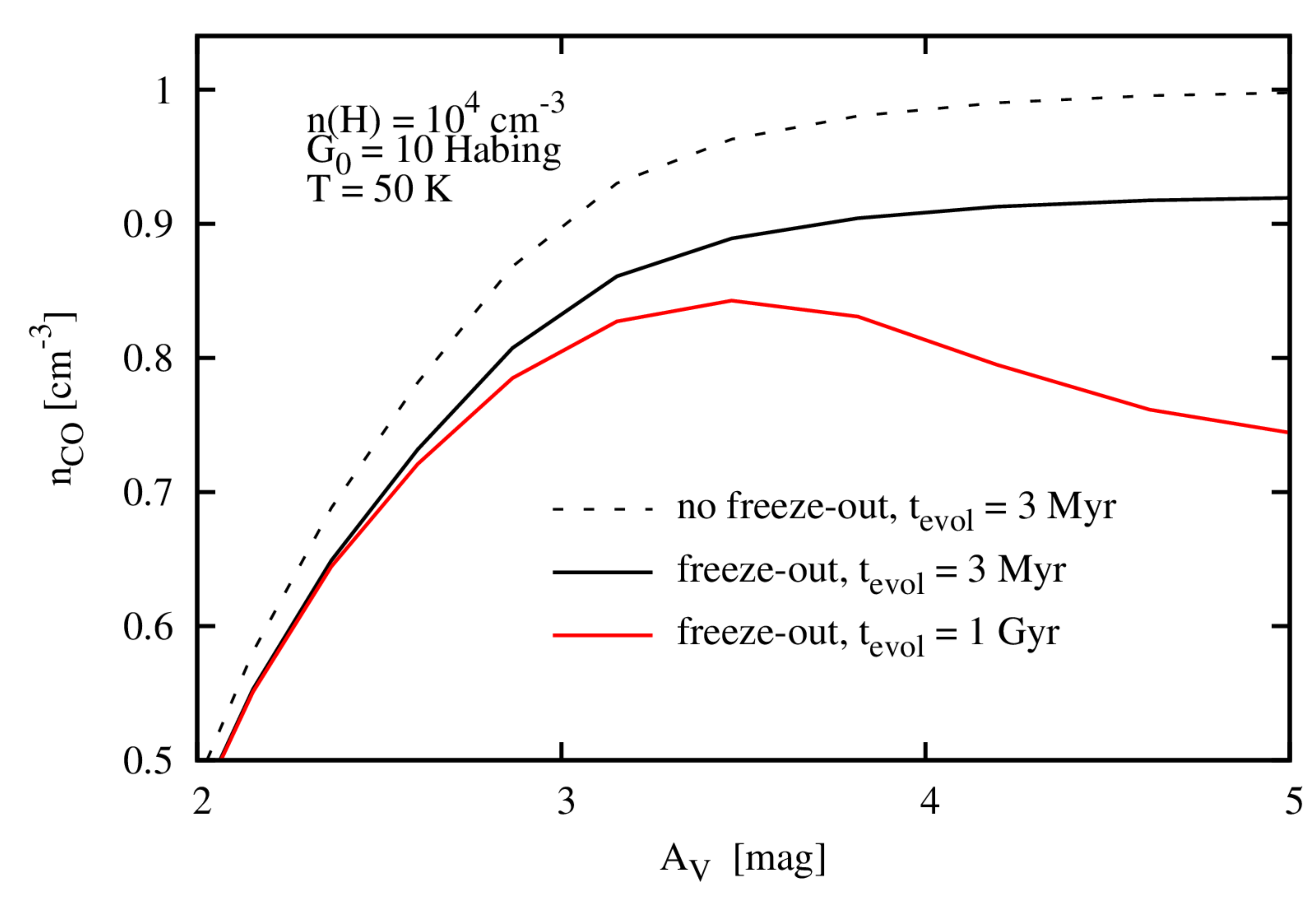}
    \caption{The carbon monoxide number density $n_\rmn{CO}$ vs. the visual extinction $A_\rmn{V}$ in a 1D-PDR setup, comparing our post-processing network with freeze-out reactions turned on (black solid line) and the same network with those reactions turned off (dashed). Freeze-out effects lead to a net decrease in $n_\rmn{CO}$ starting around $A_\rmn{V}=2$. By $A_\rmn{V}=5$, the run where freeze-out is enabled has a CO abundance about 10\% lower than the run where it is not. When the setup is run with freeze-out for $t_\rmn{evol}=1$~Gyr (red), $n_\rmn{CO}$ declines relative to the $t_\rmn{evol}=3$~Myr result starting around $A_\rmn{V}\simeq2.5$.}
    \label{fig:bench_freezeout}
\end{figure}

We show the importance of including these freeze-out effects in Fig.~\ref{fig:bench_freezeout}, where we plot the CO abundance vs. $A_\rmn{V}$ for our network (solid lines) and with the freeze-out reactions turned off (dashed). We employ a 1D-PDR setup in which $n(\rmn{H})=10^4$~cm$^{-3}$, \mbox{$G_0=10$} in Habing units, $T=50$~K, and the evolution time is \mbox{$t_\rmn{chem}=3$~Myr} (black). The inclusion of freeze-out starts to diminish the CO abundance around $A_\rmn{V}\simeq2$. By $A_\rmn{V}\simeq5$, the network with freeze-out has a CO abundance about ten percent lower than the network without freeze-out. We then repeat this test with freeze-out at $t_\rmn{chem}=1$~Gyr (red), finding a decline in $n_\rmn{CO}$ relative to the $t_\rmn{chem}=3$~Myr case beginning at $A_\rmn{V}\simeq2.5$. The different outcomes for $n_\rmn{CO}$ at different times underscore the importance of freeze-out effects to time-dependent chemistry.

\subsection{The post-processing procedure}
\label{subsection-ppprocedure}

We post-process the chemical history of each tracer particle individually, using the post-processed abundances of each particle snapshot as the input state of the subsequent post-processing step. We will show a procedure to weight these abundances relative to the local bulk density around the tracer particle, which changes as the particle advects throughout the simulation domain. Additionally, we employ a number of environmental parameters saved in the tracer histories as parameters in the chemical network’s reactions: the bulk density, the gas and dust temperatures, the 3D visual extinction $A_\rmn{V,3D}$, and the H\textsubscript{2} and CO self-shielding coefficients. When one of these environmental parameters $p$ changes between successive snapshots by more than a specified threshold percent change $s$ (in this work, 10\%), we subdivide the inter-snapshot timestep of 3.3~kyr into smaller pieces based on a linear interpolation scheme. This safeguards the rate equation solver from large environmental shifts which may produce inaccurate solutions. The steps of this algorithm are described below.

\subsubsection{Initialization}

The output of our post-processing is an array of chemical number densities covering every species included in the chemical network. This evolving abundance array must be initialized before the first post-processing step at $t_\rmn{evol}=0$. Each tracer particle history includes the on-the-fly values for the mass fractions of H, H\textsubscript{2}, C, CO, and C\textsuperscript{+} at each snapshot time (separated by 3.3~kyr). We read these mass fractions from the first particle snapshot in the history, along with the local bulk density. The mass fractions are then converted to number densities via their respective molecular weights, and are saved to the evolving abundance array as the initial values for these species. We derive an initial value for the density of ionized hydrogen by the conservation relation

\begin{equation}
n_{\textrm{H\textsuperscript{+}}}=n_{\textrm{H,tot}}-n_{\textrm{H}}-2n_{\textrm{H\textsubscript{2}}},
\end{equation}
since the tracer particles did not record this density directly.

Additionally, we set the initial number densities for helium, carbon, and oxygen to $n_{\textrm{He}}=0.1 n_{\textrm{H,tot}}$, $n_{\textrm{C}}=(1.4 \times 10^{-4})n_\rmn{H,tot}$, and $n_{\textrm{O}}=(3.2 \times 10^{-4})n_\rmn{H,tot}$, following the abundances given in \cite{sembach2000}. All other species in the evolving abundance array are initialized to number densities of $10^{-20}$~cm$^{-3}$. We find that the densities of these other species converge to reasonable values within a few post-processing timesteps, regardless of their initial value. In general, these arrays are established separately for every tracer particle in the simulation.

\subsubsection{Iterating the post-processing}

Fundamentally, our post-processing methodology tries to reach the most accurate chemical solution by advancing in incremental steps, rather than evolving the abundances over long time-scales to equilibrium. This is facilitated by always iterating the \textsc{Krome} solver by a time less than or equal to the time between two successive tracer snapshot times $t_1$ and $t_2$, which are separated by 3.3~kyr. The decision of whether to advance the chemistry by less than $t_{2}-t_{1}$, and if so, what fraction of that time to advance by instead, is made in the following way.

At $t_1$, the procedure reads in the tracer particle’s saved values for $A_\rmn{V,3D}$, the bulk density, the gas and dust temperatures, and the self-shielding coefficients of H\textsubscript{2} and CO. These parameters are required to solve various reaction rates in the chemical network. We refer to an arbitrary member of this set of six environmental parameters at $t=t_1$ by $p_1$. The parameter values for the subsequent timestep $t_2$ are then read as well. 

Next, the code checks whether any of the parameters $p_1$ undergo a percent change exceeding some user-defined value $s$, that is:

\begin{equation}
    \label{equation-subcycling-thresholds}
    p_{2} > (1+s)p_{1} \\ \textrm{or} \\ p_2 < (1-s)p_1,
\end{equation}
if $p_2>p_1$ or $p_2<p_1$ respectively. These are the subcycling threshold conditions. If neither condition is fulfilled (that is, if none of $p_1$ experienced a percent change from $t_1$ to $t_2$ greater than $s$), the post-processing works as follows. 

The \textsc{Krome} solver is passed the set of species number densities $n_{i,1}$ in its current state at $t_1$, along with the values of $p_1$. The solver then advances the chemistry for a time $\Delta t = t_2 - t_1$, keeping the parameters $p_1$ fixed, and reaches solutions for the number densities $n_{i,2}^{*}$. Then $n_{i,2}^{*}$ is multiplied by the inter-snapshot weight $W_{\Delta t}$, which is defined here as the ratio of the bulk densities $\rho_2$ and $\rho_1$ at $t_2$ and $t_1$, respectively, such that each species number density $n_i$ at $t_2$ is now given by:

\begin{equation}
    n_{i,2}=n_{i,2}^{*}W_{\Delta t}=n_{i,2}^* \frac{\rho_2}{\rho_1},
\end{equation}
where $n_{i,2}$ is the set of weighted species number densities at $t_2$. This weighting accounts for the motion of the tracer particle through regions of different density, corresponding to compression or rarefaction of the corresponding fluid element between $t_1$ and $t_2$. After this, the post-processing is triggered anew for the timestep $t_2$, and so forth, until the entire tracer history has been post-processed. 

\subsubsection{Subcycling}

If, however, either of the aforementioned threshold conditions in Eq.~\ref{equation-subcycling-thresholds} is fulfilled between two snapshots in a particle's history, a subcycling procedure is applied. This determines a smaller time over which to advance the chemistry $\Delta t< t_2 - t_1$, limiting the permissible amount of change in the parameters $p$ given that they are held constant over $\Delta t$ when solving the chemistry.

\begin{enumerate}
    \item In a first step, the code performs a linear interpolation for the values of the six environmental parameters $p$ between $t_1$ and $t_2$, and calculates the time it would take each $p$ to undergo a relative change of exactly $s$. The smallest of these times is selected as the initial subcycle time $\Delta t_\rmn{sub,1}$. 
\item Next, for any additional necessary subcycling step $j$ we determine the timestep as
\begin{equation}
    \Delta t_{\rmn{sub,}j}=(1+s)\Delta t_{\rmn{sub,}j-1} \, .
\label{eq:dtsubcycle}    
\end{equation}
For the very first step (\mbox{$j = 1$}) we use $\Delta t_\rmn{sub,1}$ (see below for an explanation of progressive increase).
\item We then advance the chemistry by the timestep $\Delta t_{\rmn{sub,}j}$ from the current starting time of the subcycling step, $t_{\rmn{current},j}$ (i.e. $t_1$ for the first subcycling step, \mbox{$j = 1$}), to \mbox{$t_{\rmn{current},j+1} = t_{\rmn{current},j} + \Delta t_{\rmn{sub,}j}$} using the values of the parameters $p$ at $t_{\rmn{current},j}$.
\item Next, using our linear interpolation from step (i), we calculate the values of the environmental parameters $p$ at \mbox{$t_{\rmn{current},j+1}$}, to be used for the next subcycling timestep.
\item The abundance array $n\textsuperscript{*}$ obtained from step (iii) is multiplied by the timestep weight:
\begin{equation}
W_{\Delta t}=\frac{\rho_{t_{\rmn{current},j+1}}}{\rho_{t_{\rmn{current},j}}} \, .
\end{equation}
This provides the properly weighted abundances for the next subcycling step.
\end{enumerate}
We repeat steps (ii) to (v) until the next proper snapshot at $t_2$ is reached. For the very last subcycling timestep we take as the timestep the difference between $t_2$ and the end-time of the previous subcycle $t_\rmn{current, 2nd-last}$, i.e. $\Delta t_\rmn{sub, last} = t_2 - t_\rmn{current, 2nd-last}$ to assure that we end up exactly at $t_2$.

We note that we apply the iterative increase in the subcycling timestep (Eq.~\ref{eq:dtsubcycle}) to avoid the following edge case. If one of the environmental parameters undergoes a particularly large change between timesteps $t_1$ and $t_2$, the linear interpolation described in step (i) will establish a particularly small $\Delta t_\rmn{sub,1}$. If we were to advance the chemistry by only $\Delta t_\rmn{sub,1}$ until $t_2$ is reached, and $\Delta t_\rmn{sub,1}$ is very small, the subcycling procedure may impose as many as dozens of additional substeps before $t_2$. In seeking a balance between small substeps which can properly shepherd the chemistry solver through rapid environmental changes, and the additional computational demands of solving the chemistry so many extra times, we found that iteratively increasing the substep size by factors of $(1+s)$ is a viable compromise. 

\begin{figure}
    \centering
    \includegraphics[width=\columnwidth]{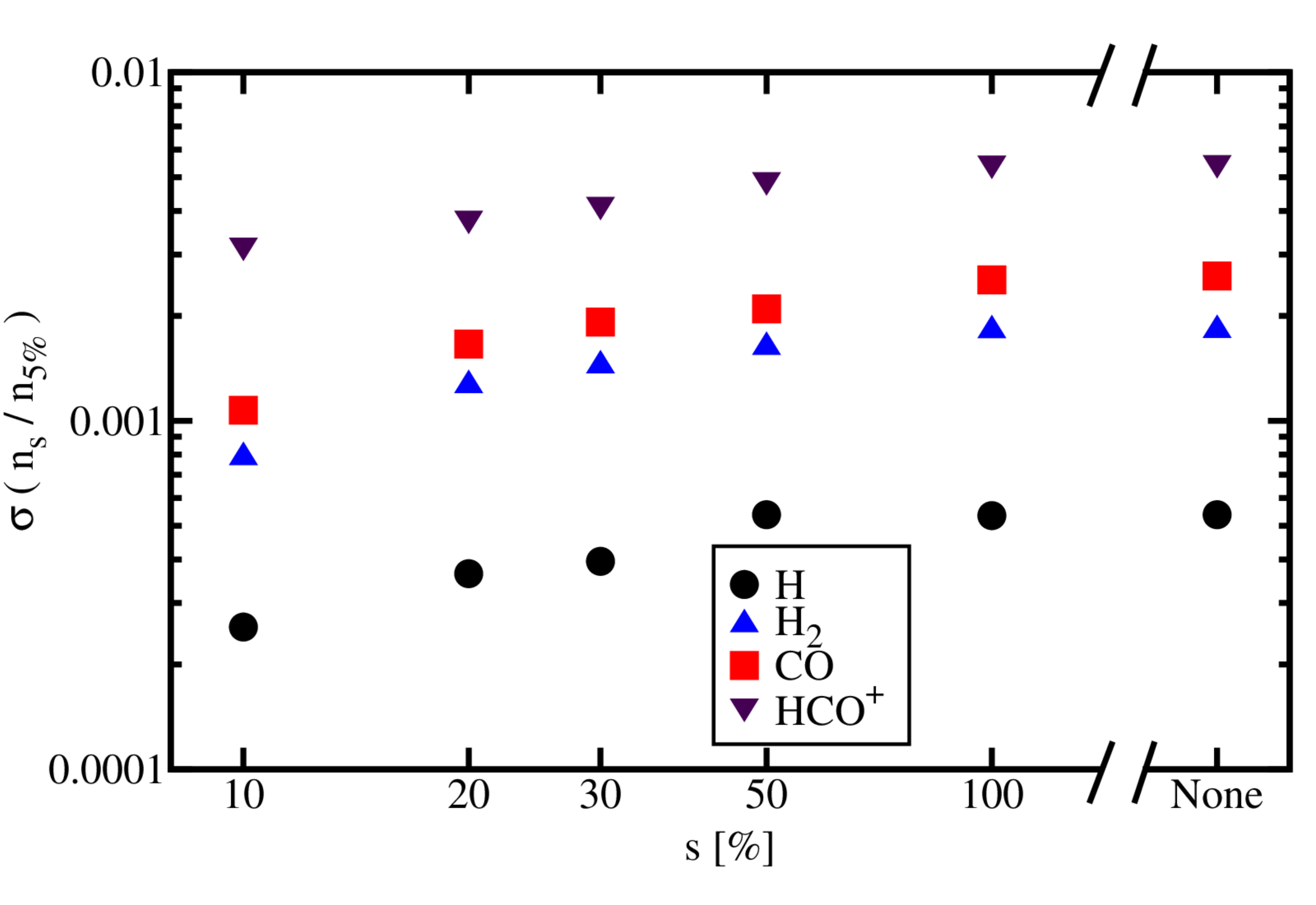}
    \caption{The standard deviations $\sigma(n_{s}/n_{5\%})$ of the distributions of the ratios of the number densities of H, H\textsubscript{2}, CO, and HCO\textsuperscript{+}, measured as a function of the selected subcycling threshold $s$, with respect to the results obtained for a threshold of $s=5$~\% (see Eq.~\ref{eq:sigmafraction}). Smaller subcycling thresholds correspond to smaller values of $\sigma(n_{s}/n_{5\%})$, but these standard deviations are all small or negligible. We therefore can select our subcycling threshold as $s=10$~\% with confidence.}
    \label{fig:sub-stds}
\end{figure}

For the results presented in this paper, we use a subcycling threshold value $s=0.1$, meaning a change of more than 10\% in any environmental parameter would trigger subcycling. To validate this number, we post-process about 40,000 randomly selected tracer particles in \mbox{MC1-HD} (5\% of its total particle population) with different values for the threshold: 5, 10, 20, 30, 50, and 100\%. Additionally, we perform a test with subcycling deactivated, so that the solver would always iterate for $\Delta t=t_2 - t_1$ regardless of any changes in the environmental parameters between timesteps.

We concatenate the chemistry results from each test run with different values of $s$ into snapshots at $t_\rmn{evol}=2.5~\textrm{Myr}$. For each value of $s$, we denote the number densities of H, H\textsubscript{2}, CO, and HCO\textsuperscript{+} as $n_s$. Next, for each particle we calculate the ratio $n_{s}/n_{5\%}$ and the standard deviation of this ratio for each species:

\begin{equation}
    \label{eq:sigmafraction}
    \sigma(n_{s}/n_{5\%}) = \left( \sum_{i}^{N}\frac{n_{i,s}}{n_{i,5\%}}-\langle \frac{n_s}{n_{5\%}}\rangle\right)^{\frac{1}{2}} N^{-\frac{1}{2}},
\end{equation}
where $N$ is the number of tracers in the set and $\langle n_s/n_{5\%}\rangle$ is the mean of all particles' values for $n_{s}/n_{5\%}$. These standard deviations for each species and value for $s$ are shown in Fig.~\ref{fig:sub-stds}. 
In general, the standard deviations are extremely small, on the order of $0.01$ when considering the number densities of H, \ce{H2}, CO, and \ce{HCO+}. The values of $\sigma(n_{s}/n_{5\%})$ decrease with decreasing $s$. Even for the run in which subcycling is deactivated, the variation is marginal. Therefore, we feel secure in selecting a subcycling threshold of $s=0.1$ as sufficient for post-processing.

\subsection{Comparing on-the-fly and post-processed abundances}
\label{subsection-ppvalidation}

\begin{figure*}
    \centering
    \includegraphics[width=\linewidth]{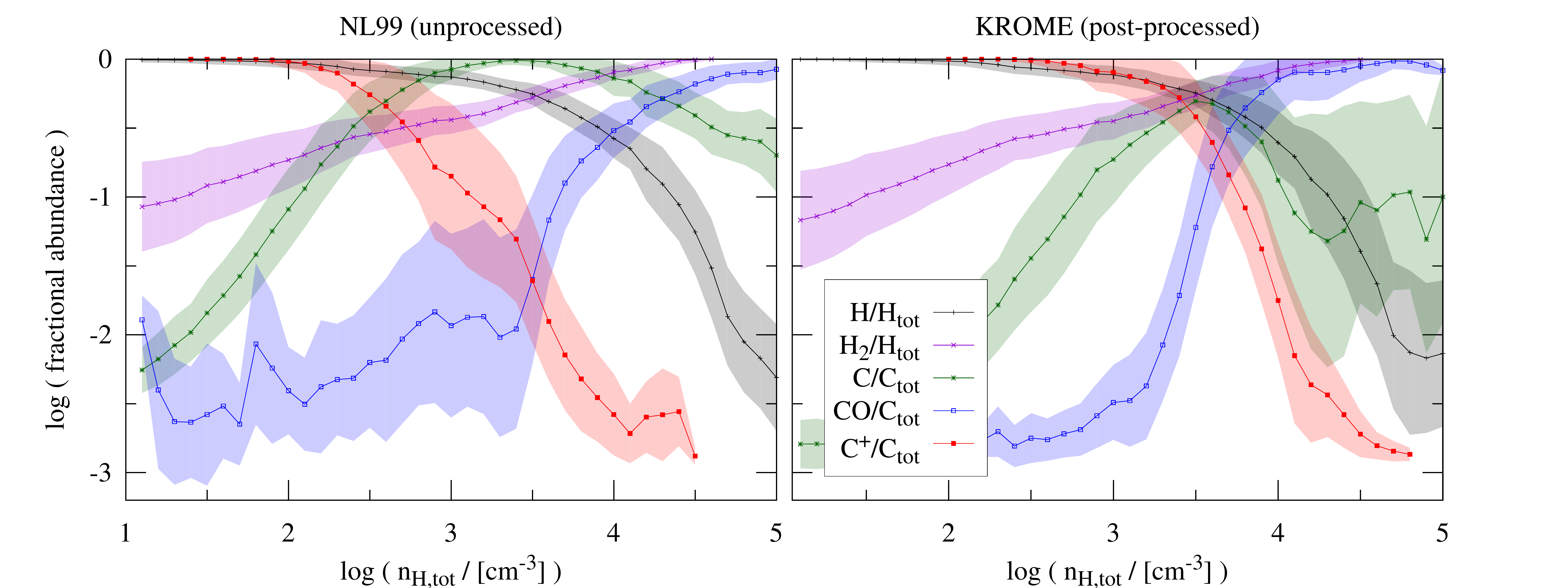}
    \caption{The average fractional abundances of H, H\textsubscript{2}, C, CO, and C\textsuperscript{+} vs. the total hydrogen number density $n_{\textrm{H,tot}}$ of the molecular cloud \mbox{MC1-HD} at $t_\rmn{evol}=2$~Myr, before post-processing (left) and after (right). The shaded areas represent one standard deviation around each species average. The abundance profiles of H (black) and H\textsubscript{2} (purple) do not change appreciably after post-processing. Post-processing reduces the saturation of atomic carbon (green) between $n_{\textrm{H,tot}}=10^3$--$10^4$~cm$^{-3}$. The bulk density at which the abundance of C\textsuperscript{+} (red) begins to decline is a magnitude higher after post-processing. CO (blue) becomes the dominant carbon species at a slightly lower bulk density after post-processing, primarily due to the reduction in atomic carbon.}
    \label{fig:fracvn-particles}
\end{figure*}

It can be instructive to compare the on-the-fly and post-processed abundances for certain hydrogen and carbon species which are present in the original simulations. In Fig.~\ref{fig:fracvn-particles}, we plot the average fractional abundances of H, H\textsubscript{2}, C, CO, and C\textsuperscript{+} for the tracer particles before post-processing (left) and after (right), for the cloud \mbox{MC1-HD} at $t_\rmn{evol} = 2$~Myr, as a function of $n_{\textrm{H,tot}}$. The ratios of the post-processed mean fractional abundances to the unprocessed mean fractional abundances (i.e., the ratios of the abundances in the right and left panels of Fig.~\ref{fig:fracvn-particles}) are plotted in Appendix~\ref{appendix:fracvn_ratios}, in the left panel of Fig.~\ref{fig:fracvn-ratios}.

The post-processed abundances for these species are broadly similar to their original abundances. In particular, the fractional abundance profiles of H (black) and H\textsubscript{2} (purple) with respect to $n_{\textrm{H,tot}}$ are almost unchanged. Atomic hydrogen dominates at lower densities, and gradually diminishes in abundance as the density increases. At $n_{\textrm{H,tot}} \simeq 3 \times 10^{3}$~cm$^{-3}$, molecular hydrogen becomes the predominant hydrogen species. Atomic hydrogen continues to decline as H\textsubscript{2} nears saturation. Saturation occurs by $n_{\textrm{H,tot}} \simeq 3 \times 10^{4}$~cm$^{-3}$ both before and after post-processing.

The fractional abundances of C, CO, and C\textsuperscript{+} (shown with respect to the summed carbon density from those three species, rather than with respect to total hydrogen) undergo some changes during the post-processing. In the on-the-fly results, atomic carbon (green) dominates around $n_{\textrm{H,tot}}=3 \times 10^{3}$~cm$^{-3}$ before declining as CO (blue) predominates. However, in the post-processed results, the peak fractional abundance of atomic carbon occurs at the same $n_{\textrm{H,tot}}$ but only reaches 0.5. Consequently, C$^+$ and CO reach higher fractional abundances in this density range for the post-processed results. We attribute this outcome to a well-known problem in the on-the-fly NL99 network, in which atomic carbon is overproduced. This problem has been discussed in a number of works \citep{glover_approximations_2012,gong_simple_2017,hu_metallicity_2021}. A solution remains elusive, but is probably linked to the limited size of the network. In any case, as this problem appears to be alleviated by post-processing with our more extensive network, we are confident that it does not affect the analysis presented in this work. 

Finally, we note that although the post-processing network includes additional carbon species (in particular, frozen-out CO), their abundances are low enough at most densities that their contribution to the total carbon can be neglected for this analysis. Around $n_\rmn{H,tot}$, the abundance of CO dips slightly, which we ascribe to an increase in the proportion of CO which has frozen onto dust at this density.

\section{The formation of HCO\textsuperscript{+}}
\label{section-tracerhistory}

By post-processing the entire chemical histories reported by the tracers, we can now examine the full time-dependent evolution of interesting species rather than merely their equilibrium abundances. Because the tracers advect passively with the gas flow, they recount the full history of localized patches of gas. Analysing their bulk motion in conjunction with the post-processed chemistry can give us, for the first time, dynamical information about species only present in more extensive chemical networks than are run on-the-fly. In particular, we can analyse the formation rate, peak production density regime, and predominant creation pathways of interesting species. 

\subsection{HCO\textsuperscript{+} evolution}
\label{subsection-hic}

As a first scientific application, we consider the temporal and dynamical evolution of the HCO\textsuperscript{+} content in both hydrodynamic and magnetohydrodynamic molecular clouds. This species is present in some of the principal formation pathways for CO. Understanding the evolution of the HCO\textsuperscript{+} abundance can therefore provide time-dependent information about the CO content of molecular clouds and answer questions about how these species are related \citep{van_dishoeck_photodissociation_1988,nikolic_hco_2007,papadopoulos_hcn_2007,gerin_co_2021}. 

\begin{figure*}
    \centering
    \includegraphics[width=\linewidth]{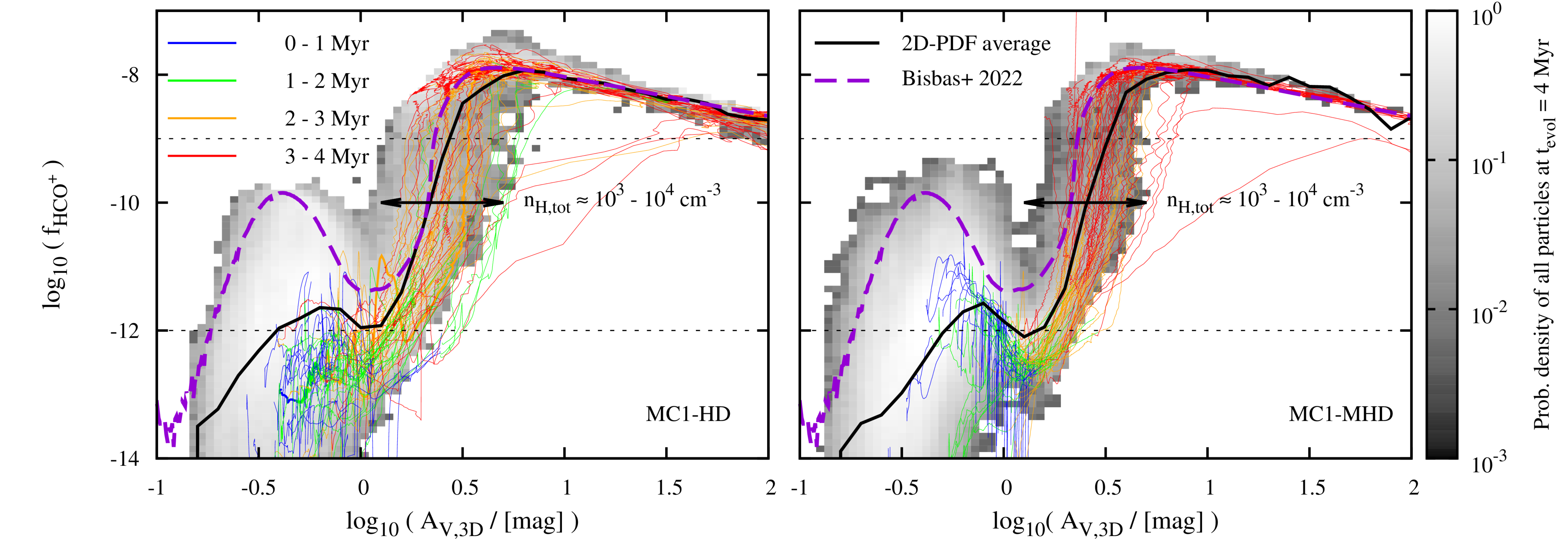}
    \caption{History tracks of a selection of 50 tracer particles that rise from a fractional abundance of $10^{-12}$ to $10^{-9}$ (indicated by horizontal black dashed lines) at some point during their lifetime, for \mbox{MC1-HD} (left) and \mbox{MC1-MHD} (right). The history tracks are colored according to the age of the particle at the time. Plotted beneath the tracks are 2D-PDFs of the HCO\textsuperscript{+} fractional abundance vs. visual extinction $A_\rmn{V,3D}$ for all tracer particles in each simulation at $t_\rmn{evol}=4$~Myr, weighted by their initial density at $t_0$. The averages of these PDFs are also provided (solid black lines). The approximate $A_\rmn{V,3D}$ range where $n_\rmn{H,tot}=10^3$--$10^4$~cm$^{-3}$ is indicated by a double-headed arrow. The $f_\rmn{\ce{HCO+}}$ obtained from running a PDF$_\rmn{CHEM}$ model \citep{bisbas_proceedings_2022} is plotted over the distribution (dashed purple line) to illustrate that the post-processed tracers comprehensively cover the extinction domain in which \ce{HCO+} is prevalent. Both molecular clouds contain similar distributions of tracer particles in this phase space. There is a sharp jump in fractional abundance between total hydrogen densities of $10^3$ and $10^4$~cm$^{-3}$, where the tracer particles display local, in situ HCO\textsuperscript{+} production rather than turbulent mixing of HCO\textsuperscript{+} from deeper (i.e. from higher $A_\rmn{V,3D}$) in the clouds. For both clouds, the selected tracers move from the lower to the upper threshold in a time-scale of about 1~Myr, corresponding to the typical HCO\textsuperscript{+} formation time. For the cloud \mbox{MC1-HD}, this growth mostly occurs some 2-3~Myr after the start of the zoom-in refinement, but for the cloud \mbox{MC1-MHD}, somewhat later, 3-4~Myr after the start of the refinement.}
    \label{fig:hic-both}
\end{figure*}

In the background of Fig.~\ref{fig:hic-both}, we plot a 2D-PDF of the fractional abundance of \ce{HCO^+}, $f_\rmn{HCO^+}$, vs. visual extinction $A_\rmn{V,3D}$ for every tracer in the clouds \mbox{MC1-HD} (left) and \mbox{MC1-MHD} (right), at $t_\rmn{evol}=4$~Myr, weighted by the initial density of each tracer at $t_0$. The average of this 2D-PDF is shown as well (solid black line). We see that the peak of $f_\rmn{HCO^+}$ for both clouds at this time is found around $\log A_\rmn{V,3D}\simeq0.5$. Because we post-process the entire history of every tracer in the clouds, we can investigate the time-dependent evolution of the particular tracers which achieve this peak $f_\rmn{HCO^+}$ value. In particular, how do these particular tracers' values of $A_\rmn{V,3D}$ change as they experience HCO\textsuperscript{+} formation, and what is the time-scale of that formation?

To answer these questions, we must first select an appropriate subset of tracer particles. We establish an HCO\textsuperscript{+} growth time-scale $\tau$, defined as the time it takes a tracer particle to rise from just below an HCO\textsuperscript{+} fractional abundance of $10^{-12}$ to just above an abundance of $10^{-9}$:
\begin{equation}
    \tau=t\left(f_\rmn{HCO^+}=10^{-9}\right)-t\left(f_\rmn{HCO^+}=10^{-12}\right) \, .
    \label{eq:tau}
\end{equation}
The upper threshold of $10^{-9}$ is chosen as it is approximately the minimum fractional abundance of \ce{HCO+} reported by any tracer with $A_\rmn{V,3D}>3$, which is roughly the $A_\rmn{V,3D}$-threshold in our post-processed chemistry above which high values of $f_\rmn{HCO^+}$ begin to appear \citep[Fig.~\ref{fig:hic-both}, and see also][]{lucasandliszt96}. The tracers which fulfill the `growth condition' of Eq.~\ref{eq:tau} and thus have a value for $\tau$ can be analysed as an ensemble. For the cloud \mbox{MC1-HD}, this comprises 27130 particles (3.1\%), and for the cloud \mbox{MC1-MHD}, only 6420 (0.3\%). The factor of ten smaller percentage of particles in \mbox{MC1-MHD} which fulfill our growth condition reflects the more diffuse nature of MHD clouds compared to pure HD ones \citep{seifried_silcc-zoom_2020}. Although the high-density regions of the HD and MHD clouds have similar properties \citep[in contrast to their differently-distributed envelopes; see][]{ganguly2022}, a smaller fraction of the MHD cloud’s gas reaches a sufficient density for HCO\textsuperscript{+} formation, with a corresponding smaller number of tracer particles in dense gas.

We randomly select 50 particles in each cloud which fulfill the growth condition of Eq.~\ref{eq:tau}, and plot the entire history of their values in the foreground of Fig.~\ref{fig:hic-both}. These history trajectories are each split into four segments, color-coded according to their values for $t_\rmn{evol}$. For both clouds, the selected tracers linger at values for $f_\rmn{HCO\textsuperscript{+}}$ below the lower threshold of $10^{-12}$ until around $t_\rmn{evol}=2$~Myr (the blue and green segments). Then, the tracers ascend over the course of about 1~Myr (the orange and red segments) beyond the upper threshold of $10^{-9}$ without significant change to their visual extinction of about $A_\rmn{V}\simeq1$--$3$. After this, the gradual and ongoing gravitational contraction of these dense regions guides many of the tracers into higher density regions where $A_\rmn{V}\gtrsim5$. However, this is after the time that the peak $f_\rmn{HCO\textsuperscript{+}}$ has been attained. By repeating this plotting procedure for both molecular clouds several times with different random batches of particles that fulfill the growth condition, we find this pattern is consistent.

In the figure, we additionally plot the result of the PDF$_\rmn{CHEM}$ model (dashed purple line) from \cite{bisbas_proceedings_2022}. This model corresponds well to the average of the tracer distribution, especially at $\log A_\rmn{V,3D}\gtrsim0.5$, validating the post-processed tracer chemistry in the extinction regime where most \ce{HCO+} is found. We can also correlate the range $\log A_\rmn{V,3D}\simeq0$--$0.5$, in which the tracers experience most of their HCO\textsuperscript{+} growth, with the total gas density. By comparing the values for $n_\rmn{H,tot}$ and $A_\rmn{V,3D}$ reported by the tracers at different times, we can establish a rough correspondence between these two values. We plot black arrows in Fig.~\ref{fig:hic-both} to indicate the approximate $A_\rmn{V,3D}$ range in each cloud where $n_\rmn{H,tot}$ ranges from $10^3$ to $10^4$~cm$^{-3}$. This result is in accordance with figure~12 in \cite{seifried_silcc-zoom_2017}.

Between $A_\rmn{V,3D} \simeq 5$ and the densest regions of the clouds where $A_\rmn{V,3D}\simeq 100$, $f_\rmn{HCO^+}$ declines by approximately one order of magnitude in both \mbox{MC1-HD} and \mbox{MC1-MHD}. We posit that this results from a corresponding decline of one magnitude in $f_\rmn{H_3^+}$ which we see over this extinction range, stemming from the constant number density of H\textsubscript{3}\textsuperscript{+} in dense molecular clouds even as $n_\rmn{H,tot}$ increases \citep[see e.g.][]{Oka2006,lepetit2016}. In Section~\ref{subsection:formationpaths}, we will show that in our post-processing network at very high $A_\rmn{V,3D}$, the primary formation pathway for HCO\textsuperscript{+} is the reaction \mbox{H\textsubscript{3}\textsuperscript{+} + CO}. A decline in $f_\rmn{H_3^+}$ would lead to a bottleneck in that reaction and therefore a matching decline in $f_\rmn{HCO^+}$, as seen in Fig.~\ref{fig:hic-both}. The possible impact of our constant value for the CRIR on the \ce{H3+} abundance is discussed in Section~\ref{section:discussion}.

\subsection{Turbulent mixing}

Recent works have discussed the role of turbulent mixing in distributing molecules like \ce{H2} throughout molecular clouds from dense sites of peak formation \citep[][and see also \citealt{glover_modelling_2010}]{seifried_silcc-zoom_2017}. For instance, \citet{valdivia_origin_2017} \citep[see also][]{valdivia_chemistry_2016} and \citet{godard2023} find that the presence of \ce{CH+} in the diffuse ISM stems from advection and thermal instability in denser gas. It is natural to ask whether \ce{HCO+} (which can be formed via \ce{CH+} among other things; see next section and Table~\ref{tab:hcoj_reactions}) is produced in dense gas and distributed around molecular clouds in the same fashion. By comparing each tracer's time-dependent \ce{HCO+} abundance to its local density, we can determine the dynamical backstory of gas which has a high \ce{HCO+} abundance at late times. We note that the tracers which fulfill the condition of Eq.~\ref{eq:tau} (of which the tracks plotted in Fig.~\ref{fig:hic-both} are a representative sample) do not exhibit this mixing action.

In general, for all clouds, about 90\% of the tracers which fulfill the growth condition experience HCO\textsuperscript{+} formation in situ around $n_{\textrm{H,tot}}\lesssim 10^4$~cm$^{-3}$ ($\log A_\rmn{V,3D}\lesssim0.7$). Of these, $\sim15\%$ are subsequently mixed into higher extinctions ($\log A_\rmn{V,3D}\gtrsim0.7$) and then back out again to the lower extinctions, where the in situ \ce{HCO+} formation first occurred. The remaining $\sim10\%$ of the high-\ce{HCO+} tracers first exceed $f_\rmn{HCO^+}=10^{-9}$ while having $\log A_\rmn{V,3D}\gtrsim0.7$ (represented by the few tracks moving diagonally to the upper right in Fig.~\ref{fig:hic-both}), and practically all ($\sim99\%$) of these subsequently remain in the high density regime, rather than mix back out to regions of lower density. Overall, we thus find a very small mixing efficiency for HCO$^+$ within molecular clouds. We note, however, that this may represent a lower limit on the actual mixing fraction, due to the propensity of tracer particles to become trapped at density peaks despite the presence of gas outflows \citep[see e.g.][]{price10,konstandin12,genel_following_2013,cadiou19}.

\subsection{The HCO\textsuperscript{+} formation time-scale}
\label{subsection-growthtimes}

The preceding analysis has shown that HCO\textsuperscript{+} is preferentially formed in the density regime $n_{\textrm{H,tot}}\simeq 10^{3}$--$10^4$~cm$^{-3}$ with a time-scale on the order of 1~Myr. We will analyse this formation time-scale in greater detail, in particular how it correlates with the abundances of various reactants in HCO\textsuperscript{+} production pathways. 

\begin{figure}
    \centering
    \includegraphics[width=\linewidth]{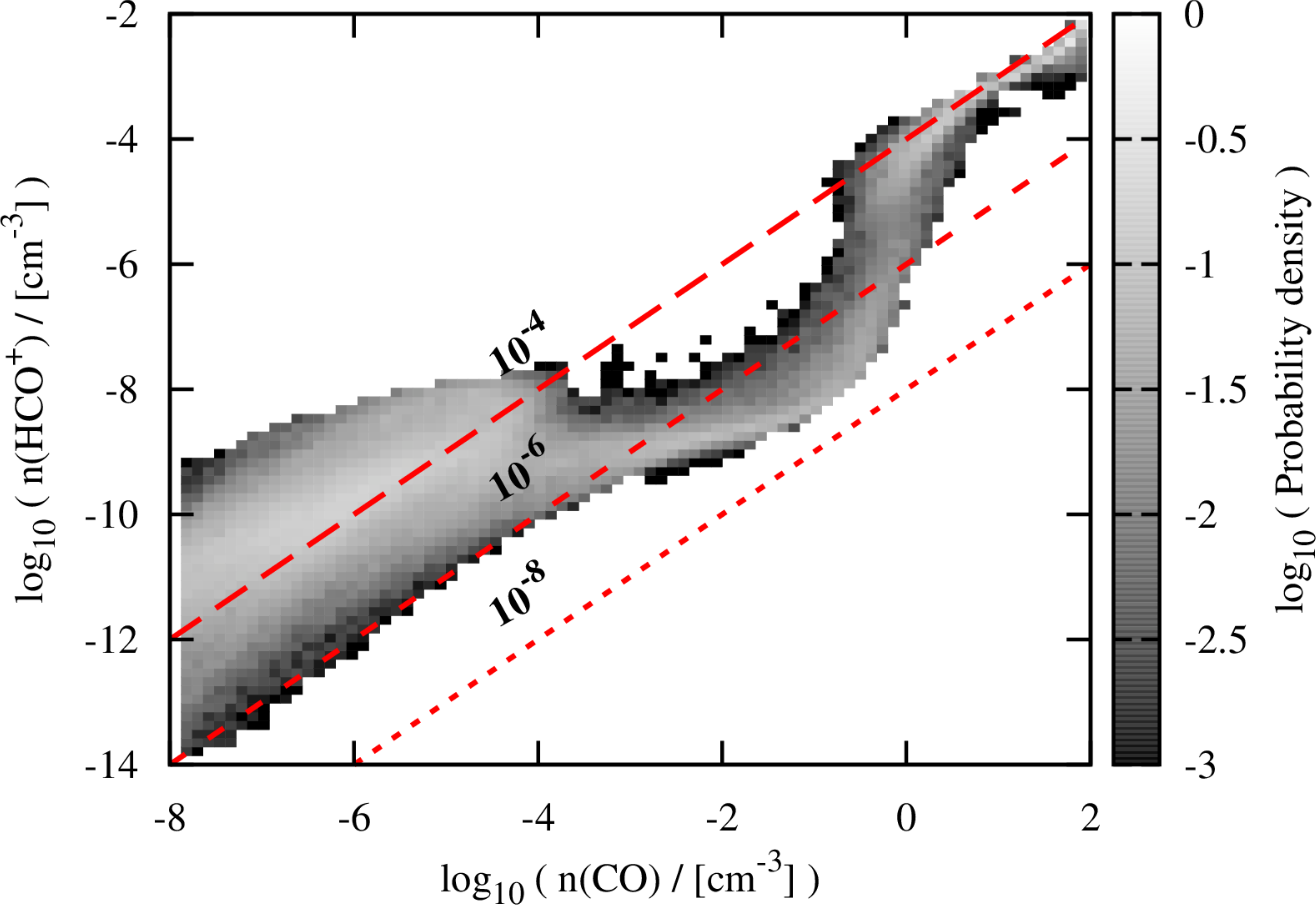}
    \caption{The number density of HCO\textsuperscript{+} vs. the number density of CO, for all post-processed tracer particles in the zoom-in region of cloud \mbox{MC1-HD} at $t_\rmn{evol}=4$~Myr, in the form of a 2D-PDF. Dashed lines indicate different constant ratios. There is a close, but highly non-linear, relationship between the two species.}
    \label{fig:hcojvsco}
\end{figure}

At a given time, if the densities of HCO\textsuperscript{+} and one of its formation reactants are well-correlated, we might posit that that particular reaction is an important source of HCO\textsuperscript{+}. For instance, in Fig.~\ref{fig:hcojvsco}, we compare $n$\textsubscript{HCO\textsuperscript{+}} and $n$\textsubscript{CO} in a 2D-PDF of cloud \mbox{MC1-HD} at $t_\rmn{evol}=4$~Myr. Several lines of constant ratio are included. The two species have a tight, non-linear relationship across a large range of densities. But although the correlation of HCO\textsuperscript{+} with CO is evident, the causal direction of this relationship is not, nor can we see how the relationship changes in time. CO is present in both production and destruction reactions of HCO\textsuperscript{+}, so we must adopt a time-dependent perspective to truly assess the relationship between CO and the growth of \ce{HCO+}.

Our chemical network includes 11 reactions which form HCO\textsuperscript{+}, listed in Table~\ref{tab:hcoj_reactions}. We now go beyond the simple comparison of $n_\rmn{HCO\textsuperscript{+}}$ with other $n_\rmn{i}$ of the various reactants (O, \ce{CO+}, \ce{H3+}, \ce{CH3+}, \ce{HOC+}, \ce{H2O}, CH, and \ce{C+}), and consider the direct dependence of the formation time-scale of HCO\textsuperscript{+} on these $n_\rmn{i}$. For this purpose, we again select the tracer particles which report over some period following $t_\rmn{evol}=0$ that their HCO\textsuperscript{+} fractional abundance ascended from below $10^{-12}$ to above $10^{-9}$, the time-scale of which we defined as $\tau$ in Eq.~\ref{eq:tau}. Over the period containing $N$ particle snapshots between these two thresholds, we calculate the logarithmic-average number density of each reactant, $\langle \log n_i \rangle$, in the following way: 

\begin{equation}
    \label{eq:logni}
    \langle \log n_i \rangle = \frac{1}{N}\sum_{j=1}^{N} \log n_{i,j} .
\end{equation}

\begin{table}
 \caption{List of the 11 reactions in our chemical network which produce HCO\textsuperscript{+}.}
\begin{center}
\begin{tabular}{ l l } 
 \hline
 1.& HOC\textsuperscript{+} + H\textsubscript{2} $\rightarrow$ HCO\textsuperscript{+} + H\textsubscript{2} \\
 2.& HOC\textsuperscript{+} + CO $\rightarrow$ HCO\textsuperscript{+} + CO \\
 3.& CO\textsuperscript{+} + H\textsubscript{2} $\rightarrow$ HCO\textsuperscript{+} + H \\
 4.& CH + O $\rightarrow$ HCO\textsuperscript{+} + e\textsuperscript{-} \\
 5.& CH\textsubscript{2}\textsuperscript{+} + O $\rightarrow$ HCO\textsuperscript{+} + H \\
 6.& CH\textsubscript{3}\textsuperscript{+} + O $\rightarrow$ HCO\textsuperscript{+} + H\textsubscript{2} \\ 
 7.& H\textsubscript{2}O + C\textsuperscript{+} $\rightarrow$ HCO\textsuperscript{+} + H \\
 8.& H\textsubscript{3}O\textsuperscript{+} + C $\rightarrow$ HCO\textsuperscript{+} + H\textsubscript{2} \\
 9.& CH\textsubscript{2}\textsuperscript{+} + O\textsubscript{2} $\rightarrow$ HCO\textsuperscript{+} + OH \\
 10.& H\textsubscript{3}\textsuperscript{+} + CO $\rightarrow$ HCO\textsuperscript{+} + H\textsubscript{2} \\
 11.& HCO + $\gamma$ $\rightarrow$ HCO\textsuperscript{+} + e\textsuperscript{-}\\
 \hline
\end{tabular}
\end{center}
 \label{tab:hcoj_reactions}
\end{table}

\begin{figure*}
    \centering
    \includegraphics[width=\linewidth]{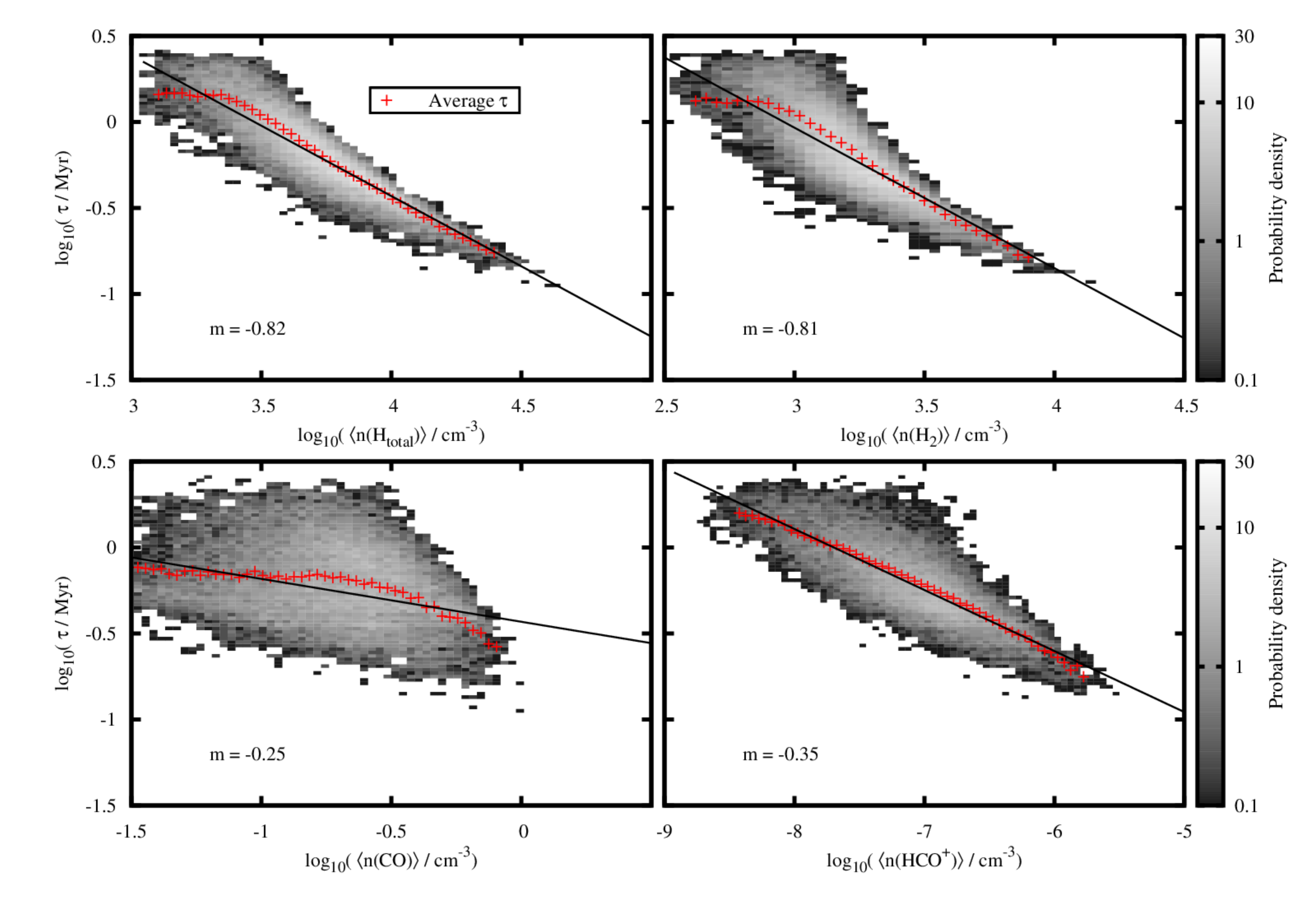}
    \caption{The 2D-PDF of the HCO\textsuperscript{+} growth time-scale $\tau$ (see Eq.~\ref{eq:tau}), vs. the logarithm of the time-averaged mean density $\langle \log n_i \rangle$ (see Eq.~\ref{eq:logni}), for the subset of tracers in \mbox{MC1-HD} which fulfill the \ce{HCO+} growth condition in Eq.~\ref{eq:tau}. The red crosses denote the average $\tau$ for each density bin. The black line denotes a fit to these average points (see Eq.~\ref{eq:tau-slope}), for which the slope $m$ is given in each panel. The higher the average density experienced by the tracer particles, the shorter the growth time-scale. The correlation is strongest with regard to $n_\rmn{H,tot}$, as HCO\textsuperscript{+} formation occurs predominantly in dense gas (see Fig.~\ref{fig:hic-both}).}
    \label{fig:tau-basic}
\end{figure*}

In Fig.~\ref{fig:tau-basic}, we plot 2D-PDFs of $\tau$ vs. $\langle \log n_i \rangle$ for H\textsubscript{tot}, H\textsubscript{2}, CO, and HCO\textsuperscript{+} itself, for the cloud \mbox{MC1-HD}. The over-plotted red points indicate the average $\tau$ for each density bin. The black over-plotted line in each 2D-PDF indicates a linear fit to the red points as follows:
\begin{equation}
    \label{eq:tau-slope}
    \log \tau=m \langle\log n_i \rangle+ C,
\end{equation}
where $C$ corresponds to the value of $\log \tau$ when $\langle\log n_i \rangle=0$. The closer the slope $m$ is to -1, the greater the correlation of a particular reactant with the overall HCO\textsuperscript{+} production rate. 

As expected, a shorter time-scale $\tau$ corresponds to higher densities for these species, and thus also with higher $A_\rmn{V,3D}$. The correlation of $\tau$ with CO abundance is the smallest of these four species, with $m=-0.25$. This is likely due to the aforementioned presence of CO in both the creation and destruction reactions of HCO\textsuperscript{+}. The correlations of $\tau$ with H\textsubscript{tot} ($m=-0.82$) and H\textsubscript{2} ($m=-0.81$) are almost identical, which is reasonable given that HCO\textsuperscript{+} forms in the extinction regime of molecular gas (see Fig.~\ref{fig:hic-both}). The correlation of the HCO\textsuperscript{+} number density with $\tau$ is weaker ($m=-0.35$), which means that the abundance of HCO\textsuperscript{+} is a less-reliable indicator of its own formation rate than are the abundances of H\textsubscript{2} and CO. Repeating this analysis for \mbox{MC1-MHD} shows almost identical correlations, with the average $\tau$ in each density bin almost unchanged, but with less spread in the underlying 2D distribution.

\begin{figure*}
    \centering
    \includegraphics[width=0.95\linewidth]{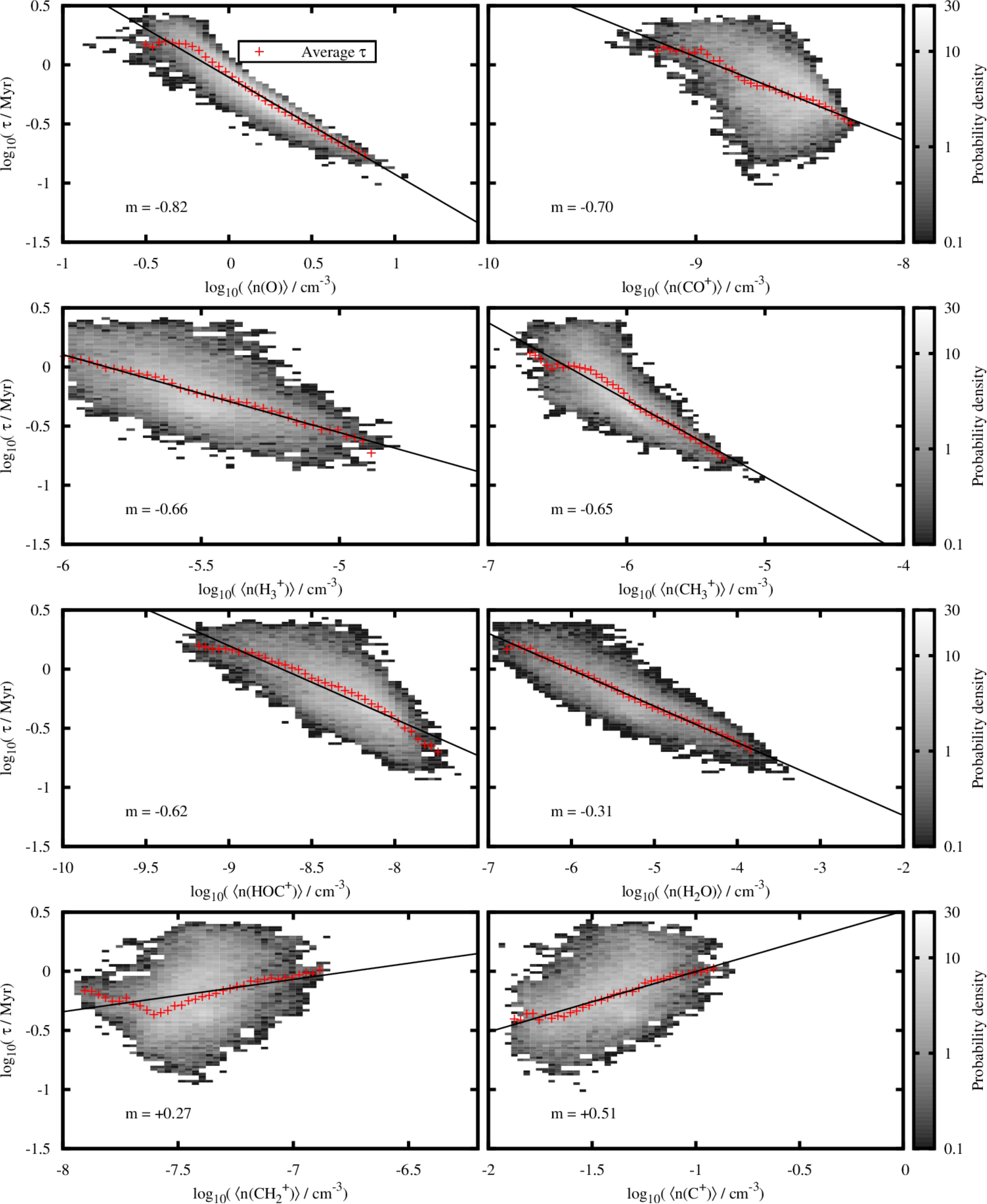}
    \caption{The 2D-PDF of the \ce{HCO+} growth time-scale $\tau$ (see Eq.~ref{eq:tau}) vs. the time-averaged mean density $\langle \log n_i \rangle$ (see Eq.~\ref{eq:logni}) of various formation reactants of \ce{HCO+} given in Table~\ref{tab:hcoj_reactions}, for the subset of tracers in \mbox{MC1-HD} which fulfill the \ce{HCO+} growth condition in Eq.~\ref{eq:tau}. The red crosses denote the average $\tau$ for each density bin. The black line denotes a fit to these average points (see Eq.~\ref{eq:tau-slope}), for which the slope $m$ is given in each panel. Reactants which are abundant in the low-density (i.e. low-$A_\rmn{V,3D}$) regime, like \ce{C+} and \ce{CH2+}, correlate poorly with $\tau$ (low values of $|m|$), partly because they imply the presence of electrons and unshielded radiation which both destroy \ce{HCO+}. However, the formation species which are found at higher extinctions, like \ce{H3+}, are more strongly correlated.}
    \label{fig:tau-species}
\end{figure*}

We can expand this analysis to other reactants listed in Table~\ref{tab:hcoj_reactions}. In Fig.~\ref{fig:tau-species}, we display the correlation of $\tau$ with $\langle n \rangle$ for O, CO\textsuperscript{+}, H\textsubscript{3}\textsuperscript{+}, CH\textsubscript{3}\textsuperscript{+}, HOC\textsuperscript{+}, H\textsubscript{2}O, CH, and C\textsuperscript{+}. Atomic oxygen, which can react to form HCO\textsuperscript{+} via several pathways, has the strongest correlation of these species with $\tau$ , with $m=-0.82$. Next we see that CO\textsuperscript{+} has a fitted slope of $m=-0.7$, corresponding to the high correlation of its reactant H\textsubscript{2} (see Table~\ref{tab:hcoj_reactions}). The cosmic ray tracer H\textsubscript{3}\textsuperscript{+} has $m=-0.66$, a stronger correlation than its reactant partner CO. CH\textsubscript{3}\textsuperscript{+} reacts with atomic oxygen and has a slope of $m=-0.65$. The isomer HOC\textsuperscript{+} could form HCO\textsuperscript{+} by reacting with either H\textsubscript{2} or CO, and has a slope of $m=-0.62$. We ascribe the weaker correlation of \ce{H2O} with $\tau$ ($m=-0.31$) to the fact that its co-reactant, \ce{C+}, connotes the presence of free electrons. Since electrons can recombine with and eliminate \ce{HCO+}, it is sensible both that the \ce{H2O} correlation is weak, and that the \ce{C+} density is in fact anti-correlated ($m=+0.51$) with the \ce{HCO+} formation timescale. This also explains the anti-correlation of \ce{CH2+} density with $\tau$ ($m=+0.27$), in spite of the high magnitude of correlation of its co-reactant O with $\tau$ (apparently due solely to oxygen's reaction with \ce{CH3+}; oxygen's third co-reactant, CH, correlates poorly to $\tau$ with $m=+0.16$).

Qualitatively similar results are seen for these correlations in the other three molecular clouds.  The ranges in the fitted slopes for these species for the four different molecular clouds, as well as for the species shown in Fig.~\ref{fig:tau-basic}, are reported in Table~\ref{tab:tauclouds}. Generally, the values for $m$ are similar for all four molecular clouds and thus the range is limited. The MHD and HD clouds differ in the correlations of $\langle \log n_\rmn{CO^+} \rangle$ and $\langle \log n_\rmn{H_3^+} \rangle$ with $\tau$. For instance, $\langle \log n_\rmn{CO^+} \rangle$ has a stronger correlation with $\tau$ in the HD clouds (e.g. $m=-0.7$ for \mbox{MC1-HD}) than in the MHD clouds (e.g. $m=-0.13$ for \mbox{MC1-MHD}). 

Conversely, the correlation of $\tau$ with $\langle \log n_\rmn{H_3^+} \rangle$ is weaker in the HD clouds (e.g. $m=-0.6$ for \mbox{MC1-HD}) than in the MHD clouds (e.g. $m=-0.86$ for \mbox{MC1-MHD}). This implies that the reaction \mbox{CO\textsuperscript{+} + H\textsubscript{2}} has greater influence during \ce{HCO^+ } formation in HD clouds than in MHD clouds. Conversely, the reaction \mbox{H\textsubscript{3}\textsuperscript{+} + CO} has a somewhat greater influence in MHD clouds than in HD ones. Corroborating the latter point, we find that the reservoir of available \ce{H3+} covers a greater spatial extent in the MHD clouds than in the HD clouds, since the MHD clouds also have a more extensive \ce{H2} envelope in which \ce{H3+} is formed via cosmic ray interactions. Thus, the \mbox{H\textsubscript{3}\textsuperscript{+} + CO} reaction is an available formation pathway for \ce{HCO+} over a wider spatial extent in the MHD clouds than in the HD clouds.

\begin{table}
    \caption{The minimum and maximum values of $m$ for each \ce{HCO+} production reactant (see Eq.~\ref{eq:tau-slope}) for all four simulated clouds.}
    \centering
    \begin{tabular}{c|c|c}
        \hline
        Species & $m_\rmn{min}$ & $m_\rmn{max}$ \\
        \hline
        H\textsubscript{tot} & -0.95 & -0.73 \\
        H\textsubscript{2} & -0.96 & -0.75 \\
        CO & -0.25 & -0.10 \\
        HCO\textsuperscript{+} & -0.43 & -0.32 \\
        O & -0.82 & -0.68 \\
        CO\textsuperscript{+} & -0.70 & -0.13 \\
        H\textsubscript{3}\textsuperscript{+} & -0.86 & -0.60 \\
        CH\textsubscript{3}\textsuperscript{+} & -0.65 & -0.56 \\
        HOC\textsuperscript{+} & -0.62 & -0.59 \\
        H\textsubscript{2}O & -0.36 & -0.30 \\
        CH & +0.11 & +0.16 \\
        CH\textsubscript{2}\textsuperscript{+} & -0.01 & +0.27  \\
        C\textsuperscript{+} & +0.30 & +0.51 \\
         \hline
    \end{tabular}
    \label{tab:tauclouds}
\end{table}

Fully explaining these correlations requires us to explore the relative importance of the different pathways for HCO\textsuperscript{+} production. For the moment, we point out that the correlations are sensible in light of our prior discussion of the particle history tracks. Reactants which are prevalent in lower-extinction gas, such as C\textsuperscript{+} and CH, have poor correlations with $\tau$. This is unsurprising given that Fig.~\ref{fig:hic-both} shows HCO\textsuperscript{+} formation is most prevalent in gas with total hydrogen density around $10^3$--$10^4$~cm$^{-3}$. Additionally, it is worth noting that for reactions where one reactant is more abundant than the other, the correlation with $\tau$ is stronger for the less-abundant reactant, as its limited supply constrains the rate of the reaction. For example, in the reaction \mbox{H\textsubscript{3}\textsuperscript{+} + CO}, the less-abundant reactant is H\textsubscript{3}\textsuperscript{+}, which has a stronger correlation with $\tau$ than CO does.

\subsection{HCO\textsuperscript{+} formation pathways}
\label{subsection:formationpaths}

Figs.~\ref{fig:tau-basic} and \ref{fig:tau-species} show the correlation of individual reactants' number densities with the HCO\textsuperscript{+} formation time-scale, over the course of the entire simulation but only for the very small subset of particles which fulfill the growth condition of Eq.~\ref{eq:tau}. It is natural to consider next all the tracer particles regardless of their lifetime peak HCO\textsuperscript{+} abundance, and what their individual chemical histories can tell us about the predominant modes of HCO\textsuperscript{+} formation: a topic that fundamentally requires time-dependent chemistry to properly explore. 

Because all the tracer particles retain not only the species number densities, but also environmental parameters like the local temperature and visual extinction, we can calculate each tracer particle's reaction rate for all 11 HCO\textsuperscript{+} formation reactions listed in Table~\ref{tab:hcoj_reactions}. We calculate the rate $C_i$ (in units of cm$^{-3}$~s$^{-1}$) of a given reaction $i$ with $R$ reactants and a temperature-dependent rate coefficient $k_i(T)$ as

\begin{equation}
    C_i=k_i(T)\prod_{j=1}^{R} n_j.
\end{equation}

We then assess the relative importance of each HCO\textsuperscript{+} formation reaction as a function of the local visual extinction $A_\rmn{V,3D}$. Rather than normalizing the rates of the formation reactions on a per-particle basis (which would inaccurately suppress the contribution of tracers in regions of high absolute production), we first allocate $C_i$ into bins of $A_\rmn{V,3D}$, and then normalize the formation rates within each bin. We calculate this average normalized reaction rate, $\langle F_\rmn{norm} \rangle$, for each reaction $i$ in each $A_\rmn{V,3D}$ bin containing $N_\rmn{bin}$ particles as follows:

\begin{equation}
    \langle F_\rmn{norm} \rangle = \frac{1}{\sum_{i} C_i} \frac{1}{N_\rmn{bin}} \sum_{j=1}^{N_\rmn{bin}} C_{i,j}.
    \label{eq:F_i}
\end{equation}
We choose to analyse $\langle F_\rmn{norm} \rangle$ at $t_\rmn{evol}=2$~Myr, rather than at the later time \mbox{$t_\rmn{evol}=4$~Myr} as with most of the other analysis in this work, in order to capture the state of the HCO\textsuperscript{+} formation reactions at the beginning of the principal formation epoch (see Fig.~\ref{fig:hic-both}), rather than after the global \ce{HCO+} quantity has reached its final state. 

\begin{figure}
    \centering
    \includegraphics[width=\linewidth]{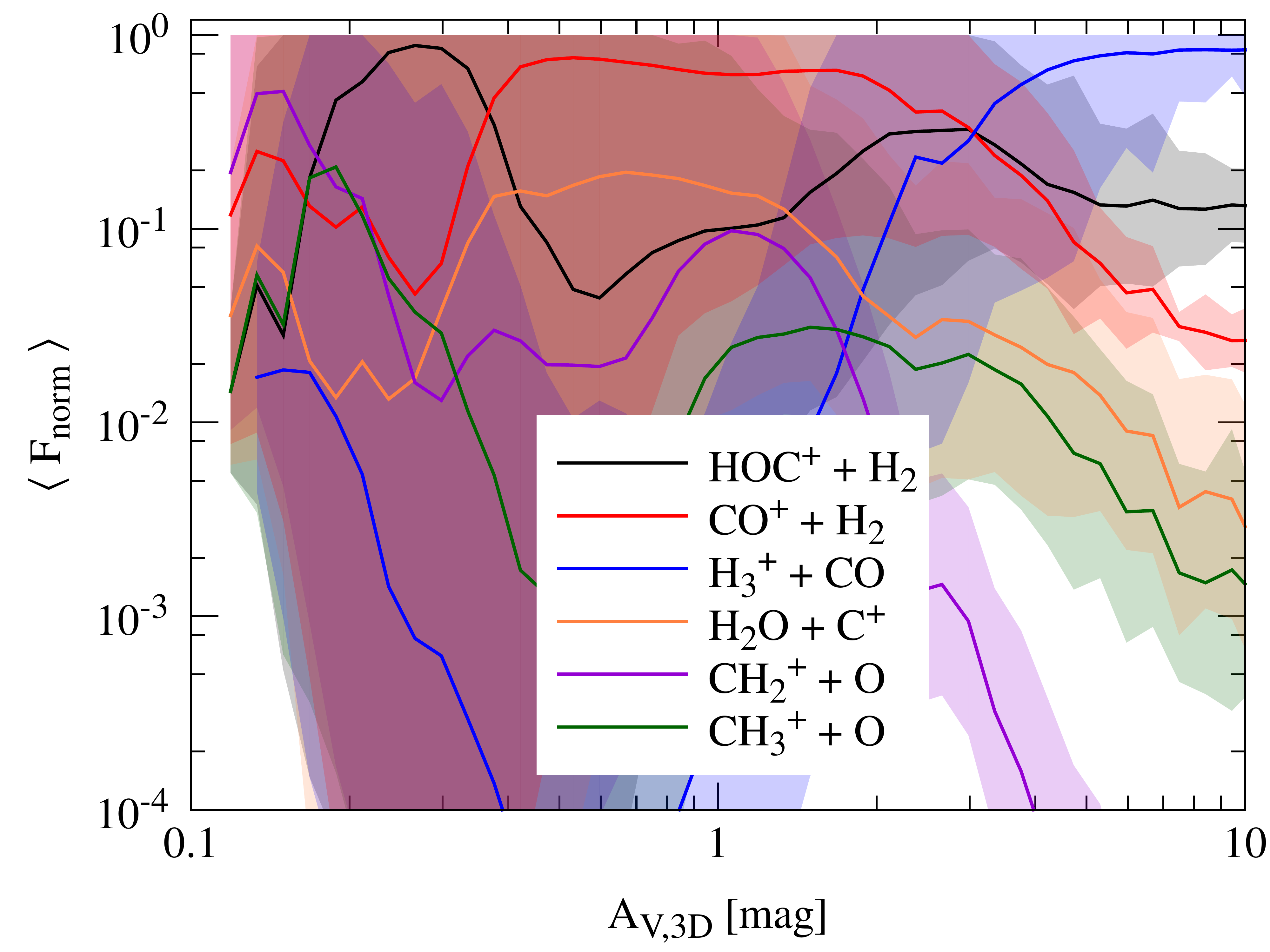}
    \caption{The average normalized reaction rates $\langle F_\rmn{norm} \rangle$ (see Eq.~\ref{eq:F_i}) of reactions in the post-processing network which produce HCO\textsuperscript{+}, vs. visual extinction, for cloud \mbox{MC1-HD} at $t_\rmn{evol}=2$~Myr. We have neglected the contributions of reactions which are generally subdominant in their impact upon HCO\textsuperscript{+} abundance. The dominant HCO\textsuperscript{+} formation reaction is heavily dependent on the extinction. Up to $A_\rmn{V,3D}\simeq0.4$, formation is dominated by \mbox{HOC\textsuperscript{+} + H\textsubscript{2}} (black). Then up to $A_\rmn{V,3D}\simeq3$, the reaction \mbox{CO\textsuperscript{+} + H\textsubscript{2}} (red) is dominant. However, the actual production of HCO\textsuperscript{+} below $A_\rmn{V,3D}\simeq3$ is minimal: the formation species which contribute the greatest fraction of the net HCO\textsuperscript{+} formation are uncommon at such low extinctions. Thus, most of the HCO\textsuperscript{+} production in the cloud stems from the reaction \mbox{H\textsubscript{3}\textsuperscript{+} + CO} (blue), which dominates above $A_\rmn{V,3D} \simeq 3$. This reaction represents more than 90\% of the total production at the high extinctions where HCO\textsuperscript{+} is actually present.}
    \label{fig:reaction-flux}
\end{figure}

In Fig.~\ref{fig:reaction-flux}, we show these normalized reaction rates, $\langle F_\rmn{norm} \rangle$, for \mbox{MC1-HD} at $t_\rmn{evol}=2$~Myr. The solid colored lines indicate the average value of each reaction in the local $A_\rmn{V,3D}$-bin, with color-matched shading imposed to indicate one standard deviation above and below the mean. The reactions \mbox{CH + O} (Reaction 4 in Table~\ref{tab:hcoj_reactions}), \mbox{CH\textsubscript{2}\textsuperscript{+} + O\textsubscript{2}} (Reaction 5), \mbox{HOC\textsuperscript{+} + CO} (Reaction 2), \mbox{H\textsubscript{3}O\textsuperscript{+} + C} (Reaction 8), and the cosmic ray interaction \mbox{HCO + $\gamma$} (Reaction 11) have been neglected due to their consistently minimal contributions to the total \ce{HCO+} production in comparison to the other six remaining reactions across multiple snapshots and simulations. The relative rates of the non-negligible reactions achieve stable values by $A_\rmn{V,3D}=10$. We have confirmed these values hold up to $A_\rmn{V,3D}=100$, and leave out the final magnitude for purposes of readability.

It is clear that particular HCO\textsuperscript{+} formation reactions predominate in different visual extinction regimes. Up to \mbox{$A_\rmn{V,3D}\simeq0.4$}, HCO\textsuperscript{+} is chiefly generated via \mbox{\ce{CH2+} + O $\rightarrow$ \ce{HCO+} + H} and the unidirectional isomerization reaction \mbox{HOC\textsuperscript{+} + H\textsubscript{2} $\rightarrow$ HCO\textsuperscript{+} + H\textsubscript{2}}. However, the absolute quantity of HCO\textsuperscript{+} remains negligible in this poorly-shielded extinction regime due to efficient photodissociation by incident radiation. 

From $A_\rmn{V,3D}\simeq0.4$ to $3$, the reaction \mbox{CO\textsuperscript{+} + H\textsubscript{2} $\rightarrow$ HCO\textsuperscript{+} + H} contributes about 75\% of the total HCO\textsuperscript{+} production. The remaining 25\% of the HCO\textsuperscript{+} contribution in this range comes from \mbox{H\textsubscript{2}O + C\textsuperscript{+} $\rightarrow$ HCO\textsuperscript{+} + H} (up to $A_\rmn{V,3D} \simeq 1.5$) or the isomerization of HOC\textsuperscript{+} ($A_\rmn{V,3D} \simeq 1.5$--$3$). HOC\textsuperscript{+} is also produced by both the water reaction and the CO\textsuperscript{+} reaction at an approximately equal rate to HCO\textsuperscript{+} \citep{gerin_molecular_2019,gerin_co_2021}. However, because of the isomerization reaction, some of this HOC\textsuperscript{+} becomes HCO\textsuperscript{+} anyway. There is no equivalent route backwards for HCO\textsuperscript{+} to isomerize to HOC\textsuperscript{+}, resulting in an abundance ratio $\rmn{n_{HCO\textsuperscript{+}}}/\rmn{n_{HOC\textsuperscript{+}}}\sim 100$ beginning around $A_\rmn{V,3D} \simeq 1$.

The preceding reactions all decline in importance around \mbox{$A_\rmn{V,3D}\simeq3$}, as the reaction \mbox{H\textsubscript{3}\textsuperscript{+} + CO $\rightarrow$ HCO\textsuperscript{+} + H\textsubscript{2}} rapidly becomes, and then remains, the dominant one. This corresponds to a transition from an $A_\rmn{V,3D}$ regime dominated by photochemistry to a regime where the chemistry is driven by cosmic ray interactions. The \mbox{H\textsubscript{3}\textsuperscript{+} + CO} reaction contributes more than 90\% of the total HCO\textsuperscript{+} production by $A_\rmn{V,3D}\simeq5$. This extinction magnitude is approximately where the tracer particles whose trajectories are plotted in Fig.~\ref{fig:hic-both} experience an epoch of in situ HCO\textsuperscript{+} formation.

The H\textsubscript{3}\textsuperscript{+} reaction can also produce HOC\textsuperscript{+} at an equal rate, the isomerization of which contributes most of the remaining HCO\textsuperscript{+} production at $A_\rmn{V,3D}\gtrsim 5$ . Comparing the reaction rates, we attribute about half of the HOC\textsuperscript{+} which is then isomerized to \ce{HCO+} at very high extinction to the \mbox{H\textsubscript{3}\textsuperscript{+} + CO} reaction. As will be shown in Section~\ref{subsection-colmaps}, more than 90\% of all HCO\textsuperscript{+} by mass is found above $A_\rmn{V,3D}\simeq5$, meaning the \mbox{H\textsubscript{3}\textsuperscript{+} + CO} reaction is by far the most important driver of the cloud's total HCO\textsuperscript{+} content.

\begin{figure}
    \centering
    \includegraphics[width=\linewidth]{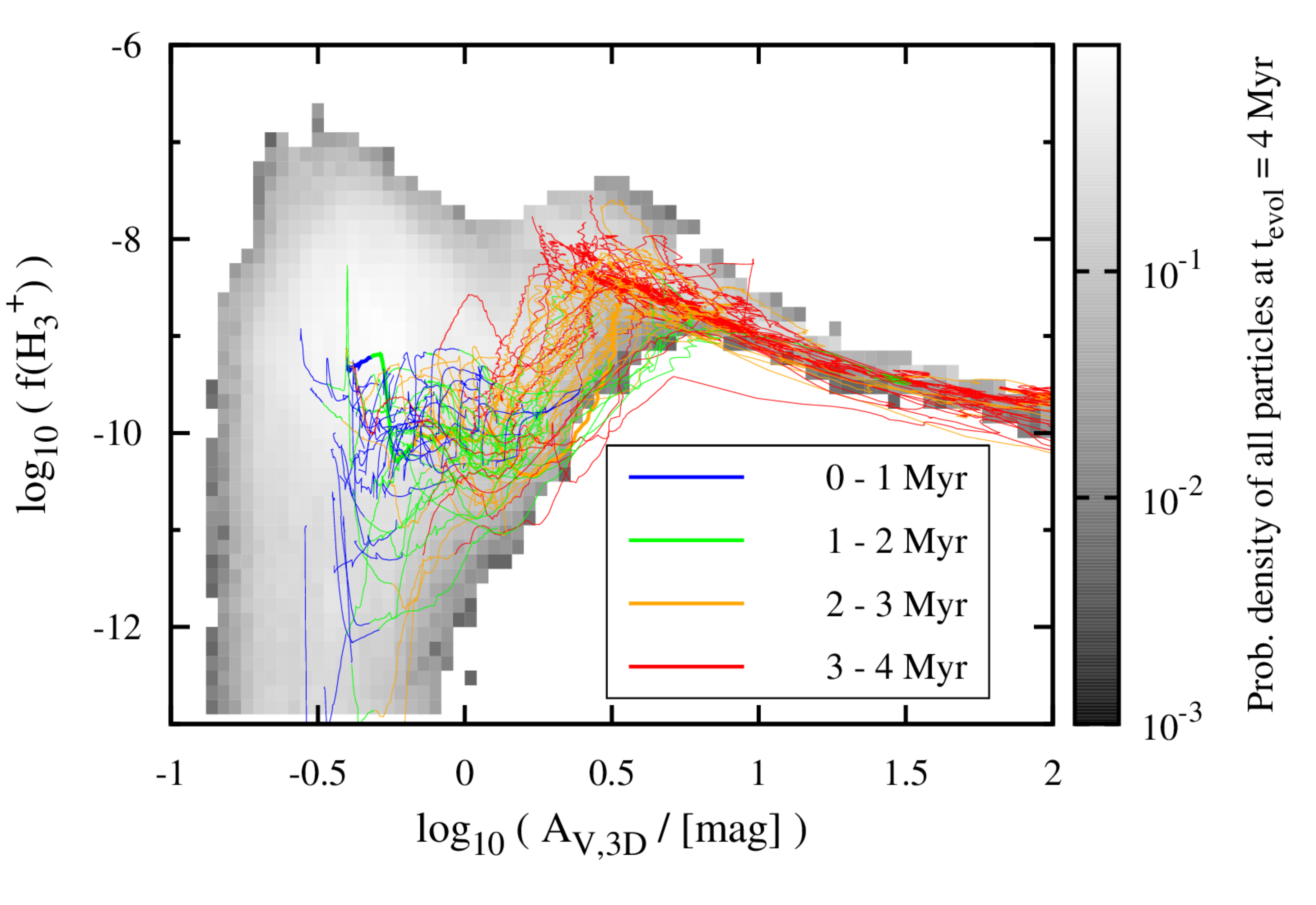}
    \caption{Same as Fig.~\ref{fig:hic-both}, but showing tracer history tracks of H\textsubscript{3}\textsuperscript{+} fractional abundance vs. $A_\rmn{V,3D}$ in cloud \mbox{MC1-HD} for the same representative random sample of tracer particles, plotted over a 2D probability density function of $f(\rmn{H_3^+})$ vs. $A_\rmn{V,3D}$ for every tracer particle in the simulation at \mbox{$t_\rmn{evol}=4$~Myr}. As the less abundant reactant in the \mbox{\ce{H3+} + CO} reaction, the available \ce{H3+} content bottlenecks the formation of \ce{HCO+} by this route.}
    \label{fig:hic_h3j}
\end{figure}

To further investigate this reaction, in Fig.~\ref{fig:hic_h3j} we analyse the distribution and evolution of \ce{H3+} in cloud MC1-HD. We again investigate the history of the \ce{H3+} distribution using the tracks followed by the same 50 tracer particles as in Fig.~\ref{fig:hic-both}. Comparing the two figures, we see that the selected tracers (which were chosen for having a high value of $f$(\ce{HCO+}) at late times) display the same bulk behavior in the growth of their \ce{H3+} content. Assuming the chemical state is near equilibrium at late times, the density of H\textsubscript{3}\textsuperscript{+} is decoupled from $n_\rmn{H,tot}$ \citep{Oka2006}, explaining why $f(\rmn{H_3^+})$ declines later in the simulation at high density ($\log A_\rmn{V,3D} \simeq 0.5$). This decline may explain some of the corresponding decline in $f(\rmn{HCO^+})$ above $\log A_\rmn{V,3D} \simeq 0.5$ seen in Fig.~\ref{fig:hic-both}, given that the reaction \mbox{H\textsubscript{3}\textsuperscript{+} + CO} contributes the most to the HCO\textsuperscript{+} formation in this $A_\rmn{V,3D}$ regime (see Fig.~\ref{fig:reaction-flux}). Since H\textsubscript{3}\textsuperscript{+} is less abundant than CO, the decline in H\textsubscript{3}\textsuperscript{+} would bottleneck this reaction. 

When the reaction rate analysis is repeated for cloud \mbox{MC1-MHD} at $t_\rmn{evol}=2$~Myr, the relative importance of the reactions is unchanged from the HD case. However, the crossover point where \mbox{H\textsubscript{3}\textsuperscript{+} + CO} becomes the dominant reaction is instead at visual extinction $A_\rmn{V,3D}\simeq4$. Later, at $t_\rmn{evol}=4$~Myr, this reaction becomes dominant at $A_\rmn{V,3D}\simeq3$, the same as cloud \mbox{MC1-HD}. We attribute this effect to the slower coalescence rate of the MHD clouds compared to the HD clouds \citep{seifried_silcc-zoom_2020}. The ongoing cloud coalescence increases the abundances of \mbox{H\textsubscript{3}\textsuperscript{+} and CO}, fueling this reaction at lower and lower extinctions as time passes. We note that we find analogous results for the rate of \mbox{H\textsubscript{3}\textsuperscript{+} + CO} vs. time in the clouds \mbox{MC2-HD} and \mbox{MC2-MHD} (not shown).

\section{Where can HCO\textsuperscript{+} be found?}
\label{section-regridding}

In the previous section, we use our non-equilibrium chemical post-processing tools to explore the formation density regime, the formation time-scale, and the visual extinction-dependent dominant formation reactions of HCO\textsuperscript{+}. Next, we wish to compare the distribution of HCO\textsuperscript{+} to molecular cloud observations \citep[e.g.][and others]{sanhueza_chemistry_2012,gerin_molecular_2019,goicoechea_molecular_2019,barnes_lego_2020,liu_atoms_2020-1,liu_atoms_2020,nayana_alma_2020,yun_times_2021,yang_search_2021}. To do so, however, we must transform our ensemble of passive tracer particles into a comprehensive, space-filling array of number densities (hereafter `density grid'). Unlike SPH particles, these tracers do not represent mass and cannot be regridded the same way as SPH particles. A different approach is necessary, which we present in Section~\ref{subsection-regridalgorithm}, followed by validations in Section~\ref{subsection:validation}. Finally, in Section~\ref{subsection-colmaps}, we present HCO\textsuperscript{+} column density projections and compare them to observations.

\subsection{The regridding algorithm}
\label{subsection-regridalgorithm}

Because passive tracer particles -- as opposed to particles in SPH simulations -- do not represent fluid elements and thus are not associated with a certain quantity of mass or volume, we have developed a novel algorithm which maps the tracer particle back on a volume-filling grid.
This procedure has several steps:

\begin{enumerate}
    \item Generation of a blank, uniformly-resolved grid whose spatial extent matches the zoom-in region.
    \item Assignment of a desired species' fractional abundance from the tracer particles to the spatially-corresponding blank cells.
    \item Interpolation and extrapolation of the fractional abundance values of cells containing tracer particles into adjacent empty cells.
    \item Repetition of the interpolation procedure (iii) until the entire grid is filled with fractional abundance values.
    \item Multiplication of the volume-filling fractional abundance grid with a corresponding, congruent grid of $n_\rmn{H,tot}$, generating a volume-filling, uniform number density grid of the considered species.
\end{enumerate}

We now describe these steps in more detail. First, we define a grid domain and a uniform cell resolution, and select a time snapshot. The natural domain choice for this study is the exact extent of each zoom-in region. To explore the relationship between the resolution and the obtained species masses of the final grid, we tested uniform cell resolutions of 1, 0.5, 0.25, and 0.125~pc, for which the regridding procedure is conceptually identical.

A blank grid of the selected shape and resolution is initialized. We assign the particle’s saved value for $f_i$ to the corresponding cell in the blank grid. We choose to assign the fractional abundance values, rather than the naively more obvious choice of the number densities $n_i$, to avoid overestimating species densities at the diffuse frontiers of the molecular cloud. This will be explained in the interpolation phase of the algorithm. When $N>1$ tracer particles occupy the same grid cell, their $f_i$ values are logarithmically averaged:
\begin{equation}
    \langle f_i \rangle = \exp\left( \frac{1}{N} \sum_j^N \ln f_{i,j} \right)
\end{equation}
This logarithmic average prevents the higher $f_i$ of two (or more) tracers from dominating the average abundance in a cell. This helps to avoid an overestimation of the total species mass in the given cell (see Section~\ref{subsection:validation}).

Merely regridding the tracer particle data is insufficient to fill the entire zoom-in domain due not only to the limited number of particles contained in the simulations, but also to the aforementioned tendency of tracer particles to congregate as the clouds contract over time. The densest regions of the clouds exhibit the best number statistics, but even here some cells lack direct tracer data. This issue worsens with increasing resolution. For instance, at a resolution of 0.125~pc, the zoom-in region for \mbox{MC1-HD} is split into more than $3\times10^8$ cells, but at an elapsed time of 2~Myr, only contains about $3\times10^5$ particles. Even neglecting the congregation of particles at density peaks, this represents a maximal filled proportion of 0.1\%.

To remedy this, we iteratively interpolate the \textit{fractional abundances} into adjacent empty cells, until the entire grid is filled. In this phase of the algorithm, each empty cell checks all 26 neighbouring cells for a nonzero value. If a single nonzero neighbour is found, that neighbour’s $f_i$ value is copied into the empty cell. If there are nonzero $N_\rmn{neighbour}>1$, we calculate the final $f_i$ value as: 
\begin{equation}
    \langle f_i \rangle = \exp\left( \frac{1}{\sum_j \frac{1}{d_j}} \sum_j^{N_\rmn{neighbour}} \frac{\ln f_{i,j}}{d_j} \right), 
\end{equation}
where $d_j$ is the distance between the centroids of the empty cell and each neighbouring cell, divided by the cell resolution. During our tests, we found that the logarithmic averages recover the masses better than simple averages. Simple averages would be dominated by local neighbouring density peaks, improperly extending their spatial size.

We emphasize that during each interpolation pass, the empty cells all assess their neighbours independently. If a pair of adjacent empty cells \textit{A} and \textit{B} share a single nonzero neighbour \textit{C}, the value interpolated into \textit{A} on this step from \textit{C} will not simultaneously be considered by \textit{B} as it looks for its own nonzero neighbours. This avoids any dependence on the sweeping order of the interpolation, i.e., whether the pixels are interpolated, for instance, in the order $x$-$y$-$z$ as opposed to $z$-$y$-$x$.

Since the density of tracer particles is lower in regions of low gas density, the interpolation procedure might need to assign values to empty grid cells from a tracer that is located several cells away, in a higher density region. Hence, if we had tried to generate a uniform grid by interpolating values for the \textit{number density} of a target species, we would have unrealistically filled diffuse zones with gas that was not present in the SILCC-Zoom simulations of the same clouds, violating mass conservation. To avoid this, we found it to be crucial to interpolate merely $f_i$, and generate the number densities in a final step as follows. 

The interpolation procedure repeats until the entire uniform grid is filled with nonzero values for $f_i$, and no empty cells remain. To convert this grid to the desired final $n_i$ distribution, we multiply the filled grid of $f_i$ values by another grid of identical size and resolution, which contains $n_\rmn{H,tot}$ obtained from the original simulation data. Since the identity of the target species is irrelevant to the operation of this algorithm, we can thus produce self-consistent density distributions for \textit{any} species whose abundances are saved to the tracer particles.

Compared to the chemical post-processing of the tracer particle data, this regridding procedure has a negligible computational cost even at our highest resolution of 0.125~pc. Critically, the regridding cost is independent of the complexity of the post-processing network. Time-dependent density distributions for even the most exotic species can therefore be computed with great efficiency. 

\subsection{Validation}
\label{subsection:validation}

A critical validation of the interpolation procedure is whether it conserves the total hydrogen and carbon in the molecular cloud. The post-processing procedure allocates the hydrogen and carbon atoms into more species than were originally present in the on-the-fly network, but the total quantity of each element is unchanged by that procedure. Failures in conservation due to the interpolation process must be well-understood and minimized.

To check this, we revisit the original SILCC-Zoom simulations and calculate the total mass of hydrogen, $M_\rmn{SILCC,H,tot}$ from the sum of the masses of H, \ce{H2}, and \ce{H+}; as well as the carbon mass, $M_\rmn{SILCC,C,tot}$, from the sum of C, C\textsuperscript{+}, and CO (subtracting the mass of the oxygen atom). The SILCC-Zoom simulation grids, which contain cells with volumes $dV_i$ that depend on the refinement level, report the mass density $\rho_i$ in each cell for a given species. We calculate the total masses of hydrogen and carbon reported by the SILCC-Zoom grids at a given time as:

\begin{equation}
    \label{eq:Mgridstart}
    M_\rmn{SILCC,H,tot} = \sum_\textrm{i}^N \left(\rho_\rmn{i,H}+\rho_\rmn{i,H_2}+\rho_\rmn{i,H^+}\right)dV_i,
\end{equation}

\begin{equation}
    M_\rmn{SILCC,C,tot} = \sum_\textrm{i}^N \left(\rho_\rmn{i,C}+\frac{12}{28}\rho_\rmn{i,CO}+\rho_\rmn{i,C^+}\right)dV_i,
\end{equation}
where the prefactor on $\rho_\rmn{i,CO}$ accounts for only considering the mass of the molecule's carbon atom. 

We wish to compare these total mass values to those of our regridded, interpolated data. Because the regridded data are in units of number density, the total regridded hydrogen and carbon mass equations are of a different form:

\begin{equation}
    M_\rmn{regrid,H,tot}=m_p\sum_\rmn{i}^N \left(n_\rmn{i,H} + 2 n_\rmn{i,H_2} + n_\rmn{i,H^+}\right)dV_i ,
\end{equation}

\begin{equation}
    \label{eq:Mgridend}
    M_\rmn{regrid,C,tot}=m_p\sum_\rmn{i}^N 12\left(n_\rmn{i,C} + n_\rmn{i,CO,gas} + n_\rmn{i,CO,frozen} + n_\rmn{i,C^+}\right)dV_i, 
\end{equation}
where the coefficients correspond to each term's molar mass contribution to the total hydrogen and carbon masses respectively, and $m_p$ is the proton mass. We include the post-processed number density of the CO frozen onto dust grains because this can be comparable to the gas-phase density of CO in the densest regions of the cloud, particularly at late $t_\rmn{evol}$. As none of the other carbon-bearing species contain more than 0.1\% of the total carbon mass, they were -- for the moment -- neglected when comparing M$_\rmn{SILCC,C,tot}$ and M$_\rmn{regrid,C,tot}$.

First, we analyse the regridded masses of CO and C as a function of the regridding resolution. We find that for resolutions coarser than 0.125~pc, the regridded CO mass  -- and to a lesser extent also the regridded mass of atomic carbon -- do not converge well for either HD or MHD clouds (not shown here), falling below their values in the 0.125~pc resolution grid  by up to a factor of 2. We attribute this to the highly concentrated nature of the CO content. If the cells are too large to resolve sub-parsec scale density peaks, the tracer averaging procedure will smooth out these peaks and report a peak value for $n_\rmn{CO}$ which is too low. As we have shown, \ce{HCO+} and CO occupy similar density regimes, and we would therefore expect errors of a similar magnitude in \ce{HCO+} at low resolutions. Thus, we restrict ourselves hereafter to our maximal regridding resolution of 0.125~pc.

\begin{figure*}
    \centering
    \includegraphics[width=\linewidth]{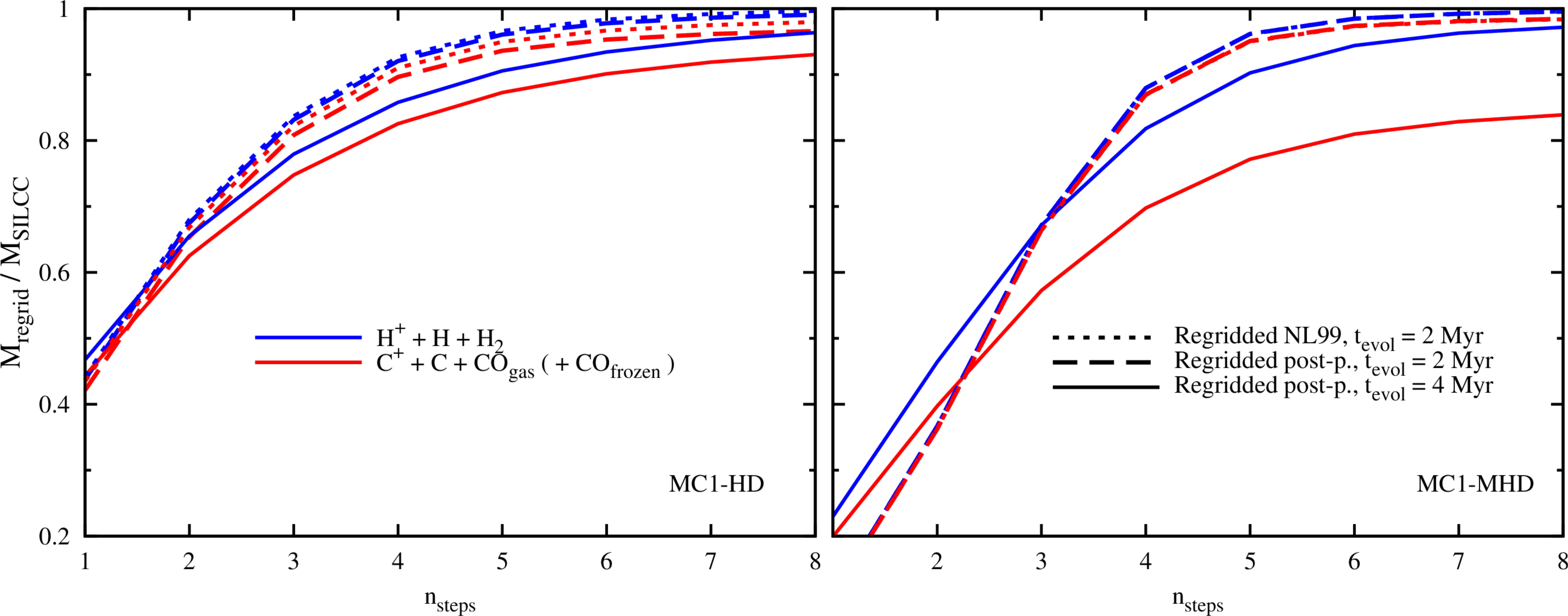}
    \caption{The ratios $M_\rmn{regrid,H,tot}/M_\rmn{SILCC,H,tot}$ (blue) and $M_\rmn{regrid,C,tot}/M_\rmn{SILCC,C,tot}$ (red) from Eqs.~\ref{eq:Mgridstart}--\ref{eq:Mgridend}, vs. the number of interpolation steps, $n_\rmn{steps}$, for \mbox{MC1-HD} (left) and \mbox{MC1-MHD} (right). Results are provided for the post-processed tracers at $t_\rmn{evol}=2$~Myr (dashed) and 4~Myr (solid), along with the result after regridding the tracer particles' unprocessed NL99 abundances (dotted). The contribution of frozen-out CO is included for $M_\rmn{regrid,C,tot}$, since a non-negligible amount of the total carbon is frozen after post-processing. For MC1-MHD, the 2~Myr results for NL99 and the post-processed data are virtually indistinguishable. In general, the regridding reproduces the masses with an accuracy of $\sim10\%$.}
    \label{fig:interpmass-convergence}
\end{figure*}

Next, we assess the effectiveness of the regridding procedure at recovering the total hydrogen and carbon masses of the simulated clouds by taking the ratio of $M_\rmn{regrid}/M_\rmn{SILCC}$ for different clouds at different timesteps. We display these results in Fig.~\ref{fig:interpmass-convergence}, as a function of the number of interpolation steps, $n_\rmn{steps}$. Results are presented at $t_\rmn{evol}=2$~Myr (dashed lines) and $t_\rmn{evol}=4$~Myr (solid lines). Finally, we include the results when the tracers' unprocessed, NL99 abundances are regridded directly (dotted lines), without any of the post-processing described in Section~\ref{section-postprocessing}. This separates the mass conservation impact of the regridding procedure from the question of redistribution of hydrogen and carbon atoms into other species that are only present in the post-processing network, and not in NL99.

For cloud \mbox{MC1-HD}, the total hydrogen mass is recovered to impressive accuracy at $t_\rmn{evol}=2$~Myr, within 1\% for both the unprocessed and post-processed tracers. The carbon mass represented by C, CO, and C\textsuperscript{+} at the same time converges to within 5\% of the original $M_\rmn{SILCC}$, and the NL99 value to within 3\%, indicating that about 1--2\% of the carbon has been distributed to other species by post-processing. This difference is not seen in the more diffuse cloud MC1-MHD at $t_\rmn{evol}=2$~Myr, where the total regridded carbon masses from both the NL99 and post-processed grids are about \%2 below the SILCC carbon total. It therefore appears that the redistribution of carbon is occurring in very dense gas, which MC1-MHD at $t_\rmn{evol}=2$~Myr almost entirely lacks. 

At $t_\rmn{evol}=4$~Myr, the accuracy of the regridding process is lower. The total regridded hydrogen mass of both \mbox{MC1-HD} and \mbox{MC1-MHD} converges to 2--3\% below the SILCC hydrogen mass. Carbon performs worse than hydrogen at this late time, with the total regridded carbon mass falling below the SILCC mass by $\sim8$\% in \mbox{MC1-HD} and $\sim17$\% in \mbox{MC1-MHD}. We attribute this to the aforementioned importance of high resolution in the neighbourhood of dense peaks, which are well-developed by this point in both the HD and MHD clouds. Even in the densest regions of the clouds, the tracers occupy only a small fraction of the cells at the 0.125~pc resolution, and the interpolation procedure may miss some dense pockets of CO.

We find that $M_\rmn{regrid}$ converges by the eighth interpolation step to within one percent of their final values when the grid is totally full. This suggests the remainder of the interpolation procedure, which is predominated by interpolation into low-density cells on the frontier of the zoom-in region (and takes between 70 and 200 more interpolation steps at a resolution of 0.125~pc), could be skipped without sacrificing precision in the final total mass. For the clouds \mbox{MC1-HD} and \mbox{MC1-MHD} at $t_\rmn{evol}=2$~Myr, this respectively amounts to 87\% and 94\% interpolation steps which we perform but which have a minimal impact upon the final result. 

\begin{figure}
    \centering
    \includegraphics[width=\linewidth]{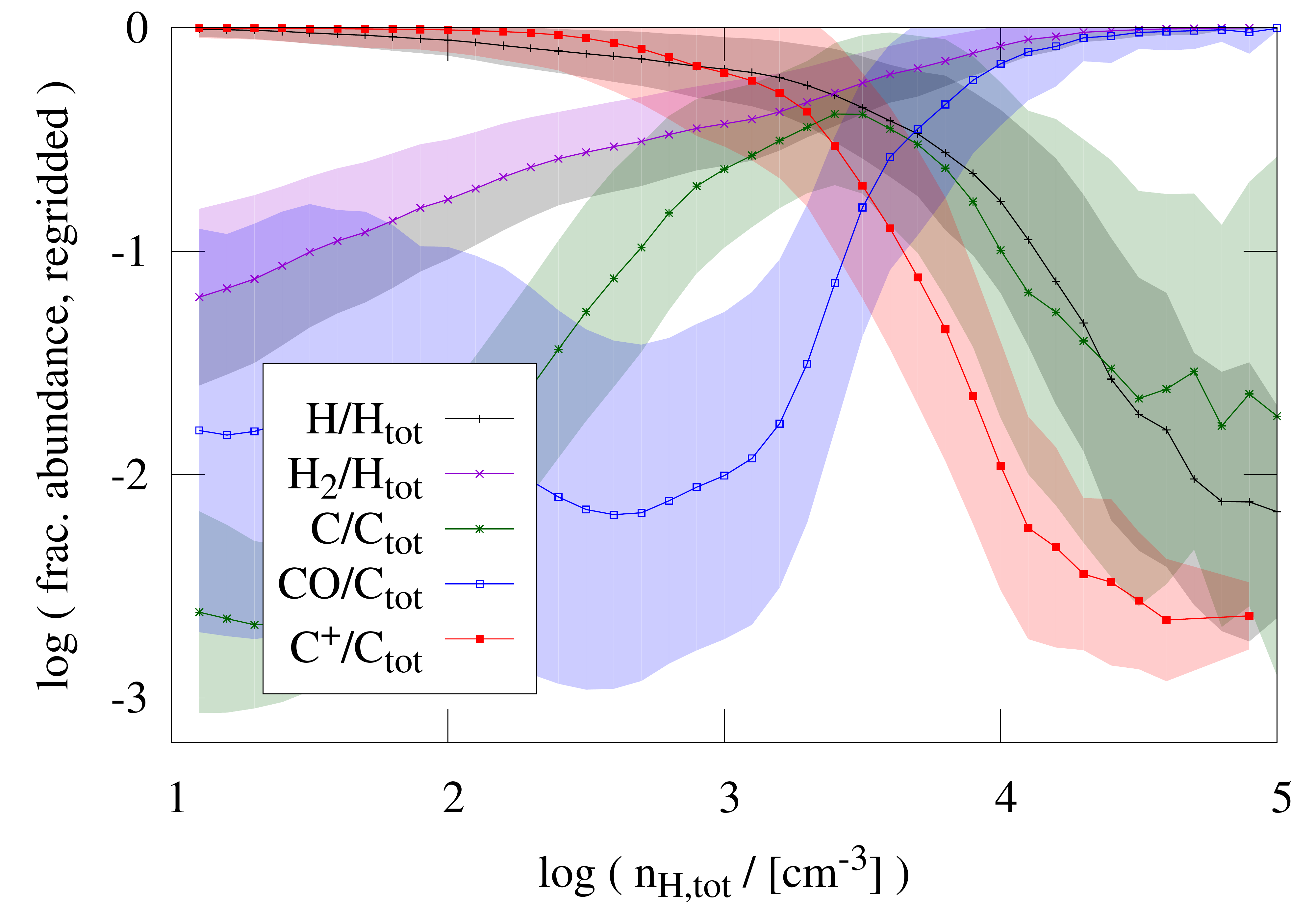}
    \caption{Same as Fig.~\ref{fig:fracvn-particles}, but for the post-processed abundances from the reconstructed grids after the interpolation is complete. Comparing to the right-hand panel of Fig.~\ref{fig:fracvn-particles}, which shows the post-processed tracer abundances, our regridding algorithm is generally successful at recovering $f_i$ for each species in the regime where that species is most abundant. However, the two data sets are less congruent for each species in regimes where the species is less abundant. In particular, CO is over-represented at lower densities, and C is under-represented at the highest densities.}
    \label{fig:fracvn-postgrid}
\end{figure}

We can validate the method further by another plot of the average fractional abundance of H, \ce{H2}, C, CO, and \ce{C+} vs. $n_\rmn{H,tot}$ for the molecular cloud \mbox{MC1-HD} at $t_\rmn{evol}=2$~Myr, this time analysing the interpolated grids (Fig.~\ref{fig:fracvn-postgrid}). In general, we find similar outcomes to the post-processed particle results shown in the right panel of Fig.~\ref{fig:fracvn-particles}. The ratios of these mean fractional abundances after regridding to the mean fractional abundances of the post-processed tracers before regridding are plotted in Appendix~\ref{appendix:fracvn_ratios}, in the right panel of Fig.~\ref{fig:fracvn-ratios}.

For each carbon species, the similarity is lowest in the regime where the species is not the dominant representative of that element. Hence, the interpolated grid reflects the tracer values for CO least well at low densities where CO is rare, and the values for atomic carbon least well at high densities, where CO is saturated. We speculate that this occurs because the interpolation process introduces a certain degree of noise into the abundance profiles, which can be commensurate in scale to the true signal of a species in a regime where its fractional abundance is low. 

The accuracy of the regridding technique at preserving species' fractional abundances in their dominant density regimes supports our approach described in Section~\ref{subsection-regridalgorithm}, in spite of the shortcomings we have discussed. The computational cost of simulating larger chemical networks on-the-fly is simply prohibitive at this time. Therefore, we choose to accept a certain degree of inaccuracy and uncertainty as this is the only way to obtain filled, 3D density data for complex species that are not present in smaller chemical networks. 

\begin{figure}
    \centering
    \includegraphics[width=0.9\linewidth]{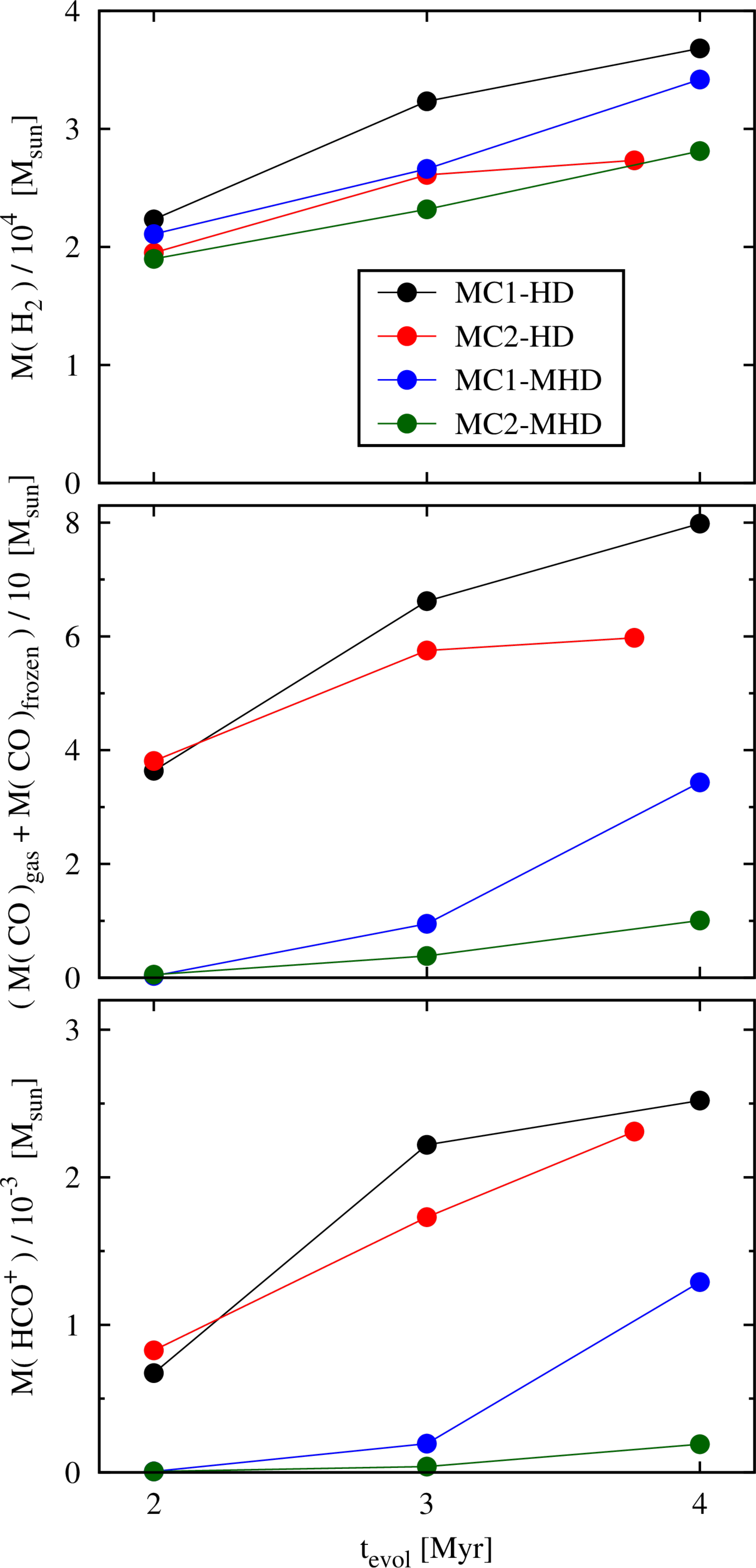}
    \caption{The masses of \ce{H2} (top), all CO (middle), and \ce{HCO+} (bottom) in the zoom-in regions of the four simulated clouds, vs. $t_\rmn{evol}$. These values are calculated by summing over the density grids produced by regridding the post-processed tracer particles. The CO mass represents the sum of the gaseous and frozen-out states. The masses of \ce{H2} and CO generally correspond to the values in the SILCC-Zoom clouds themselves \citep{seifried_silcc-zoom_2017,seifried_silcc-zoom_2020}. The character of the \ce{HCO+} growth resembles that of the CO growth, emphasizing that they exist in the same extinction regime.}
    \label{fig:massesvst}
\end{figure}

Finally, in Fig.~\ref{fig:massesvst}, we show the time-dependent total mass of \ce{H2} (top), \ce{CO} (middle, the sum of the gaseous and frozen-out states), and \ce{HCO+} (bottom) inside each simulation's zoom-in region. The masses are given at \mbox{$t_\rmn{evol}=2$, 3, and 4~Myr}. The exception is MC2-HD, which terminated at a final time of \mbox{$t_\rmn{evol}=3.76$~Myr}. The total masses of \ce{H2} and CO at the different time snapshots correspond to the masses given for the clouds in their originating papers \citep[][especially figure~2 of the latter]{seifried_silcc-zoom_2017,seifried_silcc-zoom_2020}, with deviations of $\sim10\%$ ascribed to the effects of post-processing and the uncertainties introduced in the regridding process.

In the bottom panel of Fig.~\ref{fig:massesvst}, we report the time-dependent total mass of \ce{HCO+} in the four simulated clouds, a novel result. The increase in \ce{HCO+} mass resembles the increase in CO mass, underscoring the close link between the two species. The HD clouds already possess some \ce{HCO+} at $t_\rmn{evol}=2$~Myr, but the MHD clouds have negligible \ce{HCO+} content. Only at later times have the MHD clouds condensed enough that their cores are sufficiently well-shielded for the formation of CO, and also of \ce{HCO+}. The fractional abundance of \ce{HCO+} (with respect to all hydrogen nuclei) averaged over each cloud is as low as $\sim10^{-12}$ (the MHD clouds at early times) and as high as $\sim 1.3 \times10^{-9}$ (the HD clouds at late times).

\subsection{The HCO$^+$ column density}
\label{subsection-colmaps}

\begin{figure*}
    \centering
    \includegraphics[width=\linewidth]{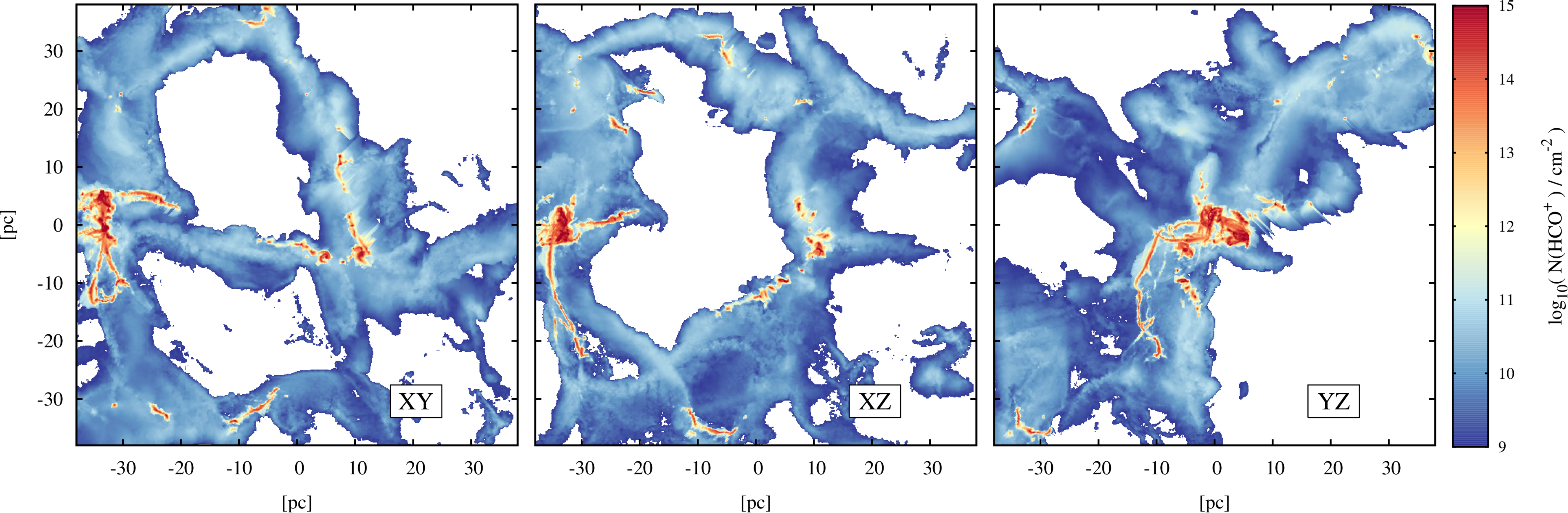}
    \includegraphics[width=\linewidth]{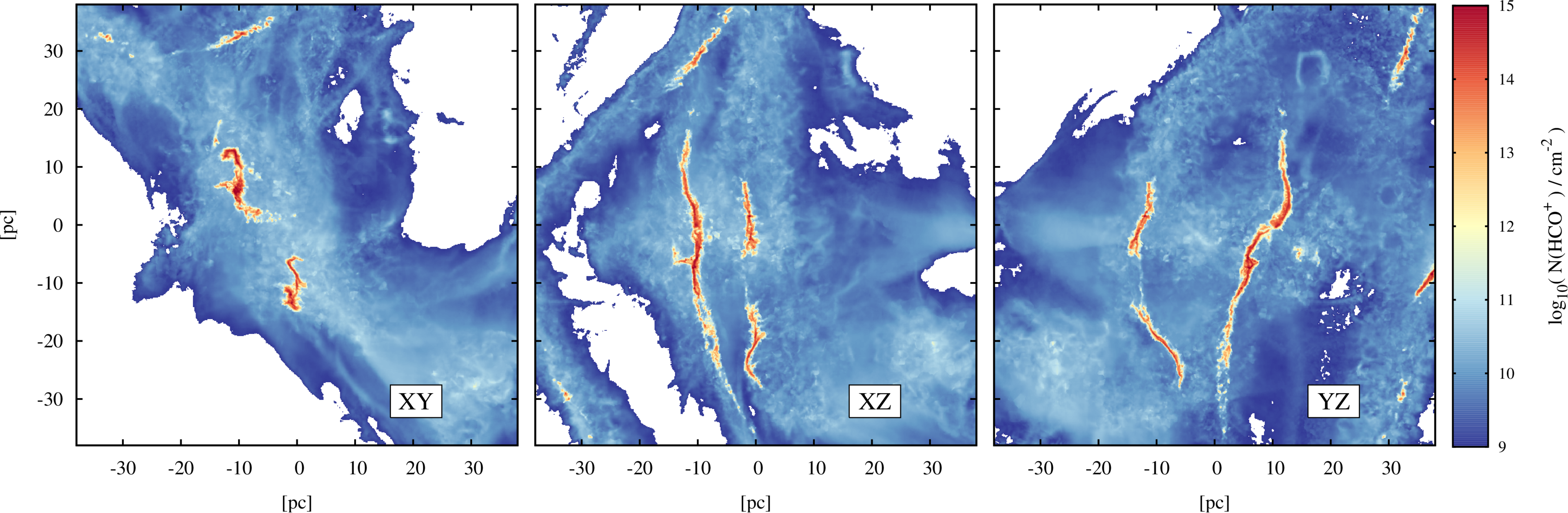}
    \caption{Maps of the column density of HCO\textsuperscript{+} for clouds \mbox{MC1-HD} (top row) and \mbox{MC1-MHD} (bottom row) at $t_\rmn{evol}=4$~Myr. The three columns show projections in the $x$-$y$, $x$-$z$, and $y$-$z$ planes respectively. Both clouds reach maximal values of $N\rmn{(HCO^{+})}\simeq 10^{15}$~cm$^{-2}$. The dense regions traced by HCO\textsuperscript{+} in \mbox{MC1-HD} are clumpy in shape, while the core distributions in \mbox{MC1-MHD} are more filamentary. Regions with \mbox{$N\rmn{(HCO^+)}<10^9$}~cm$^{-2}$ have been masked in white, underlining the more diffuse distribution of the molecular gas in the MHD simulation compared to the HD simulation.}
    \label{fig:maps_hcoj}
\end{figure*}

With these tracer-derived density grids, we can produce column density maps of species that were not present in the on-the-fly network. In Appendix \ref{appendix:Nmaps}, we compare column density maps of the post-processed H\textsubscript{tot}, H, H\textsubscript{2}, and CO abundances in cloud \mbox{MC1-HD} to the results shown in \cite{seifried_silcc-zoom_2017}. Our results are in good agreement with the original maps. As predicted in that work, post-processing the on-the-fly results for these species did not have a large impact upon the abundances.

Next, in Fig.~\ref{fig:maps_hcoj}, we show -- to our knowledge -- the first-ever maps of the \ce{HCO+} column density, $N$(\ce{HCO+}), in simulated molecular clouds. Since we show in Fig.~\ref{fig:hic-both} that much of the HCO\textsuperscript{+} formation in both \mbox{MC1-HD} and \mbox{MC1-MHD} takes place around $t_\rmn{evol}\simeq2$--3~Myr with a formation time-scale of $\tau \approx 1$~Myr, we choose to examine the column density maps at $t_\rmn{evol}=4$~Myr, i.e. after the principal epoch of HCO\textsuperscript{+} formation. The distributions of HCO\textsuperscript{+} in \mbox{MC1-HD} (top row) and \mbox{MC1-MHD} (bottom row) showcase the more diffuse molecular distribution seen in MHD simulations compared to hydrodynamic ones \citep{seifried_silcc-zoom_2020,ganguly2022}. Regions with $N\rmn{(HCO\textsuperscript{+})}<10^9$~cm$^{-2}$ have been masked in white. The maximal HCO\textsuperscript{+} column density in both clouds is on the order of $10^{15}$~cm$^{-2}$. The dense regions ($N\rmn{(HCO\textsuperscript{+})}>10^{12}$~cm$^{-2}$) in \mbox{MC1-HD} are clumpy, with lower-density regions (where $N\rmn{(HCO\textsuperscript{+})}=10^9$--$10^{11}$~cm$^{-2}$) only extending short distances from the peak sites. On the other hand, in \mbox{MC1-MHD}, the HCO\textsuperscript{+} is far more extended, forming a diffuse envelope tens of parsecs out from the filamentary structures where $N$(HCO\textsuperscript{+}) is maximal.

\begin{figure*}
    \centering
    \includegraphics[width=\linewidth]{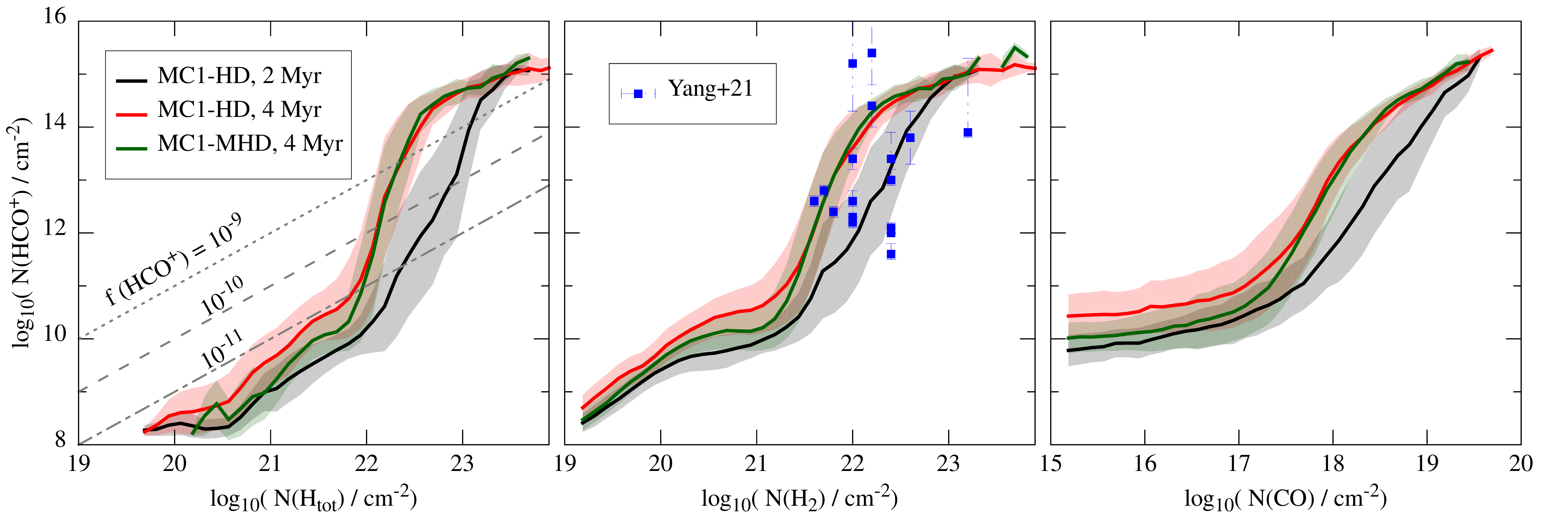}
    \caption{The average column densities of \ce{HCO+} for the $y$-$z$ projection of the molecular clouds \mbox{MC1-HD} at 2~Myr (black) and 4~Myr (red) and \mbox{MC1-MHD} at 4~Myr (green), vs. the column densities of H\textsubscript{tot} (left), H\textsubscript{2} (middle), and CO (right). The shaded areas represent one standard deviation from the respective average. In the middle panel, we overplot the observations of \citet{yang_search_2021} of infalling prestellar cores. Overall, we can see a significant increase in $N$(\ce{HCO+}) for $N$(H$_\rmn{tot})\sim N$(\ce{H2}) $\gtrsim 10^{21}$~cm$^{-2}$, and $N\rmn{(CO)}\gtrsim 10^{17}$~cm$^{-2}$. Clouds at later evolutionary stages have somewhat more \ce{HCO+} at given H$_\rmn{tot}$, \ce{H2}, and CO column densities.}
    \label{fig:NvsLots}
\end{figure*}

In Fig.~\ref{fig:NvsLots}, we contextualize our findings for $N$(HCO\textsuperscript{+}) in relation to the column densities of other species. The average values of $N$(HCO\textsuperscript{+}) from the $y$-$z$ projections of \mbox{MC1-HD} (red) and \mbox{MC1-MHD} (green) at $t_\rmn{evol}=4$~Myr are plotted against $N$(H\textsubscript{tot}) (left), $N$(H\textsubscript{2}) (middle), and $N$(CO) (right). The results for \mbox{MC1-HD} at $t_\rmn{evol}=2$~Myr (black) are plotted as well. We include several lines of constant fractional abundance (dashed), and find that the relationship between $f$(\ce{HCO+}) and the gas density shown for the tracers in Fig.~\ref{fig:hic-both} is preserved through the regridding process, with $f$(\ce{HCO+}) exceeding $10^{-9}$ at high gas density.

Around $N\rmn{(H_{tot})}\simeq N\rmn{(H_{2})}\simeq10^{21}$--$10^{22}$~cm$^{-2}$, the average $N$(HCO\textsuperscript{+}) for all clouds varies between $10^9$--$10^{11}$~cm$^{-2}$. For \mbox{$N\rmn{(H\textsubscript{tot})}~\gtrsim10^{22}$~cm$^{-2}$} and \mbox{$N$(CO)~$\gtrsim10^{18}$~cm$^{-2}$}, $N$(HCO\textsuperscript{+}) increases to maximal values of around $10^{15}$~cm$^{-2}$. Comparing the results at $t_\rmn{evol}=2$~Myr and $t_\rmn{evol}=4$~Myr for \mbox{MC1-HD}, we see that $N$(HCO\textsuperscript{+}) increases over time, as expected following the results in Section~\ref{section-tracerhistory}.

A systematic comparison of our results with observations is reserved for Section~\ref{subsub:w49}, but we provide some measurements by \cite{yang_search_2021} in the middle panel of Fig.~\ref{fig:NvsLots}. They measured $N$(\ce{HCO+}) for infalling cores, vs. $N$(\ce{H2}) derived from dust continuum measurements. We find that our data match their observations well, with peak values of $N\rmn{(HCO^+)}\simeq10^{15}$~cm$^{-2}$ around $N\rmn{(H_2)}\simeq10^{22}$~cm$^{-2}$.

\begin{figure*}
    \centering
    \includegraphics[width=\linewidth]{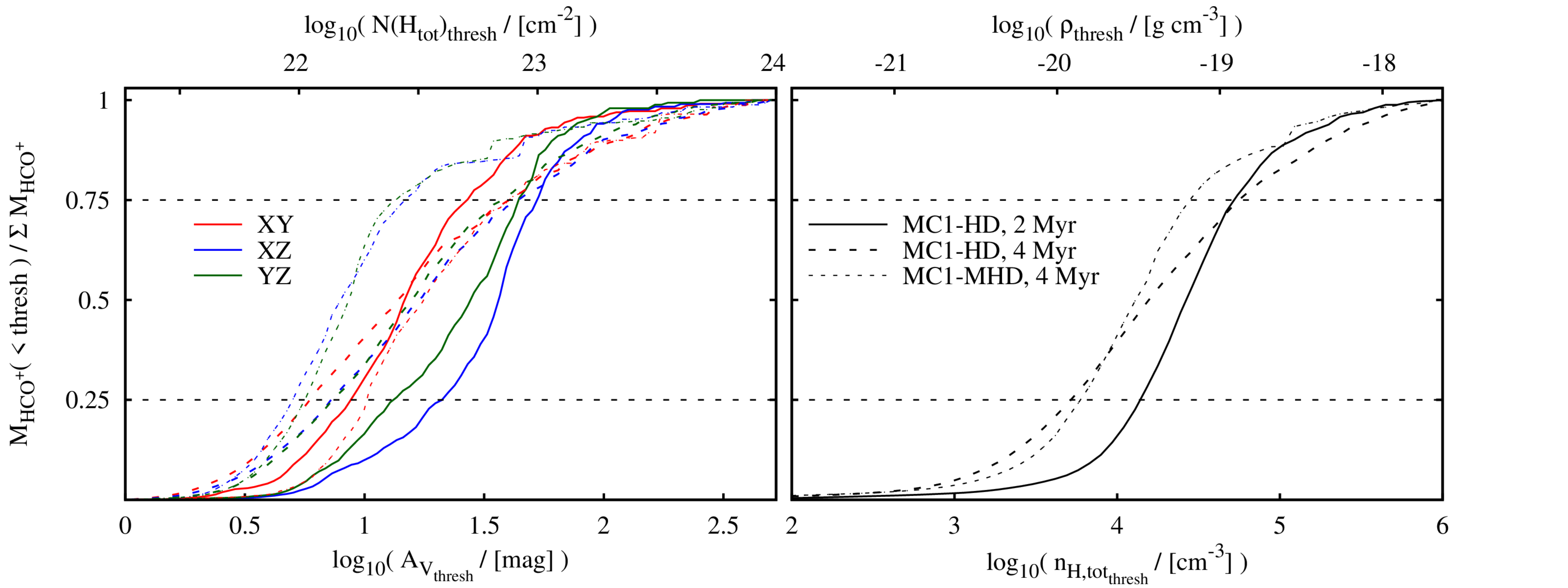}
    \caption{The cumulative mass fraction of HCO\textsuperscript{+} vs. $A_\rmn{V}$ (left) and $n$\textsubscript{H,tot} (right) for molecular cloud \mbox{MC1-HD} at $t_\rmn{evol}=2$~Myr (solid lines) and $t_\rmn{evol}=4$~Myr (large dashes), and cloud \mbox{MC1-MHD} at $t_\rmn{evol}=4$~Myr (small dashes). The three colors in the left-hand plot indicate different projections. The values for $A_\rmn{V}$ are calculated from projections of $N$(H\textsubscript{tot}) via the relation $N\rmn{(H_{tot})}=(1.87\times10^{21}$~cm$^{-2}$)$A_\rmn{V}$ \citep{draine_structure_1996}, and the corresponding HCO\textsuperscript{+} mass is calculated from the projections of $N$(HCO\textsuperscript{+}). In the right-hand plot, the values for $n$\textsubscript{H,tot} are given by a 3D density grid, and the corresponding HCO\textsuperscript{+} mass is calculated from a 3D grid of $n_\rmn{HCO^+}$. At $t_\rmn{evol}=2$~Myr, the cumulative distribution of the HCO\textsuperscript{+} is highly dependent on the viewing angle. At $t_\rmn{evol}=4$~Myr, the HCO\textsuperscript{+} distributions have become considerably more similar. The average interquartile range (containing 50\% of the HCO$^+$ mass) for $A_\rmn{V}$ ranges from $\sim10$ to $\sim30$ increasing slightly over time. Corresponding relations between the HCO\textsuperscript{+} mass and $n_\rmn{H,tot}$ are seen in the right-hand plot. Cloud \mbox{MC1-MHD}, which condenses more slowly than the hydrodynamic cloud \mbox{MC1-HD}, contains a greater fraction of its total HCO\textsuperscript{+} at lower $A_\rmn{V}$ or $n_\rmn{H,tot}$.}
    \label{fig:cumulmass}
\end{figure*}

\subsubsection{The distribution of \ce{HCO+}}

Next, we study the cumulative mass distribution of HCO\textsuperscript{+} as a function of the observed visual extinction using the relation $N\rmn{(H_{tot})}=(1.87\times10^{21}$~cm$^{-2}$)$A_\rmn{V}$ \citep[][also as used in the SILCC-Zoom chemical network; see Section~\ref{subsection-reference-simulation}]{draine_structure_1996}. In
Fig.~\ref{fig:cumulmass} we show the fraction of mass sitting below a certain $A_\rmn{V}$-threshold for the cloud \mbox{MC1-HD} at $t_\rmn{evol}=2$~Myr (solid lines) and $t_\rmn{evol}=4$~Myr (long dashes), and cloud \mbox{MC1-MHD} at $t_\rmn{evol}=4$~Myr (short dashes). Each color represents a different projection. We see that for \mbox{MC1-HD} at $t_\rmn{evol}=2$~Myr, the HCO\textsuperscript{+} distribution varies for different projections, with the first quartile being reached at $A_\rmn{V}$ values of 8--20. The third quartile is typically reached around $A_\rmn{V}=20$ --50. At $t_\rmn{evol}=4$~Myr, the different projections correspond much more closely. The third quartile value remains almost unchanged, but the first quartile value is now systematically lower, around $A_\rmn{V}~\simeq~5$. This indicates that a significant amount of HCO\textsuperscript{+} has formed between the two snapshots at lower extinctions, widening the average interquartile range. We repeat this procedure for the other simulated clouds, and found similar results. The lines for the three projections for the MHD clouds are still dissimilar at $t_\rmn{evol}=4$~Myr, as expected given these clouds' longer time-scale of gravitational collapse compared to the HD case. On average, we find that 50\% of the HCO\textsuperscript{+} mass -- corresponding to the average interquartile range -- lies between $A_\rmn{V}\sim10$ and $\sim30$. 

In the right-hand panel of Fig.~\ref{fig:cumulmass}, we plot again the cumulative mass of HCO\textsuperscript{+}, but this time using the values of $n_\rmn{HCO^+}$ and $n_\rmn{H,tot}$ from the 3D density grids rather than column density projections. The same time-dependent qualitative relationship is seen as in the left-hand panel. This indicates again that HCO\textsuperscript{+} is being formed over time outside the very dense regions. Overall, we find that 50\% of the HCO\textsuperscript{+} is located at $n_\rmn{H,tot}\sim10^{3.5}$ -- $10^{4.5}$~cm$^{-3}$.

\subsubsection{Resolution effects}
\label{subsub:reseffects}

Our resolution in the maps in Fig.~\ref{fig:maps_hcoj} is 0.125~pc, up to one or two orders of magnitude higher than what is available in many observations \citep[e.g.][]{barnes_lego_2020,sanhueza_chemistry_2012,nayana_alma_2020}. To improve the comparison of our data to such observations, we convolve our maps of $N$(HCO\textsuperscript{+}) with a Gaussian filter. By selecting the size of the filter, we can emulate any coarser resolution.

\begin{figure}
    \centering
    \includegraphics[width=\linewidth]{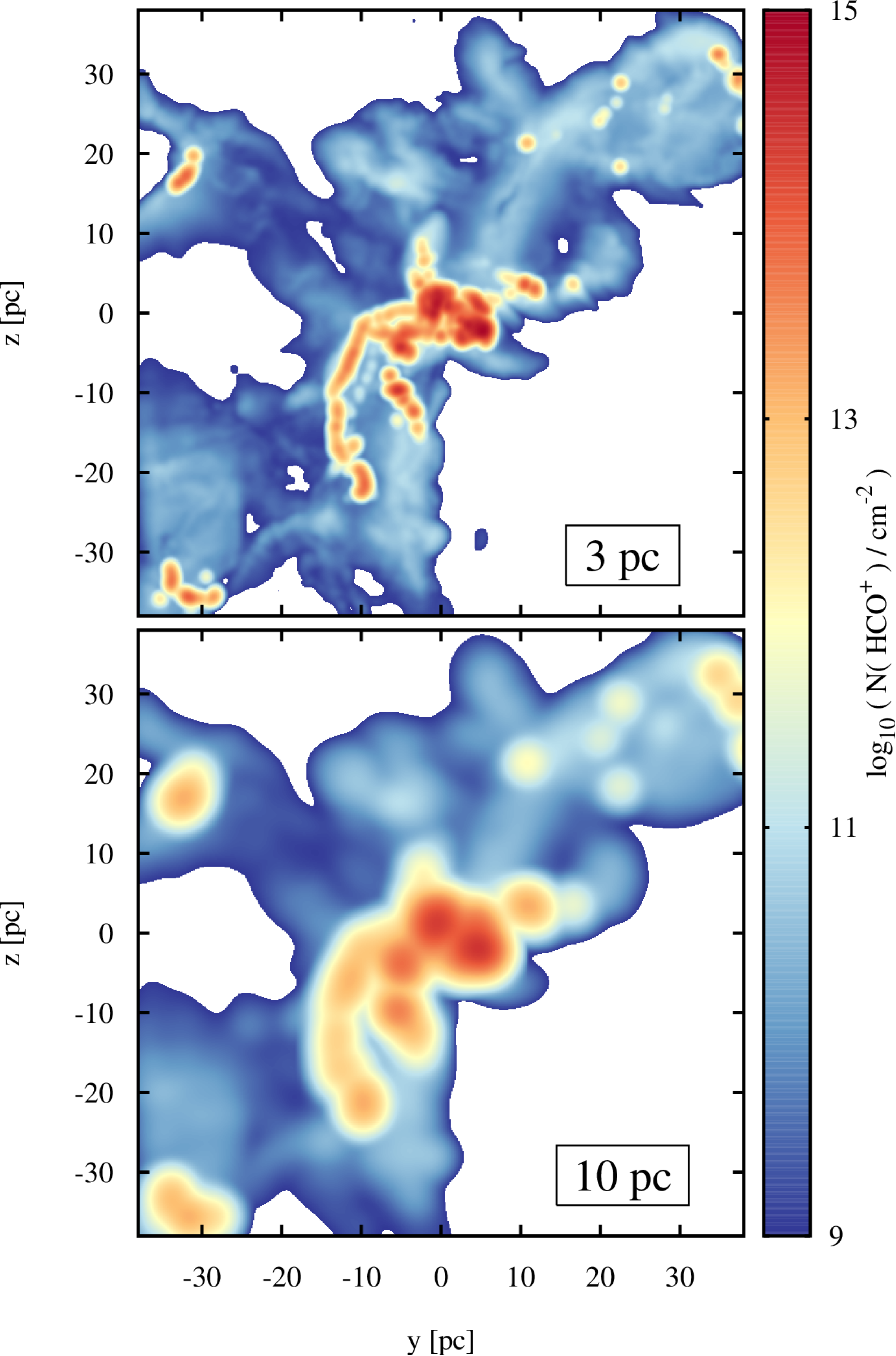}
    \caption{Column density maps of $N$(\ce{HCO+}) from cloud MC1-HD at $t_\rmn{evol}=4$~Myr (same as the top right panel of Fig.~\ref{fig:maps_hcoj}), but convolved with Gaussian filters of increasing beam size to emulate resolutions of 3~pc (top), and 10~pc (bottom). As resolution decreases, filamentary-scale structures become unresolved, and the peak $N\rmn{(HCO^+)}$ decreases as the beam samples surrounding areas of lower column density.}
    \label{fig:maps_convolved}
\end{figure}

\begin{figure*}
    \centering
    \includegraphics[width=\linewidth]{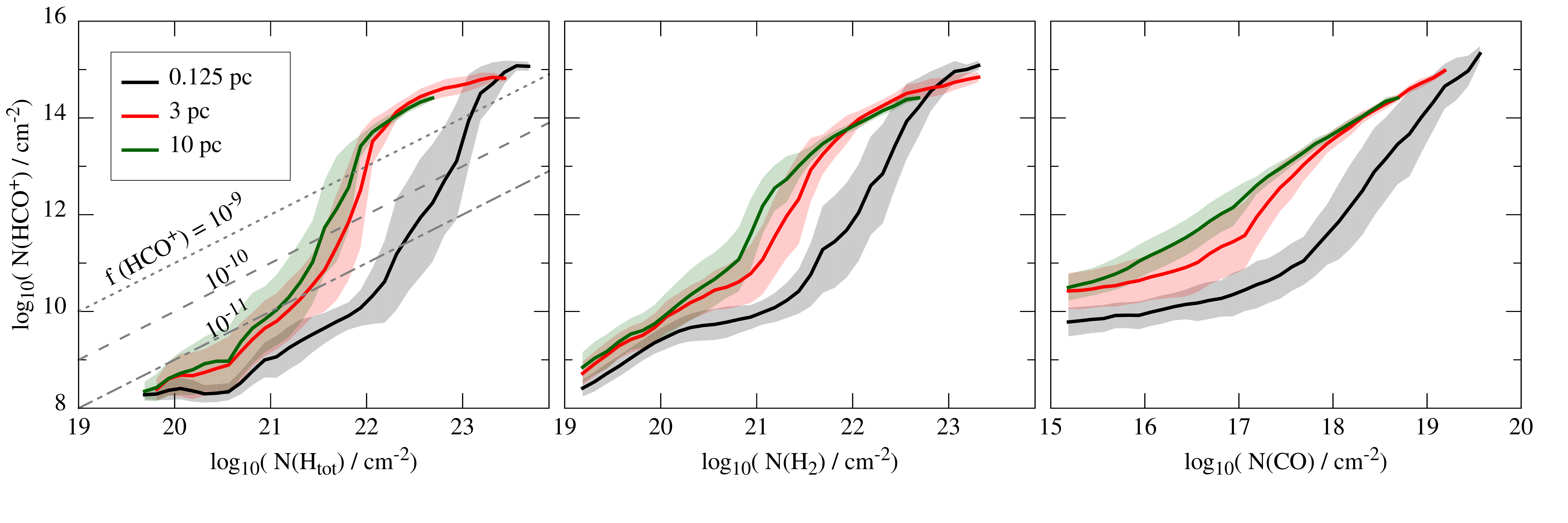}
    \caption{Same as Fig.~\ref{fig:NvsLots}, but only for the $y$-$z$ projection of \mbox{MC1-HD} at $t_\rmn{evol}=4$~Myr, convolved with Gaussian filters of increasing beam size to emulate resolutions of 3~pc (red) and 10~pc (green). The original resolution of 0.125~pc is repeated for comparison (black). As resolution decreases, higher values of $N\rmn{(HCO^+)}$ are found at given values of $N\rmn{(H_{tot})}$, $N\rmn{(H_{2})}$, and $N$(CO).}
    \label{fig:scatters_convolved}
\end{figure*}

In Fig.~\ref{fig:maps_convolved}, we show one of our 0.125~pc resolution $N$(HCO\textsuperscript{+}) maps altered in this way to emulate resolutions of 3~pc (top) and 10~pc (bottom). We apply these convolution to the $y$-$z$ projection of \mbox{MC1-HD} at $t_\rmn{evol}=4$~Myr, corresponding to the top right panel of Fig.~\ref{fig:maps_hcoj}. As the resolution decreases, the filamentary-scale structures \mbox{($\sim0.1$~pc)} lose their intricate detail, and adjacent density peaks (for instance the two small neighbouring peaks in the upper left of the cloud) become unresolved. The complex distribution of HCO\textsuperscript{+} in the more diffuse areas where $N\rmn{(HCO\textsuperscript{+})}=10^9$--$10^{11}$~cm$^{-2}$ becomes smoother as well. The peak $N$(HCO\textsuperscript{+}) value remains around $10^{15}$~cm$^{-2}$ for all resolutions. However, at a resolution of 10~pc, only the very densest and clumpiest regions retain this peak column density, which is found along the lengths of the filamentary structures in the 0.125~pc resolution map.

Next, in Fig.~\ref{fig:scatters_convolved} we repeat the comparison of $N$(HCO\textsuperscript{+}) with $N$(H\textsubscript{tot}), $N$(H\textsubscript{2}), and $N$(CO) which we performed in Fig.~\ref{fig:NvsLots}. This time, however, we restrict ourselves to the $y$-$z$ projection of \mbox{MC1-HD} at $t_\rmn{evol}=4$~Myr, and compare the average column densities at a resolution of 0.125~pc (black) to the column densities from maps emulating resolutions of 3~pc (red) and 10~pc (green). Decreasing the resolution causes the ratios between $N$(HCO\textsuperscript{+}) and the other column densities to increase, due to the Gaussian broadening of the central features with high HCO\textsuperscript{+} density. This effect implies that low-resolution observations outside the densest regions of a molecular cloud would measure higher values for $N$(HCO\textsuperscript{+}) than are physically present.

\subsubsection{Comparison to observations of W49A}
\label{subsub:w49}

Transforming our column density results to match the resolution of observations allows us to make direct comparison to measurements of $N$(HCO\textsuperscript{+}) in nature. For this purpose, we present values for $N$(HCO\textsuperscript{+}) in the massive star-forming region W49A, as observed during the LEGO project (\citealp{barnes_lego_2020}; see also \citealp{kauffmann_molecular_2017}).

The W49A observations were performed by the IRAM 30m telescope at a resolution of $\sim30''$, then smoothed to $60''$, which corresponds to physical scales of $\sim3$~pc at the cloud's distance of 11~kpc \citep{Zhang_2013}. 
We calculate the values of $N$(HCO\textsuperscript{+}) from the HCO\textsuperscript{+} $J=1\rightarrow0$ emission (Neumann et al. in prep.), which has a high signal-to-noise ratio across much of the W49 region.
We calculate the optical depth of the line using corresponding observations of H\textsuperscript{13}CO\textsuperscript{+}(1-0) emission. 
Where the H\textsuperscript{13}CO\textsuperscript{+} emission is below the noise level (corresponding to $N\rmn{(H_2)}\lesssim10^{22}~\rmn{cm^{-2}}$), we make the assumption that the line is optically thin to approximate the lower limit of the column density. 
We determine the excitation temperature of HCO\textsuperscript{+} by minimizing the column density equation when using both the $J=1\rightarrow0$ line from LEGO-IRAM \citep{barnes_lego_2020} and $J=3\rightarrow2$ line from LEGO-APEX (Neumann et al. in prep), since $N$(HCO\textsuperscript{+}) should be the same when determined from both lines. 
We complement these data with dust continuum observations from the {\it Herschel Space Observatory} large program Hi-Gal \citep{molinari2011}, also smoothed to a matched resolution of $60''$, to recover the molecular hydrogen column density (see \citealp{barnes_lego_2020}). 
The complete details of the APEX observations and full calculation of the column densities will be presented in a future work (Neumann et al. in prep.). 

\begin{figure}
    \centering
    \includegraphics[width=\linewidth]{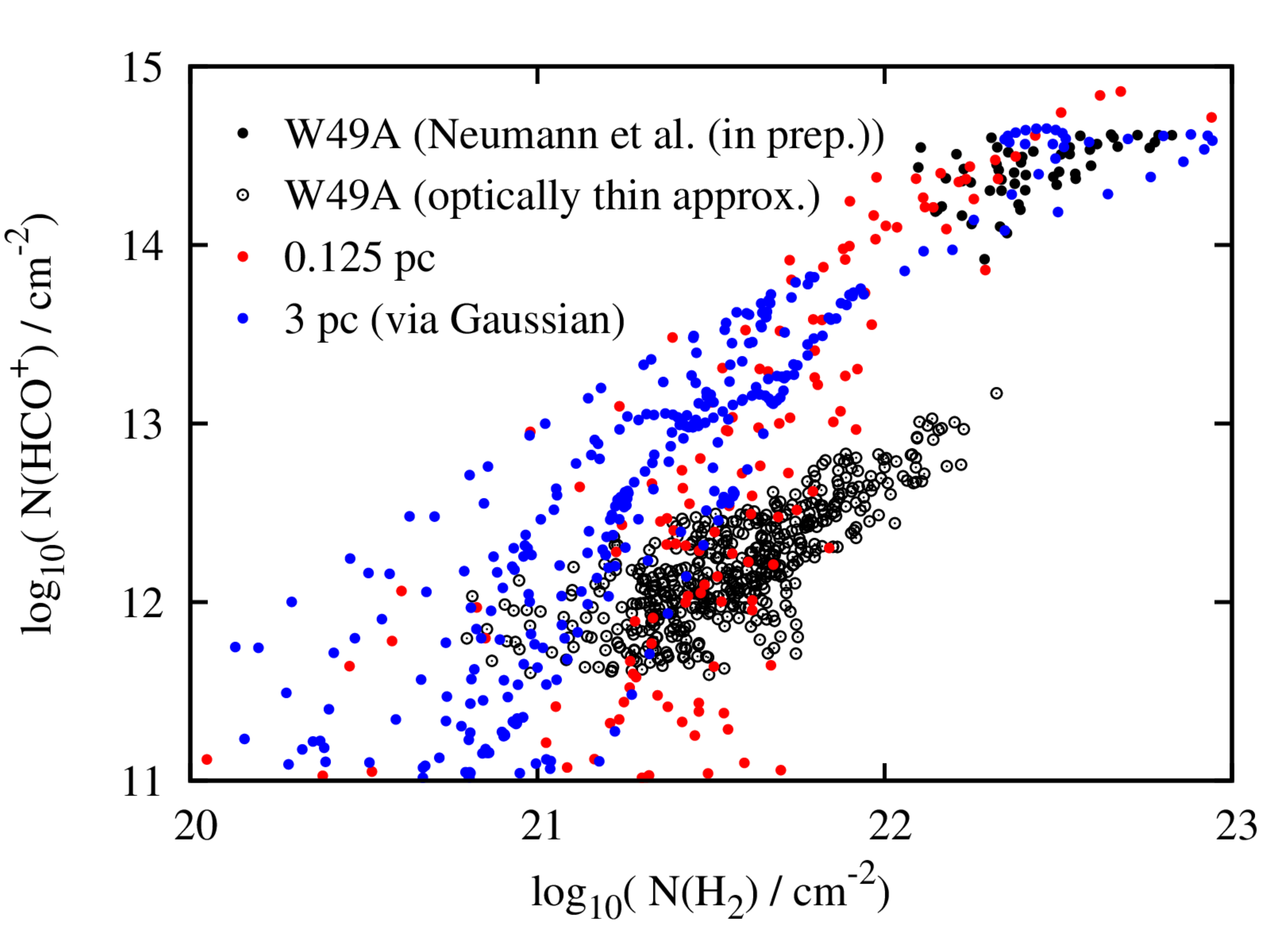}
    \caption{$N$(HCO\textsuperscript{+}) vs. $N$(H\textsubscript{2}) for the $y$-$z$ projection of \mbox{MC1-HD} at \mbox{$t_{evol}=4$~Myr} at the original resolution of 0.125~pc (red) and convolved with a Gaussian filter to emulate a resolution of 3~pc (blue), compared to observations of the star-forming region W49A \citep[Neumann et al. in prep.; see also][]{barnes_lego_2020} at a physical resolution of $\sim3$~pc. Above $N\rmn{(H_2)}\simeq10^{22}$~cm$^{-2}$, the intensity of H\textsuperscript{13}CO\textsuperscript{+} emission is sufficient to constrain the optical depth for the calculation of $N$(HCO\textsuperscript{+}) (solid black). Below $N\rmn{(H_2)}\simeq10^{22}$~cm$^{-2}$, we approximate the emission as optically thin (empty black). The approximation represents a lower limit to the actual value of $N$(HCO\textsuperscript{+}), and gradually conforms better to reality as $N$(H\textsubscript{2}) decreases further. Our simulations are in very good agreement with observations above $N\rmn{(H_2)}\simeq10^{22}$~cm$^{-2}$ where the optical depth is calculated explicitly, as well as around $N\rmn{(H_2)}\simeq10^{21}$~cm$^{-2}$ where the optically thin approximation is relatively accurate. Reducing the resolution of the modeled map to 3~pc (the same as the observations) with Gaussian convolution slightly improves the correspondence between the modeled and observed column densities.}
    \label{fig:w49}
\end{figure}

In Fig.~\ref{fig:w49}, we plot $N$(HCO\textsuperscript{+}) vs. $N\rmn{(H_2)}$ for these observations, and for our simulations at resolutions of 0.125~pc and 3~pc (using Gaussian convolution as previously explained). For $N$(H$_2)\gtrsim10^{22}~\rmn{cm^{-2}}$, our simulations match the observed values quite well. At the peak values of $N$(H\textsubscript{2}), our simulated data at a resolution of 3~pc (matching the physical resolution of the observations) are in slightly better agreement than at the original 0.125~pc resolution. The regions where $N\rmn{(H_2)}\lesssim10^{22}$~cm$^{-2}$ are in less good agreement with our simulations. We attribute this to the fact that below $N\rmn{(H_2)}\simeq10^{22}$~cm$^{-2}$, the observed $N$(HCO\textsuperscript{+}) were calculated with an optically thin assumption due to the lack of significant H\textsuperscript{13}CO\textsuperscript{+} emission in that column density regime. An improved approximation for the optical depth where $N\rmn{(H_2)}\lesssim10^{22}$~cm$^{-2}$ would likely lead to higher values of $N$(HCO\textsuperscript{+}) there, thus presumably in better agreement with our simulated data.

\section{Caveats and future directions}
\label{section:discussion}

Via comparison to column density observations of HCO\textsuperscript{+}, we have shown that our post-processing and regridding algorithms can reconstruct the HCO\textsuperscript{+} abundance in molecular clouds. Moreover, by post-processing the tracer particle abundances over individual timesteps, we consider the chemistry in a non-equilibrium approach and utilize the momentary environmental parameters like the density and temperature in our solutions to the rate equations. These factors are essential to a truly time-dependent chemistry. Despite our post-processing and regridding routines giving robust results when compared with theoretical benchmark results and observations, some caveats should be kept in mind.

When post-processing the chemistry of the ISM, the local temperature and shielding must be handled carefully. In principle, changes to the abundance of e.g. CO could shift the thermal state of the gas. This could lead to dynamic motions, altering the shielding profile, which would alter the chemistry again. Our method fixes the temperature and shielding at each timestep using the values from the underlying MHD reference simulation, and does not attempt to update or re-model them based on the post-processed abundances. This might  in general lead to inconsistencies. However, we believe that the usage of the NL99 chemistry network will provide us with temperatures and basic abundances, which are reliable enough to be used in the subsequent post-processing step. E.g. the NL99 network includes an extensive list of cooling and heating processes \citep[see Table~1 in][]{glover_modelling_2010}. Furthermore, post-processing the NL99 chemistry does not much impact the abundances of species like C, \ce{C+}, and CO which were already modeled in the original network (see Fig.~\ref{fig:fracvn-particles}). Thus, we are confident that chemically post-processing the simulations should not radically alter the abundances of these thermally-relevant species, nor the associated shielding properties. However, applying our post-processing method to a simulation with an on-the-fly network that has less-robust thermal modeling may lead to inconsistencies.

Modeling the CRIR is also important to self-consistent post-processing. To further ensure the continuity of the thermal environment, we decided to copy the H$_2$-CRIR value, $\zeta=6\times10^{-17}$~s$^{-1}$, used in the reference SILCC-Zoom simulations. This is set constant everywhere, whereas \citet{Padovani2018} suggest a decrease in deeply embedded structures. A model for the CRIR which decreases with density would, for instance, impact the abundance profile of \ce{H3+}. As discussed in Section~\ref{subsection:formationpaths}, the \ce{H3+} abundance bottlenecks the high-extinction \ce{HCO+} abundance in our post-processing network via the dominant reaction \mbox{\ce{H3+} + CO}. Attenuation of the cosmic rays might diminish the \ce{H3+} balance in dense gas and thus decrease the \ce{HCO+} density. Testing the importance of this effect will require new simulations run on-the-fly with an attenuated CRIR model.

The chemical network employed here to showcase the post-processing is still small. Containing only 37 gas-phase species, the network particularly lacks nitrogen-bearing molecules like HCN which are used as dense gas tracers in observations \citep{papadopoulos_hcn_2007,godard_molecular_2010,kauffmann_molecular_2017,goicoechea_molecular_2019}. This does not impact the results of \ce{HCO+} shown in this work, as our network already models \ce{HCO+} comprehensively. However, when considering how to expand this network for future uses, one must balance the comprehensiveness of a chemical network with its practical usability in astrophysical simulations, which are already computationally expensive even before considering chemistry. Developing and validating this post-processing scheme required a chemical network with a reasonably short convergence time \citep{seifried_modelling_2016}. Future works, for instance analysis of the time-dependent nature of deuterium fractionation, will need larger and more comprehensive networks, which will need to be validated in turn.

We include a freeze-out approximation in our network, which creates a noticeable difference in CO and H\textsubscript{2}O abundance when compared to networks lacking these approximations \citep[see Fig.~\ref{fig:bench_freezeout}, and][]{borchert2022}, but the impact of grain chemistry on chemical abundances is complex \citep{flower_freeze-out_2005,bovino_h2_2017}. In our network, the grains themselves are treated in a simple manner neglecting subtleties like ionization of the grains or changing sticking coefficients due to, e.g., the time-dependent variation in the composition of the ice mantles.

Looking ahead, we intend to create synthetic emission maps from our regridded species data. This would provide us with a more direct comparison to observations, and properly account for the optical depth effects discussion in Section~\ref{subsub:w49}. Moreover, in a forthcoming paper, we will apply these techniques to analysis of CO-dark gas, in particular focusing on OH.

\section{Conclusions}
\label{section-conclusion}

We present our novel chemical post-processing methodology for 3D-MHD simulations of the ISM and molecular clouds. This methodology provides non-equilibrium abundances for any species present in a chemical network of arbitrary complexity. In this work, we have applied our methods to investigating the time-dependent evolution of HCO\textsuperscript{+} in molecular clouds modeled in the SILCC-Zoom project. We summarize the methodology itself as follows:
\begin{enumerate}
    \item Rather than post-process the instantaneous abundances of simulation snapshots, we post-process the abundances on tracer particles which are injected into the simulation. Because these tracers follow the gas flow, they report the time-dependent chemistry and dynamics of the local fluid environment.

    \item We use a chemical network of 39 species and 301 reactions to post-process the tracer particles' chemical abundances. This network includes many species of astrochemical interest, in particular HCO\textsuperscript{+}. It also models the freeze-out of CO and H\textsubscript{2}O on to dust grains.
    
    \item We use the chemical rate equation solver \textsc{Krome} \citep{grassi_krome_2014} to post-process the tracer abundances from the SILCC-Zoom simulations. The on-the-fly abundances of the hydrogen and carbon species are used to initialize the post-processing of each tracer particle. From that point on, we post-process the updated abundances directly at each time step. 
    
    \item By advancing the chemistry over timesteps matching those of the simulation's particle snapshots, and using the time-dependent environmental parameters of the simulated cloud as inputs to the chemical network, we recover the non-equilibrium chemical state of the tracer particles over the entire history of the simulation. 
    
    \item We implement a subcycling routine to correct for large changes in the environmental parameters. This subdivides a tracer’s evolution timestep if the local environmental parameters experience more than a user-defined percent change $s$. We show that the post-processed abundances are generally converged for any value of $s$, and select a value of $s=10\%$.
    
    \item We present a novel algorithm to regrid a snapshot of the post-processed tracer abundances, using an iterative scheme to recover a volume-filling density grid, from which we make column density maps.
    We benchmark this algorithm against the masses of hydrogen and carbon in the original simulations, indicating an overall accuracy of better than $\sim10$\%.
\end{enumerate}

We can thus calculate non-equilibrium abundances for any species in a chemical network of arbitrary size, for a fraction of the computational cost of running that network on-the-fly in 3D-MHD simulations. Throughout this paper, we have explored the evolution of the HCO\textsuperscript{+} abundance in HD and MHD molecular clouds simulated in the SILCC-Zoom project. Our results include:
\begin{enumerate}
    \item We find that HCO\textsuperscript{+} predominantly forms at densities of $n_{\textrm{H,tot}}=10^3$--$10^4$~cm$^{-3}$. The formation of HCO\textsuperscript{+} occurs in situ in this density range, rather than in a high density regime followed by turbulent mixing into lower-density regions. The typical time-scale of HCO\textsuperscript{+} formation is on the order of 1~Myr.
    
    \item We show that the HCO\textsuperscript{+} formation time $\tau$ is inversely correlated with the abundances of species that are present in the high density regime and also participate in a formation pathway of HCO\textsuperscript{+}. 
    
    \item We show that different formation pathways of HCO\textsuperscript{+} predominate in different $A_\rmn{V}$ regimes. Up to $A_\rmn{V,3D}\simeq0.4$, the dominant reaction is \mbox{HOC\textsuperscript{+} + H\textsubscript{2}}, although it contributes very little to the total HCO\textsuperscript{+} mass due to the rarity of these reactants in the low extinction environment. From $A_\rmn{V,3D}\simeq0.4$--$3$, the dominant reaction is \mbox{CO\textsuperscript{+} + H\textsubscript{2}}. Above $A_\rmn{V,3D}\simeq3$, the dominant reaction is \mbox{H\textsubscript{3}\textsuperscript{+} + CO}, contributing more than 90\% of the total HCO\textsuperscript{+} production. This system of dominant reactions is established very quickly for HD clouds, but takes a few Myrs longer for more slowly evolving clouds containing magnetic fields.
    
    \item
    We produce the to-date first column density maps of HCO\textsuperscript{+} of simulated molecular clouds. We show that around \mbox{$N$(H$_\rmn{tot}) \sim 10^{22}$ cm$^{-2}$} and \mbox{$N$(CO) $\sim 10^{18}$ cm$^{-2}$},  $N$(HCO\textsuperscript{+}) rises quickly from values of \mbox{$10^{10}$--$10^{11}$ cm$^{-2}$} to peak values as high as $10^{15}$~cm$^{-2}$.
 
    \item We find that in MHD clouds, the distribution of HCO\textsuperscript{+} is more diffuse than in the HD clouds. The results match well with recent observations of HCO\textsuperscript{+}. 
    \item We find that 50\% of the HCO$^+$ mass is found at visual extinctions between $\sim10$ and $\sim30$, or at values of $n_\rmn{H,tot}$ between $\sim10^{3.5}$ and $\sim10^{4.5}$~cm$^{-3}$.
\end{enumerate}

Because our post-processing method is much faster than directly coupling large chemical networks to MHD simulations, it can be profitably applied to astrophysical problems which require complex, time-dependent chemical modeling. It is our hope that these tools can be used to support and guide future observational campaigns.

\section*{Acknowledgements}

The authors thank the referee T. Grassi for his helpful and constructive report, which greatly improved the paper.
MP, DS, SW, and BG would like to acknowledge funding support from the Deutsche Forschungsgemeinschaft (DFG) via the
Sonderforschungsbereich (SFB) 956, \textit{Conditions and Impact of Star Formation} (projects C5 and C6). Furthermore, the project is receiving funding from the programme “Profilbildung 2020", an initiative of the Ministry of Culture and Science of the State of Northrhine Westphalia. The sole responsibility for the content of this publication lies with the authors. The SILCC-Zoom simulations were performed on SuperMUC at the Leibniz Computing Centre, and the post-processing and additional analysis were performed on ODIN at the Regionales Rechenzentrum der Universität zu Köln (RRZK). The FLASH code was developed partly by the DOE-supported Alliances Center for Astrophysical Thermonuclear Flashes (ASC) at the University of Chicago.
ATB and FB would like to acknowledge funding from the European Research Council (ERC) under the European Union’s Horizon 2020 research and innovation programme (grant agreement No.726384/Empire).

\section*{Data Availability}

The data and post-processing code underlying this paper can be shared for scientific purposes after request to the authors.

\bibliographystyle{mnras}
\bibliography{bibliography.bib}

\begin{thebibliography}{}
\makeatletter
\relax
\def\mn@urlcharsother{\let\do\@makeother \do\$\do\&\do\#\do\^\do\_\do\%\do\~}
\def\mn@doi{\begingroup\mn@urlcharsother \@ifnextchar [ {\mn@doi@}
  {\mn@doi@[]}}
\def\mn@doi@[#1]#2{\def\@tempa{#1}\ifx\@tempa\@empty \href
  {http://dx.doi.org/#2} {doi:#2}\else \href {http://dx.doi.org/#2} {#1}\fi
  \endgroup}
\def\mn@eprint#1#2{\mn@eprint@#1:#2::\@nil}
\def\mn@eprint@arXiv#1{\href {http://arxiv.org/abs/#1} {{\tt arXiv:#1}}}
\def\mn@eprint@dblp#1{\href {http://dblp.uni-trier.de/rec/bibtex/#1.xml}
  {dblp:#1}}
\def\mn@eprint@#1:#2:#3:#4\@nil{\def\@tempa {#1}\def\@tempb {#2}\def\@tempc
  {#3}\ifx \@tempc \@empty \let \@tempc \@tempb \let \@tempb \@tempa \fi \ifx
  \@tempb \@empty \def\@tempb {arXiv}\fi \@ifundefined
  {mn@eprint@\@tempb}{\@tempb:\@tempc}{\expandafter \expandafter \csname
  mn@eprint@\@tempb\endcsname \expandafter{\@tempc}}}

\bibitem[\protect\citeauthoryear{{Aikawa}, {Miyama}, {Nakano}  \&
  {Umebayashi}}{{Aikawa} et~al.}{1996}]{Aikawa96}
{Aikawa} Y.,  {Miyama} S.~M.,  {Nakano} T.,   {Umebayashi} T.,  1996, \mn@doi
  [\apj] {10.1086/177644}, \href
  {https://ui.adsabs.harvard.edu/abs/1996ApJ...467..684A} {467, 684}

\bibitem[\protect\citeauthoryear{Barnes et~al.,}{Barnes
  et~al.}{2020}]{barnes_lego_2020}
Barnes A.~T.,  et~al., 2020, \mn@doi [\mnras] {10.1093/mnras/staa1814}, 497,
  1972

\bibitem[\protect\citeauthoryear{Beck \& Wielebinski}{Beck \&
  Wielebinski}{2013}]{beck_magnetic_2013}
Beck R.,  Wielebinski R.,  2013, in Oswalt T.~D.,  Gilmore G.,  eds, , Planets,
  {Stars} and {Stellar} {Systems}: {Volume} 5: {Galactic} {Structure} and
  {Stellar} {Populations}.
Springer Netherlands, Dordrecht, pp 641--723,
  \mn@doi{10.1007/978-94-007-5612-0_13}

\bibitem[\protect\citeauthoryear{Bergin, Hartmann, Raymond  \&
  Ballesteros‐Paredes}{Bergin et~al.}{2004}]{bergin_molecular_2004}
Bergin E.~A.,  Hartmann L.~W.,  Raymond J.~C.,   Ballesteros‐Paredes J.,
  2004, \mn@doi [ApJ] {10.1086/422578}, 612, 921

\bibitem[\protect\citeauthoryear{Bisbas, Tan  \& Tanaka}{Bisbas
  et~al.}{2021}]{Bisbas2021}
Bisbas T.~G.,  Tan J.~C.,   Tanaka K. E.~I.,  2021, \mn@doi [MNRAS]
  {10.1093/mnras/stab121}, 502, 2701

\bibitem[\protect\citeauthoryear{{Bisbas}, {Van Dishoeck}, {Hu}  \&
  {Schruba}}{{Bisbas} et~al.}{2022}]{bisbas_proceedings_2022}
{Bisbas} T.~G.,  {Van Dishoeck} E.,  {Hu} C.-Y.,   {Schruba} A.,  2022, in
  European Physical Journal Web of Conferences. p. 00013,
  \mn@doi{10.1051/epjconf/202226500013}

\bibitem[\protect\citeauthoryear{{Borchert}, {Walch}, {Seifried}, {Clarke},
  {Franeck}  \& {N{\"u}rnberger}}{{Borchert} et~al.}{2022}]{borchert2022}
{Borchert} E.~M.~A.,  {Walch} S.,  {Seifried} D.,  {Clarke} S.~D.,  {Franeck}
  A.,   {N{\"u}rnberger} P.~C.,  2022, \mn@doi [\mnras]
  {10.1093/mnras/stab3354}, \href
  {https://ui.adsabs.harvard.edu/abs/2022MNRAS.510..753B} {510, 753}

\bibitem[\protect\citeauthoryear{Bovino, Grassi, Schleicher  \& Caselli}{Bovino
  et~al.}{2017}]{bovino_h2_2017}
Bovino S.,  Grassi T.,  Schleicher D. R.~G.,   Caselli P.,  2017, \mn@doi [ApJ]
  {10.3847/2041-8213/aa95b7}, 849, L25

\bibitem[\protect\citeauthoryear{{Cadiou}, {Dubois}  \& {Pichon}}{{Cadiou}
  et~al.}{2019}]{cadiou19}
{Cadiou} C.,  {Dubois} Y.,   {Pichon} C.,  2019, \mn@doi [\aap]
  {10.1051/0004-6361/201834496}, \href
  {https://ui.adsabs.harvard.edu/abs/2019A&A...621A..96C} {621, A96}

\bibitem[\protect\citeauthoryear{Capelo, Bovino, Lupi, Schleicher  \&
  Grassi}{Capelo et~al.}{2018}]{capelo_effect_2018}
Capelo P.~R.,  Bovino S.,  Lupi A.,  Schleicher D. R.~G.,   Grassi T.,  2018,
  \mn@doi [MNRAS] {10.1093/mnras/stx3355}, 475, 3283

\bibitem[\protect\citeauthoryear{Clark, Glover, Klessen  \& Bonnell}{Clark
  et~al.}{2012}]{clark_how_2012}
Clark P.~C.,  Glover S. C.~O.,  Klessen R.~S.,   Bonnell I.~A.,  2012, \mn@doi
  [MNRAS] {10.1111/j.1365-2966.2012.21259.x}, 424, 2599

\bibitem[\protect\citeauthoryear{Draine}{Draine}{1978}]{draine_photoelectric_1978}
Draine B.~T.,  1978, \mn@doi [ApJS] {10.1086/190513}, 36, 595

\bibitem[\protect\citeauthoryear{Draine \& Bertoldi}{Draine \&
  Bertoldi}{1996}]{draine_structure_1996}
Draine B.~T.,  Bertoldi F.,  1996, \mnras, 468, 269

\bibitem[\protect\citeauthoryear{Dubey et~al.,}{Dubey
  et~al.}{2008}]{dubey_challenges_2008}
Dubey A.,  et~al., 2008, Astronomical Society of the Pacific Conference Series,
  385, 145

\bibitem[\protect\citeauthoryear{{Ebagezio}, {Seifried}, {Walch},
  {N{\"u}rnberger}, {Rathjen}  \& {Naab}}{{Ebagezio}
  et~al.}{2022}]{ebagezio2022}
{Ebagezio} S.,  {Seifried} D.,  {Walch} S.,  {N{\"u}rnberger} P.~C.,  {Rathjen}
  T.~E.,   {Naab} T.,  2022, arXiv e-prints, \href
  {https://ui.adsabs.harvard.edu/abs/2022arXiv220606393E} {p. arXiv:2206.06393}

\bibitem[\protect\citeauthoryear{{Ferrada-Chamorro}, {Lupi}  \&
  {Bovino}}{{Ferrada-Chamorro} et~al.}{2021}]{ferrada-chamorro_chemical_2021}
{Ferrada-Chamorro} S.,  {Lupi} A.,   {Bovino} S.,  2021, \mn@doi [\mnras]
  {10.1093/mnras/stab1525}, \href
  {https://ui.adsabs.harvard.edu/abs/2021MNRAS.505.3442F} {505, 3442}

\bibitem[\protect\citeauthoryear{Flower, {G. Pineau des Forêts}  \&
  Walmsley}{Flower et~al.}{2005}]{flower_freeze-out_2005}
Flower D.~R.,  {G. Pineau des Forêts}  Walmsley C.~M.,  2005, \mn@doi [A\&A]
  {10.1051/0004-6361:20042481}, 436, 933

\bibitem[\protect\citeauthoryear{Fryxell et~al.,}{Fryxell
  et~al.}{2000}]{fryxell_flash_2000}
Fryxell B.,  et~al., 2000, \mn@doi [ApJS] {10.1086/317361}, 131, 273

\bibitem[\protect\citeauthoryear{Gaches \& Offner}{Gaches \&
  Offner}{2018}]{gaches_model_2018}
Gaches B. A.~L.,  Offner S. S.~R.,  2018, \mn@doi [ApJ]
  {10.3847/1538-4357/aaaae2}, 854, 156

\bibitem[\protect\citeauthoryear{{Ganguly}, {Walch}, {Clarke}  \&
  {Seifried}}{{Ganguly} et~al.}{2022}]{ganguly2022}
{Ganguly} S.,  {Walch} S.,  {Clarke} S.~D.,   {Seifried} D.,  2022, arXiv
  e-prints, \href {https://ui.adsabs.harvard.edu/abs/2022arXiv220402511G} {p.
  arXiv:2204.02511}

\bibitem[\protect\citeauthoryear{{Garrod} \& {Herbst}}{{Garrod} \&
  {Herbst}}{2006}]{Garrod06}
{Garrod} R.~T.,  {Herbst} E.,  2006, \mn@doi [\aap]
  {10.1051/0004-6361:20065560}, \href
  {https://ui.adsabs.harvard.edu/abs/2006A&A...457..927G} {457, 927}

\bibitem[\protect\citeauthoryear{Gatto et~al.,}{Gatto
  et~al.}{2017}]{gatto_silcc_2017}
Gatto A.,  et~al., 2017, \mn@doi [MNRAS] {10.1093/mnras/stw3209}, 466, 1903

\bibitem[\protect\citeauthoryear{Genel, Vogelsberger, Nelson, Sijacki, Springel
   \& Hernquist}{Genel et~al.}{2013}]{genel_following_2013}
Genel S.,  Vogelsberger M.,  Nelson D.,  Sijacki D.,  Springel V.,   Hernquist
  L.,  2013, \mn@doi [MNRAS] {10.1093/mnras/stt1383}, 435, 1426

\bibitem[\protect\citeauthoryear{{Gerin} \& {Liszt}}{{Gerin} \&
  {Liszt}}{2021}]{gerin_co_2021}
{Gerin} M.,  {Liszt} H.,  2021, \mn@doi [\aap] {10.1051/0004-6361/202039915},
  \href {https://ui.adsabs.harvard.edu/abs/2021A&A...648A..38G} {648, A38}

\bibitem[\protect\citeauthoryear{Gerin, Liszt, Neufeld, Godard, Sonnentrucker,
  Pety  \& Roueff}{Gerin et~al.}{2019}]{gerin_molecular_2019}
Gerin M.,  Liszt H.,  Neufeld D.,  Godard B.,  Sonnentrucker P.,  Pety J.,
  Roueff E.,  2019, \mn@doi [A\&A] {10.1051/0004-6361/201833661}, 622, A26

\bibitem[\protect\citeauthoryear{Girichidis et~al.,}{Girichidis
  et~al.}{2016}]{girichidis_silcc_2016}
Girichidis P.,  et~al., 2016, \mn@doi [MNRAS] {10.1093/mnras/stv2742}, 456,
  3432

\bibitem[\protect\citeauthoryear{Glover \& Clark}{Glover \&
  Clark}{2012}]{glover_approximations_2012}
Glover S. C.~O.,  Clark P.~C.,  2012, \mn@doi [MNRAS]
  {10.1111/j.1365-2966.2011.20260.x}, 421, 116

\bibitem[\protect\citeauthoryear{Glover \& Mac~Low}{Glover \&
  Mac~Low}{2007a}]{glover_simulating_2007}
Glover S. C.~O.,  Mac~Low M.,  2007a, \mn@doi [ApJS] {10.1086/512238}, 169, 239

\bibitem[\protect\citeauthoryear{Glover \& Mac~Low}{Glover \&
  Mac~Low}{2007b}]{glover_simulating_2007-1}
Glover S. C.~O.,  Mac~Low M.,  2007b, \mn@doi [ApJ] {10.1086/512227}, 659, 1317

\bibitem[\protect\citeauthoryear{Glover \& Mac~Low}{Glover \&
  Mac~Low}{2011}]{glover_relationship_2011}
Glover S. C.~O.,  Mac~Low M.-M.,  2011, \mn@doi [MNRAS]
  {10.1111/j.1365-2966.2010.17907.x}, 412, 337

\bibitem[\protect\citeauthoryear{Glover, Federrath, Mac~Low  \& Klessen}{Glover
  et~al.}{2010}]{glover_modelling_2010}
Glover S. C.~O.,  Federrath C.,  Mac~Low M.-M.,   Klessen R.~S.,  2010, \mn@doi
  [MNRAS] {10.1111/j.1365-2966.2009.15718.x}, 404, 2

\bibitem[\protect\citeauthoryear{Gnedin, Tassis  \& Kravtsov}{Gnedin
  et~al.}{2009}]{gnedin_modeling_2009}
Gnedin N.~Y.,  Tassis K.,   Kravtsov A.~V.,  2009, \mn@doi [ApJ]
  {10.1088/0004-637X/697/1/55}, 697, 55

\bibitem[\protect\citeauthoryear{Godard, Falgarone, Gerin, Hily-Blant  \&
  De~Luca}{Godard et~al.}{2010}]{godard_molecular_2010}
Godard B.,  Falgarone E.,  Gerin M.,  Hily-Blant P.,   De~Luca M.,  2010,
  \mn@doi [A\&A] {10.1051/0004-6361/201014283}, 520, A20

\bibitem[\protect\citeauthoryear{{Godard}, {Pineau des For{\^e}ts},
  {Hennebelle}, {Bellomi}  \& {Valdivia}}{{Godard} et~al.}{2023}]{godard2023}
{Godard} B.,  {Pineau des For{\^e}ts} G.,  {Hennebelle} P.,  {Bellomi} E.,
  {Valdivia} V.,  2023, \mn@doi [\aap] {10.1051/0004-6361/202243902}, \href
  {https://ui.adsabs.harvard.edu/abs/2023A&A...669A..74G} {669, A74}

\bibitem[\protect\citeauthoryear{Goicoechea, Santa-Maria, Bron, Teyssier,
  Marcelino, Cernicharo  \& Cuadrado}{Goicoechea
  et~al.}{2019}]{goicoechea_molecular_2019}
Goicoechea J.~R.,  Santa-Maria M.~G.,  Bron E.,  Teyssier D.,  Marcelino N.,
  Cernicharo J.,   Cuadrado S.,  2019, \mn@doi [A\&A]
  {10.1051/0004-6361/201834409}, 622, A91

\bibitem[\protect\citeauthoryear{Gong, Ostriker  \& Wolfire}{Gong
  et~al.}{2017}]{gong_simple_2017}
Gong M.,  Ostriker E.~C.,   Wolfire M.~G.,  2017, \mn@doi [ApJ]
  {10.3847/1538-4357/aa7561}, 843, 36

\bibitem[\protect\citeauthoryear{{Gong}, {Ostriker}  \& {Kim}}{{Gong}
  et~al.}{2018}]{Gong2018}
{Gong} M.,  {Ostriker} E.~C.,   {Kim} C.-G.,  2018, \mn@doi [\apj]
  {10.3847/1538-4357/aab9af}, \href
  {https://ui.adsabs.harvard.edu/abs/2018ApJ...858...16G} {858, 16}

\bibitem[\protect\citeauthoryear{{Gong}, {Ostriker}, {Kim}  \& {Kim}}{{Gong}
  et~al.}{2020}]{gongm-2020}
{Gong} M.,  {Ostriker} E.~C.,  {Kim} C.-G.,   {Kim} J.-G.,  2020, \mn@doi
  [\apj] {10.3847/1538-4357/abbdab}, \href
  {https://ui.adsabs.harvard.edu/abs/2020ApJ...903..142G} {903, 142}

\bibitem[\protect\citeauthoryear{{G{\'o}rski} \& {Hivon}}{{G{\'o}rski} \&
  {Hivon}}{2011}]{gorski_2011}
{G{\'o}rski} K.~M.,  {Hivon} E.,  2011, {HEALPix: Hierarchical Equal Area
  isoLatitude Pixelization of a sphere} (\mn@eprint {ascl} {1107.018})

\bibitem[\protect\citeauthoryear{Grassi, Bovino, Schleicher, Prieto, Seifried,
  Simoncini  \& Gianturco}{Grassi et~al.}{2014}]{grassi_krome_2014}
Grassi T.,  Bovino S.,  Schleicher D. R.~G.,  Prieto J.,  Seifried D.,
  Simoncini E.,   Gianturco F.~A.,  2014, \mn@doi [MNRAS]
  {10.1093/mnras/stu114}, 439, 2386

\bibitem[\protect\citeauthoryear{Grassi, Bovino, Haugbølle  \&
  Schleicher}{Grassi et~al.}{2017}]{grassi_detailed_2017}
Grassi T.,  Bovino S.,  Haugbølle T.,   Schleicher D. R.~G.,  2017, \mn@doi
  [MNRAS] {10.1093/mnras/stw2871}, 466, 1259

\bibitem[\protect\citeauthoryear{Habing}{Habing}{1968}]{habing_interstellar_1968}
Habing H.~J.,  1968, Bull. Astr. Inst. Netherlands, 19, 421

\bibitem[\protect\citeauthoryear{{Hasegawa} \& {Herbst}}{{Hasegawa} \&
  {Herbst}}{1993}]{Hasegawa93}
{Hasegawa} T.~I.,  {Herbst} E.,  1993, \mn@doi [\mnras]
  {10.1093/mnras/261.1.83}, \href
  {https://ui.adsabs.harvard.edu/abs/1993MNRAS.261...83H} {261, 83}

\bibitem[\protect\citeauthoryear{{Herbst} \& {Cuppen}}{{Herbst} \&
  {Cuppen}}{2006}]{Herbst06}
{Herbst} E.,  {Cuppen} H.~M.,  2006, \mn@doi [Proceedings of the National
  Academy of Science] {10.1073/pnas.0601556103}, \href
  {https://ui.adsabs.harvard.edu/abs/2006PNAS..10312257H} {103, 12257}

\bibitem[\protect\citeauthoryear{Hollenbach, Kaufman, Bergin  \&
  Melnick}{Hollenbach et~al.}{2009}]{hollenbach_water_2009}
Hollenbach D.,  Kaufman M.~J.,  Bergin E.~A.,   Melnick G.~J.,  2009, \mn@doi
  [ApJ] {10.1088/0004-637X/690/2/1497}, 690, 1497

\bibitem[\protect\citeauthoryear{{Hu}, {Sternberg}  \& {Van Dishoeck}}{{Hu}
  et~al.}{2021}]{hu_metallicity_2021}
{Hu} C.-Y.,  {Sternberg} A.,   {Van Dishoeck} E.~F.,  2021, \mn@doi [\apj]
  {10.3847/1538-4357/ac0dbd}, \href
  {https://ui.adsabs.harvard.edu/abs/2021ApJ...920...44H} {920, 44}

\bibitem[\protect\citeauthoryear{{Jacob} et~al.,}{{Jacob}
  et~al.}{2022}]{Jacob-Hygal2022}
{Jacob} A.~M.,  et~al., 2022, \mn@doi [\apj] {10.3847/1538-4357/ac5409}, \href
  {https://ui.adsabs.harvard.edu/abs/2022ApJ...930..141J} {930, 141}

\bibitem[\protect\citeauthoryear{Kauffmann, Goldsmith, Melnick, Tolls, Guzman
  \& Menten}{Kauffmann et~al.}{2017}]{kauffmann_molecular_2017}
Kauffmann J.,  Goldsmith P.~F.,  Melnick G.,  Tolls V.,  Guzman A.,   Menten
  K.~M.,  2017, \mn@doi [A\&A] {10.1051/0004-6361/201731123}, 605, L5

\bibitem[\protect\citeauthoryear{{Keating} et~al.,}{{Keating}
  et~al.}{2020}]{keating2020}
{Keating} L.~C.,  et~al., 2020, \mn@doi [\mnras] {10.1093/mnras/staa2839},
  \href {https://ui.adsabs.harvard.edu/abs/2020MNRAS.499..837K} {499, 837}

\bibitem[\protect\citeauthoryear{{Konstandin}, {Federrath}, {Klessen}  \&
  {Schmidt}}{{Konstandin} et~al.}{2012}]{konstandin12}
{Konstandin} L.,  {Federrath} C.,  {Klessen} R.~S.,   {Schmidt} W.,  2012,
  \mn@doi [Journal of Fluid Mechanics] {10.1017/jfm.2011.503}, \href
  {https://ui.adsabs.harvard.edu/abs/2012JFM...692..183K} {692, 183}

\bibitem[\protect\citeauthoryear{{Le Petit}, {Ruaud}, {Bron}, {Godard},
  {Roueff}, {Languignon}  \& {Le Bourlot}}{{Le Petit}
  et~al.}{2016}]{lepetit2016}
{Le Petit} F.,  {Ruaud} M.,  {Bron} E.,  {Godard} B.,  {Roueff} E.,
  {Languignon} D.,   {Le Bourlot} J.,  2016, \mn@doi [\aap]
  {10.1051/0004-6361/201526658}, \href
  {https://ui.adsabs.harvard.edu/abs/2016A&A...585A.105L} {585, A105}

\bibitem[\protect\citeauthoryear{{Leger}, {Jura}  \& {Omont}}{{Leger}
  et~al.}{1985}]{Leger85}
{Leger} A.,  {Jura} M.,   {Omont} A.,  1985, \aap, \href
  {https://ui.adsabs.harvard.edu/abs/1985A&A...144..147L} {144, 147}

\bibitem[\protect\citeauthoryear{{Li}, {Narayanan}, {Dav{\`e}}  \&
  {Krumholz}}{{Li} et~al.}{2018}]{liq-DMGsims-2018}
{Li} Q.,  {Narayanan} D.,  {Dav{\`e}} R.,   {Krumholz} M.~R.,  2018, \mn@doi
  [\apj] {10.3847/1538-4357/aaec77}, \href
  {https://ui.adsabs.harvard.edu/abs/2018ApJ...869...73L} {869, 73}

\bibitem[\protect\citeauthoryear{Liu et~al.,}{Liu
  et~al.}{2020a}]{liu_atoms_2020-1}
Liu T.,  et~al., 2020a, \mn@doi [MNRAS] {10.1093/mnras/staa1577}, 496, 2790

\bibitem[\protect\citeauthoryear{Liu et~al.,}{Liu
  et~al.}{2020b}]{liu_atoms_2020}
Liu T.,  et~al., 2020b, \mn@doi [MNRAS] {10.1093/mnras/staa1501}, 496, 2821

\bibitem[\protect\citeauthoryear{{Lucas} \& {Liszt}}{{Lucas} \&
  {Liszt}}{1996}]{lucasandliszt96}
{Lucas} R.,  {Liszt} H.,  1996, \aap, \href
  {https://ui.adsabs.harvard.edu/abs/1996A&A...307..237L} {307, 237}

\bibitem[\protect\citeauthoryear{Lupi \& Bovino}{Lupi \&
  Bovino}{2020}]{lupi_c_2020}
Lupi A.,  Bovino S.,  2020, \mn@doi [MNRAS] {10.1093/mnras/staa048}, 492, 2818

\bibitem[\protect\citeauthoryear{Lupi, Bovino, Capelo, Volonteri  \& Silk}{Lupi
  et~al.}{2018}]{lupi_natural_2018}
Lupi A.,  Bovino S.,  Capelo P.~R.,  Volonteri M.,   Silk J.,  2018, \mn@doi
  [MNRAS] {10.1093/mnras/stx2874}, 474, 2884

\bibitem[\protect\citeauthoryear{Mackey, Walch, Seifried, Glover, Wunsch  \&
  Aharonian}{Mackey et~al.}{2019}]{mackey_non-equilibrium_2019}
Mackey J.,  Walch S.,  Seifried D.,  Glover S. C.~O.,  Wunsch R.,   Aharonian
  F.,  2019, \mn@doi [MNRAS] {10.1093/mnras/stz902}, 486, 1094

\bibitem[\protect\citeauthoryear{{Mathis}, {Rumpl}  \& {Nordsieck}}{{Mathis}
  et~al.}{1977}]{Mathis77}
{Mathis} J.~S.,  {Rumpl} W.,   {Nordsieck} K.~H.,  1977, \mn@doi [\apj]
  {10.1086/155591}, \href
  {https://ui.adsabs.harvard.edu/abs/1977ApJ...217..425M} {217, 425}

\bibitem[\protect\citeauthoryear{{Molinari} et~al.,}{{Molinari}
  et~al.}{2011}]{molinari2011}
{Molinari} S.,  et~al., 2011, \mn@doi [\apjl] {10.1088/2041-8205/735/2/L33},
  \href {https://ui.adsabs.harvard.edu/abs/2011ApJ...735L..33M} {735, L33}

\bibitem[\protect\citeauthoryear{Nayana et~al.,}{Nayana
  et~al.}{2020}]{nayana_alma_2020}
Nayana A.~J.,  et~al., 2020, \mn@doi [ApJ] {10.3847/1538-4357/abb466}, 902, 140

\bibitem[\protect\citeauthoryear{Nelson \& Langer}{Nelson \&
  Langer}{1997}]{nelson_dynamics_1997}
Nelson R.~P.,  Langer W.~D.,  1997, ApJ, 482, 796

\bibitem[\protect\citeauthoryear{Nelson \& Langer}{Nelson \&
  Langer}{1999}]{nelson_stability_1999}
Nelson R.~P.,  Langer W.~D.,  1999, \mn@doi [ApJ] {10.1086/307823}, 524, 923

\bibitem[\protect\citeauthoryear{Nikolic}{Nikolic}{2007}]{nikolic_hco_2007}
Nikolic S.,  2007, \mn@doi [Serbian Astro. J.] {10.2298/SAJ0775001N}, 175, 1

\bibitem[\protect\citeauthoryear{{Oka}}{{Oka}}{2006}]{Oka2006}
{Oka} T.,  2006, \mn@doi [Proceedings of the National Academy of Science]
  {10.1073/pnas.0601242103}, \href
  {https://ui.adsabs.harvard.edu/abs/2006PNAS..10312235O} {103, 12235}

\bibitem[\protect\citeauthoryear{{Padovani}, {Galli}, {Ivlev}, {Caselli}  \&
  {Ferrara}}{{Padovani} et~al.}{2018}]{Padovani2018}
{Padovani} M.,  {Galli} D.,  {Ivlev} A.~V.,  {Caselli} P.,   {Ferrara} A.,
  2018, \mn@doi [\aap] {10.1051/0004-6361/201834008}, \href
  {https://ui.adsabs.harvard.edu/abs/2018A&A...619A.144P} {619, A144}

\bibitem[\protect\citeauthoryear{Papadopoulos}{Papadopoulos}{2007}]{papadopoulos_hcn_2007}
Papadopoulos P.~P.,  2007, \mn@doi [ApJ] {10.1086/510186}, 656, 792

\bibitem[\protect\citeauthoryear{{Price} \& {Federrath}}{{Price} \&
  {Federrath}}{2010}]{price10}
{Price} D.~J.,  {Federrath} C.,  2010, \mn@doi [\mnras]
  {10.1111/j.1365-2966.2010.16810.x}, \href
  {https://ui.adsabs.harvard.edu/abs/2010MNRAS.406.1659P} {406, 1659}

\bibitem[\protect\citeauthoryear{Richings \& Schaye}{Richings \&
  Schaye}{2016}]{richings_effects_2016}
Richings A.~J.,  Schaye J.,  2016, \mn@doi [MNRAS] {10.1093/mnras/stw327}, 458,
  270

\bibitem[\protect\citeauthoryear{Sanhueza, Jackson, Foster, Garay, Silva  \&
  Finn}{Sanhueza et~al.}{2012}]{sanhueza_chemistry_2012}
Sanhueza P.,  Jackson J.~M.,  Foster J.~B.,  Garay G.,  Silva A.,   Finn S.~C.,
   2012, \mn@doi [ApJ] {10.1088/0004-637X/756/1/60}, 756, 31

\bibitem[\protect\citeauthoryear{Seifried \& Walch}{Seifried \&
  Walch}{2016}]{seifried_modelling_2016}
Seifried D.,  Walch S.,  2016, \mn@doi [MNRAS] {10.1093/mnrasl/slw035}, 459,
  L11

\bibitem[\protect\citeauthoryear{Seifried, S\'{a}nchez-Monge, Suri  \&
  Walch}{Seifried et~al.}{2017a}]{seifried_modelling_2017}
Seifried D.,  S\'{a}nchez-Monge A.,  Suri S.,   Walch S.,  2017a, \mn@doi
  [MNRAS] {10.1093/mnras/stx399}, 467, 4467

\bibitem[\protect\citeauthoryear{Seifried et~al.,}{Seifried
  et~al.}{2017b}]{seifried_silcc-zoom_2017}
Seifried D.,  et~al., 2017b, \mn@doi [MNRAS] {10.1093/mnras/stx2343}, 472, 4797

\bibitem[\protect\citeauthoryear{Seifried, Haid, Walch, Borchert  \&
  Bisbas}{Seifried et~al.}{2020}]{seifried_silcc-zoom_2020}
Seifried D.,  Haid S.,  Walch S.,  Borchert E. M.~A.,   Bisbas T.~G.,  2020,
  \mn@doi [MNRAS] {10.1093/mnras/stz3563}, 492, 1465

\bibitem[\protect\citeauthoryear{{Seifried}, {Beuther}, {Walch}, {Syed},
  {Soler}, {Girichidis}  \& {W{\"u}nsch}}{{Seifried}
  et~al.}{2022}]{seifried_accuracy_2021}
{Seifried} D.,  {Beuther} H.,  {Walch} S.,  {Syed} J.,  {Soler} J.~D.,
  {Girichidis} P.,   {W{\"u}nsch} R.,  2022, \mn@doi [\mnras]
  {10.1093/mnras/stac607}, \href
  {https://ui.adsabs.harvard.edu/abs/2022MNRAS.512.4765S} {512, 4765}

\bibitem[\protect\citeauthoryear{{Sembach}, {Howk}, {Ryans}  \&
  {Keenan}}{{Sembach} et~al.}{2000}]{sembach2000}
{Sembach} K.~R.,  {Howk} J.~C.,  {Ryans} R. S.~I.,   {Keenan} F.~P.,  2000,
  \mn@doi [\apj] {10.1086/308173}, \href
  {https://ui.adsabs.harvard.edu/abs/2000ApJ...528..310S} {528, 310}

\bibitem[\protect\citeauthoryear{Smith, Glover, Clark, Klessen  \&
  Springel}{Smith et~al.}{2014}]{smith_co-dark_2014}
Smith R.~J.,  Glover S. C.~O.,  Clark P.~C.,  Klessen R.~S.,   Springel V.,
  2014, \mn@doi [MNRAS] {10.1093/mnras/stu616}, 441, 1628

\bibitem[\protect\citeauthoryear{Teague, Semenov, Guilloteau, Henning, Dutrey,
  Wakelam, Chapillon  \& Pietu}{Teague et~al.}{2015}]{teague_chemistry_2015}
Teague R.,  Semenov D.,  Guilloteau S.,  Henning T.,  Dutrey A.,  Wakelam V.,
  Chapillon E.,   Pietu V.,  2015, \mn@doi [A\&A]
  {10.1051/0004-6361/201425268}, 574, A137

\bibitem[\protect\citeauthoryear{{Valdivia}, {Godard}, {Hennebelle}, {Gerin}
  \& {Lesaffre}}{{Valdivia} et~al.}{2016a}]{valdivia_chemistry_2016}
{Valdivia} V.,  {Godard} B.,  {Hennebelle} P.,  {Gerin} M.,   {Lesaffre} P.,
  2016a, in {Reyl{\'e}} C.,  {Richard} J.,  {Cambr{\'e}sy} L.,  {Deleuil} M.,
  {P{\'e}contal} E.,  {Tresse} L.,   {Vauglin} I.,  eds, SF2A-2016: Proceedings
  of the Annual meeting of the French Society of Astronomy and Astrophysics. pp
  163--167

\bibitem[\protect\citeauthoryear{Valdivia, Hennebelle, Gérin  \&
  Lesaffre}{Valdivia et~al.}{2016b}]{valdivia_h2_2016}
Valdivia V.,  Hennebelle P.,  Gérin M.,   Lesaffre P.,  2016b, \mn@doi [A\&A]
  {10.1051/0004-6361/201527325}, 587, A76

\bibitem[\protect\citeauthoryear{Valdivia, Godard, Hennebelle, Gerin, Lesaffre
  \& Le~Bourlot}{Valdivia et~al.}{2017}]{valdivia_origin_2017}
Valdivia V.,  Godard B.,  Hennebelle P.,  Gerin M.,  Lesaffre P.,   Le~Bourlot
  J.,  2017, \mn@doi [A\&A] {10.1051/0004-6361/201629905}, 600, A114

\bibitem[\protect\citeauthoryear{{Van Dishoeck} \& {Black}}{{Van Dishoeck} \&
  {Black}}{1988}]{van_dishoeck_photodissociation_1988}
{Van Dishoeck} E.~F.,  {Black} J.~H.,  1988, \mn@doi [\apj] {10.1086/166877},
  \href {https://ui.adsabs.harvard.edu/abs/1988ApJ...334..771V} {334, 771}

\bibitem[\protect\citeauthoryear{Walch et~al.,}{Walch
  et~al.}{2015}]{walch_silcc_2015}
Walch S.,  et~al., 2015, \mn@doi [MNRAS] {10.1093/mnras/stv1975}, 454, 238

\bibitem[\protect\citeauthoryear{Wolﬁre, Hollenbach  \& McKee}{Wolﬁre
  et~al.}{2010}]{wolre_dark_2010}
Wolﬁre M.~G.,  Hollenbach D.,   McKee C.~F.,  2010, \mn@doi [ApJ]
  {10.1088/0004-637X/716/2/1191}, 716, 1191

\bibitem[\protect\citeauthoryear{Wünsch, Walch, Dinnbier  \&
  Whitworth}{Wünsch et~al.}{2018}]{wunsch_tree-based_2018}
Wünsch R.,  Walch S.,  Dinnbier F.,   Whitworth A.,  2018, \mn@doi [MNRAS]
  {10.1093/mnras/sty015}, 475, 3393

\bibitem[\protect\citeauthoryear{{Yang}, {Jiang}, {Chen}, {Ao}  \& {Yu}}{{Yang}
  et~al.}{2021}]{yang_search_2021}
{Yang} Y.,  {Jiang} Z.,  {Chen} Z.,  {Ao} Y.,   {Yu} S.,  2021, \mn@doi [\apj]
  {10.3847/1538-4357/ac22ab}, \href
  {https://ui.adsabs.harvard.edu/abs/2021ApJ...922..144Y} {922, 144}

\bibitem[\protect\citeauthoryear{Yun et~al.,}{Yun
  et~al.}{2021}]{yun_times_2021}
Yun H.-S.,  et~al., 2021, \mn@doi [ApJS] {10.3847/1538-4365/ac090e}, 256, 16

\bibitem[\protect\citeauthoryear{Zhang, Reid, Menten, Zheng, Brunthaler, Dame
  \& Xu}{Zhang et~al.}{2013}]{Zhang_2013}
Zhang B.,  Reid M.~J.,  Menten K.~M.,  Zheng X.~W.,  Brunthaler A.,  Dame
  T.~M.,   Xu Y.,  2013, \mn@doi [\apj] {10.1088/0004-637x/775/1/79}, 775, 79

\makeatother
\end{thebibliography}

\newpage
%%%
\appendix

\section{Validating the tracer particle count}
\label{appendix:tenpercent}

Tracer particles are injected into the SILCC-Zoom simulations in a uniform lattice with a 1~pc spacing. To confirm that the resulting tracer count ($\sim10^6$ for all four clouds) rigorously samples all density regimes, we select 10\% of the tracer particles from each simulation at random, and perform upon this subset the regridding procedure described in Section~\ref{section-regridding}. We will show in the following that the 10\% subset of tracers gives the same results at each stage as the full population, confirming that our methodology is statistically rigorous with regard to the count of tracers.

First, we compare the average densities of H, \ce{H2}, CO, \ce{C+}, and \ce{HCO+} vs. $n_\rmn{H,tot}$ as reported by the full tracer population and the 10\% subset, for each cloud at several different time snapshots  (not shown). These average number densities are essentially identical between the two tracer sets for all clouds and times, in accordance with a similar investigation by \citet{ferrada-chamorro_chemical_2021}. 

\begin{figure*}
    \centering
    \includegraphics[width=0.9\linewidth]{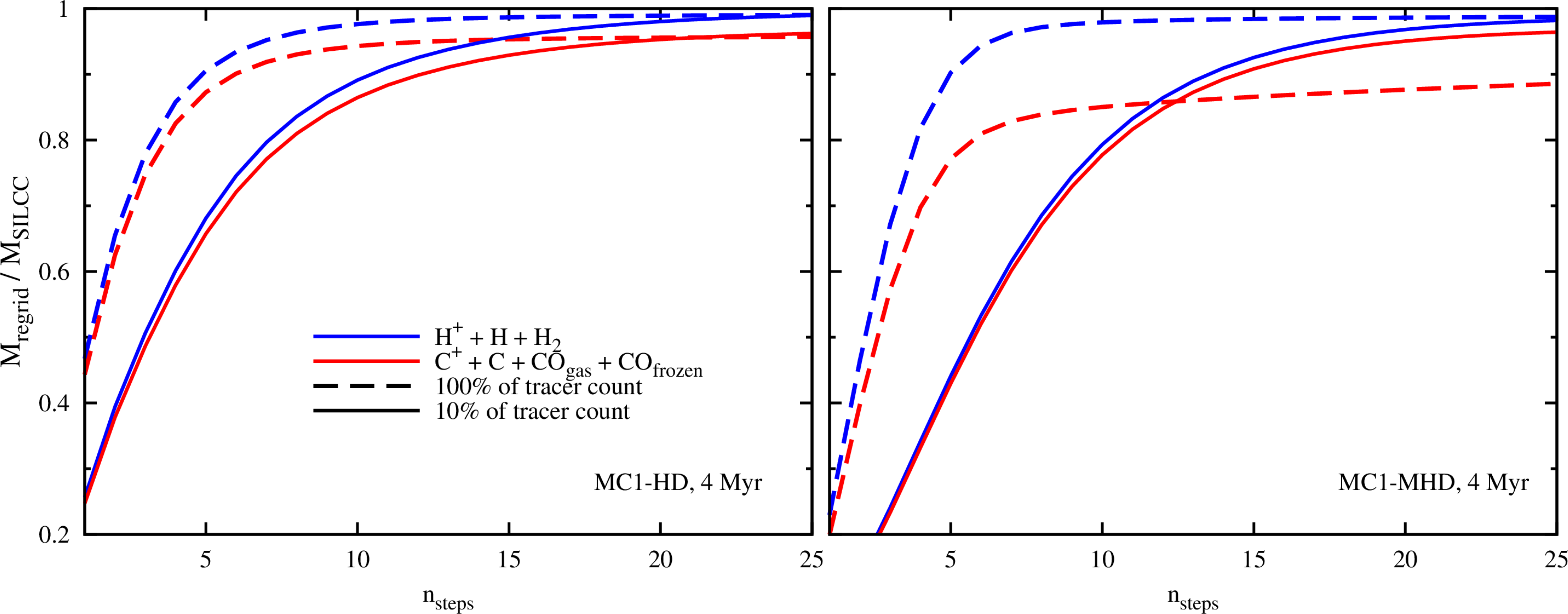}
    \caption{The ratios $M_\rmn{regrid,H,tot}/M_\rmn{SILCC,H,tot}$ (blue) and $M_\rmn{regrid,C,tot}/M_\rmn{SILCC,C,tot}$ (red) from Eqs.~\ref{eq:Mgridstart}--\ref{eq:Mgridend}, vs. the number of interpolation steps, $n_\rmn{steps}$, for \mbox{MC1-HD} (left) and \mbox{MC1-MHD} (right), at $t_\rmn{evol}=4$~Myr. Results are provided for the full tracer population (dashed) and a 10\% subset (solid). The 10\% tracer subsets mostly converge to the same values as the full population, but in about twice as many interpolation steps.}
    \label{fig:interpmass-appendix}
\end{figure*}

Next, we regrid the tracers as described in Section~\ref{section-regridding}. In Fig.~\ref{fig:interpmass-appendix}, we plot the convergence of the total hydrogen and carbon masses as a function of the running number of interpolation steps for both the full and reduced tracer sets. In both clouds, the 10\% set's masses take about twice as many interpolation steps to converge as the full set's masses. Reducing the tracer count by a factor of ten increases the average inter-tracer separation by $10^{1/3}\sim2$, explaining this change. 

For the cloud MC1-HD at $t_\rmn{evol}=4$~Myr, both the 100\% (dashed lines) and 10\% (solid) tracer sets converge to the same total hydrogen (blue) and carbon (red) masses. For the cloud MC1-MHD, the hydrogen masses both converge to the same value. The carbon mass of the 10\% set in the MHD cloud exceeds the value from the full tracer set, but as a result, the 10\% set's carbon mass is closer to the original value from the SILCC simulation. To explain this, we posit that removing 90\% of the tracers will have the greatest impact on tracer statistics in the intermediate density regime. Regions with denser gas are still well-sampled by tracers after the reduction, while low-density cells already frequently lacked tracers even using the full population. But in medium-density gas, many cells that would be sampled by $\sim1$ tracer are now empty. The grid-filling procedure interpolates abundance values into these empty cells from the nearest cells which still have tracers, which will tend to be cells with denser gas. This will inflate the total mass of e.g. carbon ascribed to those cells, with respect to the mass calculated using the full tracer population. A shallower density gradient (as in an MHD cloud) will worsen this effect, explaining why reducing the tracer population increases the carbon mass for MC1-MHD but not MC1-HD.

\begin{figure}
    \centering
    \includegraphics[width=0.9\linewidth]{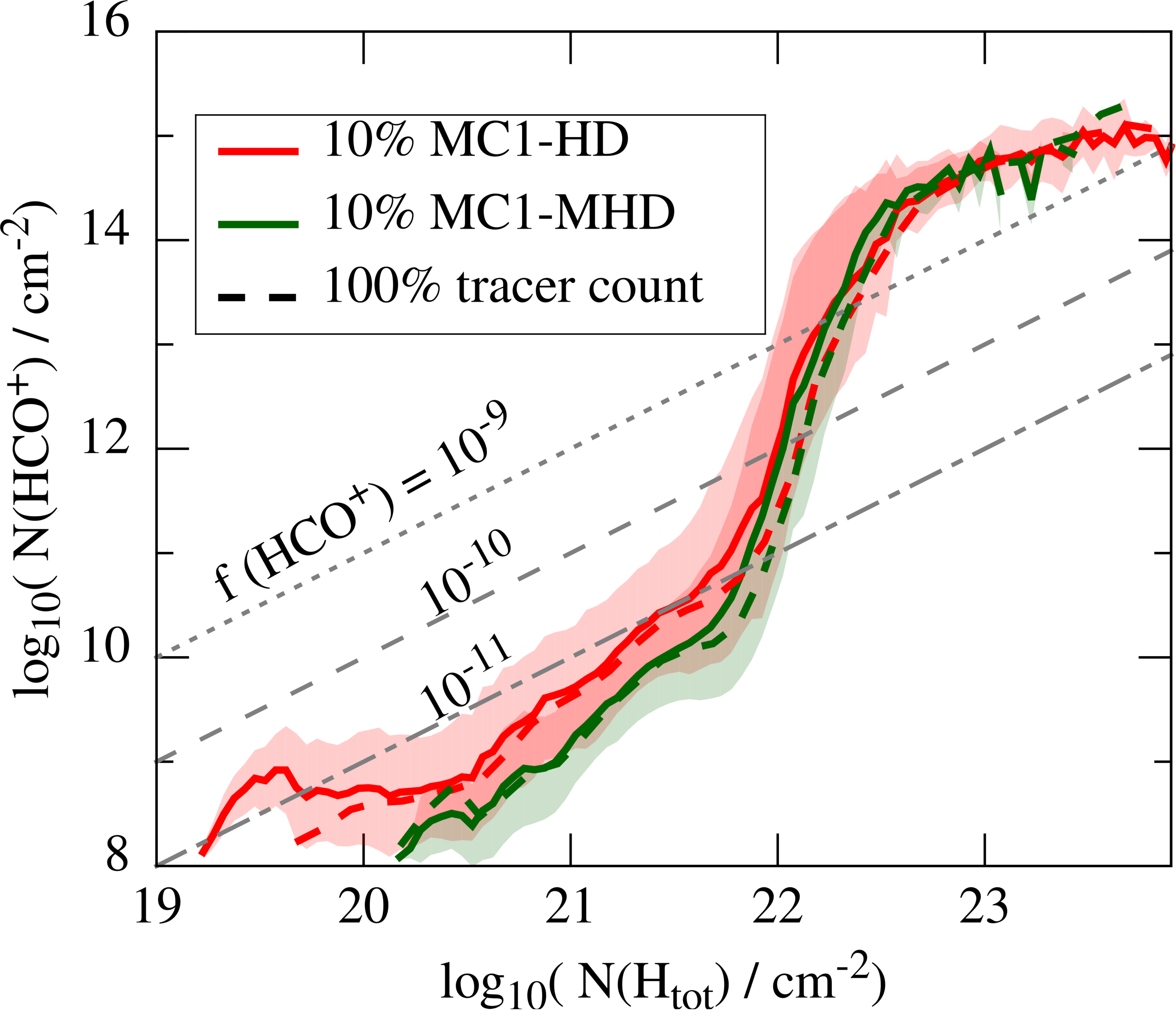}
    \caption{The average $N$(\ce{HCO+}) vs. $N$(H$_\rmn{tot}$) for the $y$-$z$ projection of the molecular clouds \mbox{MC1-HD} (red) and MC1-MHD (green) at $t_\rmn{evol}=4$~Myr, after regridding the full tracer population (dashed lines) and a 10\% subset (solid). The shaded areas represent one standard deviation from the respective average. The column density distribution from the 10\% subset agrees very well with the distribution from the full tracer population.}
    \label{fig:HCOjvslots-appendix}
\end{figure}

Finally, we compare the column density projections of the regridded 10\% tracer subset to those of the full tracer population, in Fig.~\ref{fig:HCOjvslots-appendix}. We plot $N$(\ce{HCO+}) vs. $N$(H$_\rmn{tot}$) for clouds MC1-HD (red) and MC1-MHD (green) at $t_\rmn{evol}=4$~Myr. $N$(\ce{HCO+}) from the 10\% subset (solid lines) agrees very well with the results using 100\% of the tracers (dashed lines) at all density regimes. The greatest discrepancy is found around $N\rmn{(H_{tot})}\simeq10^{22}$~cm$^{-2}$ in both clouds, though the difference is never more than $\sim1\%$. We ascribe this slight overestimation in $N$ at medium density to the same cause as the carbon mass overestimation described in the previous paragraph. The close correspondence of the 10\% tracer subset with the full population indicates that our post-processing approach is well-converged using a tracer population size that corresponds to an initial uniform spacing of 1~pc.

\section{Contents of chemical networks}
\label{appendix:networks}

\subsection{The post-processing network}
\label{appendix:networks_pp}

Our post-processing method was tested and validated using a modified version of the \texttt{react\_COthin} chemical network \citep{grassi_detailed_2017}, updated to include reactions that model the adsorption and desorption of CO and H\textsubscript{2} from dust grains. The chemical species included in this network are listed in Table~\ref{table:RC_species}, consisting of 37 gas-phase species and 2 proxy species for frozen-out CO and H\textsubscript{2}O. We refer the reader to \cite{grassi_detailed_2017} or the Appendix materials of \cite{seifried_modelling_2016} for additional details regarding the performance of and the reactions included in this network. The freeze-out modeling in our modified version of this network is described below.

\begin{table}
 \caption{The 39 chemical species included in our post-processing network, which is modified from the \texttt{react\_COthin} network included in the \textsc{Krome} distribution \citep{grassi_detailed_2017}. The last two species (in parentheses) represent H\textsubscript{2}O and CO which have frozen out onto dust grains, and have no further chemical interactions with the medium until they have thawed again.}
\begin{center}
\begin{tabular}{ c c c c c c c } 
 \hline
 e\textsuperscript{-} & H & H\textsuperscript{+} & H\textsuperscript{-} & H\textsubscript{2} & H\textsubscript{2}\textsuperscript{+} & H\textsubscript{3}\textsuperscript{+} \\
 He & He\textsuperscript{+} & He\textsuperscript{2+} & C & C\textsuperscript{+} & C\textsuperscript{-} & C\textsubscript{2} \\ 
 CH & CH\textsuperscript{+} & CH\textsubscript{2} & CH\textsubscript{2}\textsuperscript{+} & CH\textsubscript{3}\textsuperscript{+} & O & O\textsuperscript{+} \\
 O\textsuperscript{-} & O\textsubscript{2} & O\textsubscript{2}\textsuperscript{+} & OH & OH\textsuperscript{+} & H\textsubscript{2}O & H\textsubscript{2}O\textsuperscript{+}  \\
 H\textsubscript{3}O\textsuperscript{+} & HCO & HCO\textsuperscript{+} & HOC\textsuperscript{+} & CO & CO\textsuperscript{+} & Si \\ Si\textsuperscript{+} & Si\textsuperscript{2+} & (f-H\textsubscript{2}O) & (f-CO)\\
 \hline
\end{tabular}
\end{center}
 \label{table:RC_species}
\end{table}

\subsection{Freeze-out}
\label{appendix:networks_freeze}

We model the freeze-out of CO and H$_2$O as well as its desorption from dust grains as follows. The adsorption (freeze-out) occurs with a rate of
\begin{equation}
    k_{\rmn{ads,}i} = \sigma_\rmn{d} \, n_\rmn{d} \, c_{\rmn{s},i} \, n_i S\, ,
\label{eq:desorption}
\end{equation}
where $c_{\rmn{s},i}$ and $n_i$ are the sound speed and particle density of the considered species $i$, respectively, and $S$ is the sticking coefficient. For the product $\sigma_\rmn{d} n_\rmn{d}$, representing the cross-section and density of dust particles, we adopt the value of \mbox{$2 \times 10^{-21} n_\rmn{H, tot}$} given by \citet{hollenbach_water_2009} for a standard dust grain size distribution \citep{Mathis77}. The sticking coefficient $S$ is taken to be unity.

For desorption from dust grains, we consider thermal and cosmic-ray induced desorption. The thermal desorption rate is given by 
\begin{equation}
    k_{\rmn{des, therm,}i} = \nu_i e^{-E_{\rmn{D},i}/T_\rmn{d}} \, ,
\label{eq:therm_des}
\end{equation}
where $\nu_i$ is the vibrational frequency of the species $i$ in the surface potential well, and $E_{\rmn{D},i}$ is its adsorption binding energy \citep[e.g.][]{Hasegawa93}. For $E_{\rmn{D},i}$ we used the values of 1150~K and 5700~K for CO and H$_2$O, respectively, given by \citet{Garrod06}.
Using eq.~6 of \citet{Aikawa96}, we obtain $\nu_i$ of \mbox{$1.01593 \times 10^{12}$ s$^{-1}$} and \mbox{$2.82095 \times 10^{12}$ s$^{-1}$} for CO and H$_2$O, respectively. For the desorption by cosmic rays we follow the approach of \citet{Hasegawa93}, giving a rate of $k_\rmn{des,cr} = f(70 \, \rmn{K}) \cdot k_\rmn{des, therm}(T_\rmn{d} = 70 \, \rmn{K}) \cdot \rmn{CRIR}$. This is based on the fraction of time that a dust grain reaches a temperature of 70~K due to heating by cosmic ray impacts \citep{Leger85}, with updates for CO desorption as given by \citet{Herbst06}.

\subsection{The on-the-fly network `NL99'}
\label{appendix:networks_nl99}

The SILCC-Zoom simulations on which we employ our post-processing method were run coupled to the chemical network `NL99.' This network combines a model for CO chemistry by \cite{nelson_stability_1999} with a hydrogen model by \cite{glover_simulating_2007,glover_simulating_2007-1}. The full network was first advanced by \cite{glover_approximations_2012} and then modified by \cite{mackey_non-equilibrium_2019}. The chemical species whose abundances are calculated in this network are listed in Table~\ref{table:NL99_species}. Most of these species' abundances are calculated on-the-fly using an ODE solver or through conservation equations. The exceptions are H\textsubscript{2}\textsuperscript{+}, which is assumed to immediately react further; and O, O\textsuperscript{+}, and H\textsubscript{3}\textsuperscript{+}, which are evolved to equilibrium values.

For simplification, the proxy species CH\textsubscript{x} and OH\textsubscript{x} have been introduced. These represent families of carbon- and oxygen-bearing species with different numbers of hydrogen atoms, e.g., CH, CH\textsubscript{2}, and CH\textsubscript{3}, or OH, H\textsubscript{2}O, and H\textsubscript{3}O, and ionized states of these species as appropriate.

Rather than treat metal elements separately, the network combines various metals which contribute non-negligibly to the electron density into a proxy species M, as well as its singly-ionized state M\textsuperscript{+}. By number density, this mostly comprises Si, but also N, Mg, S, and Fe. 

The primary purpose of including the NL99 network in the SILCC-Zoom simulations is to properly model the heating and cooling effects of \ce{C+}, CO, and atomic oxygen, which thermally impact the bulk gas distribution. As such, the limited other species in the network are not really modeled comprehensively. In particular, we stress that NL99 does not model the full set of creation and destruction reactions for \ce{HCO+} in a self-consistent, trustworthy way. This necessitates our post-processing, and precludes us from comparing the NL99 and post-processed abundances of \ce{HCO+} in this work.

\begin{table}
 \caption{The chemical species included in the NL99 network, which originated with \citet{nelson_stability_1999} with updates by \citet{glover_approximations_2012} and \citet{mackey_non-equilibrium_2019}. The species CH\textsubscript{x} and OH\textsubscript{x} are defined as proxies for, respectively, simple hydrocarbons like CH, CH\textsubscript{2}, CH\textsubscript{3}, and similarly OH, H\textsubscript{2}O, and so forth. The species M and its ionized state M\textsuperscript{+} are proxies for metals like N, Mg, Si, S, and Fe.}
\begin{center}
\begin{tabular}{ c c c c c c c c } 
 \hline
 e\textsuperscript{-} & H & H\textsuperscript{+} & H\textsubscript{2} & He &
 He\textsuperscript{+} \\ H\textsubscript{2}\textsuperscript{+} & H\textsubscript{3}\textsuperscript{+} & O & O\textsuperscript{+} & OH\textsubscript{x} & HCO\textsuperscript{+} \\
 C & C\textsuperscript{+} & CO & CH\textsubscript{x} & M & M\textsuperscript{+} \\
 \hline
\end{tabular}
\end{center}
 \label{table:NL99_species}
\end{table}

\section{Ratios of unprocessed and post-processed abundances}
\label{appendix:fracvn_ratios}

\begin{figure*}
    \centering
    \includegraphics[width=0.9\linewidth]{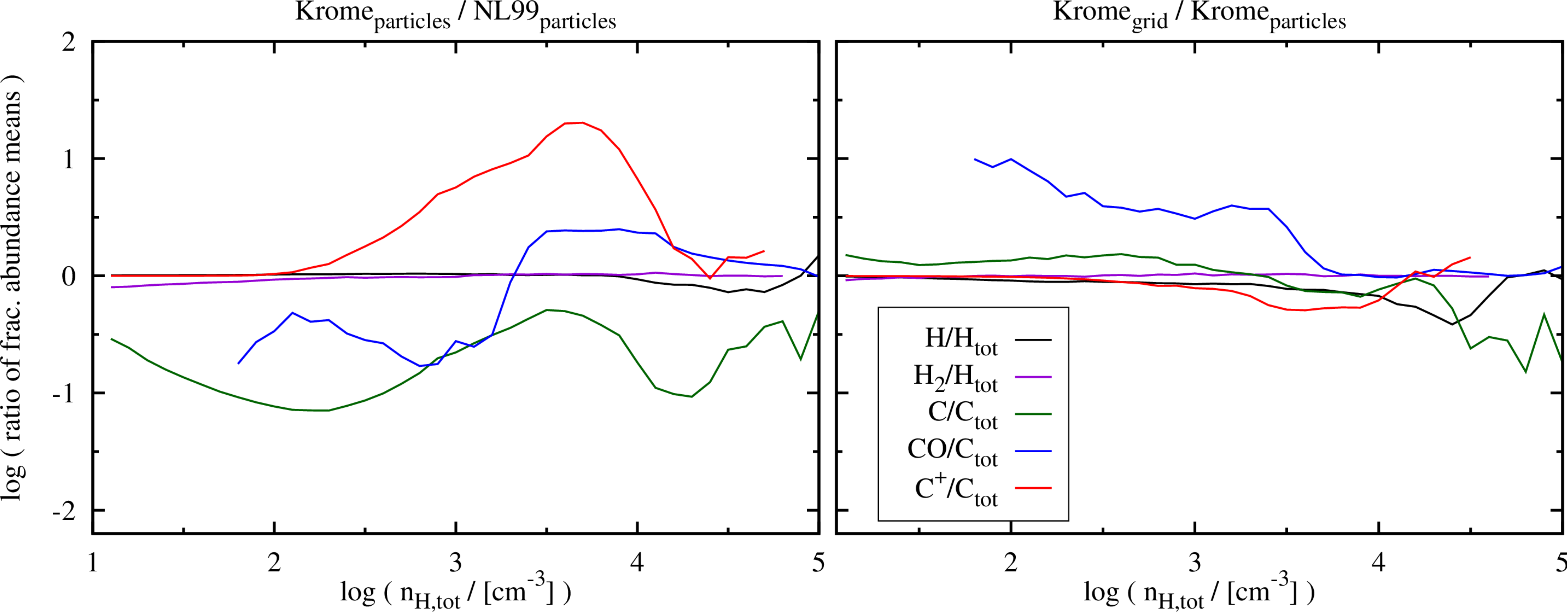}
    \caption{Left panel: the ratio of post-processed to unprocessed (NL99) mean fractional abundance reported by the tracer particles in MC1-HD at $t_\rmn{evol}=2$~Myr, vs. $n_\rmn{Htot}$. This corresponds to the ratio of the right and left panels of Fig.~\ref{fig:fracvn-particles}. Right panel: the ratio of post-processed  mean fractional abundances after regridding to the post-processed mean fractional abundances from the tracer particles, vs. $n_\rmn{H,tot}$. This corresponds to the ratio of Fig.~\ref{fig:fracvn-postgrid} with the right panel of Fig.~\ref{fig:fracvn-particles}.}
    \label{fig:fracvn-ratios}
\end{figure*}

In Fig.~\ref{fig:fracvn-ratios}, we present two sets of ratios of the means of fractional abundances at different stages in the post-processing algorithm, for cloud MC1 at $t_\rmn{evol}=2$~Myr.

In the left panel, we show the ratio of the mean post-processed fractional abundances (of H, \ce{H2}, C, CO, and \ce{C+}) reported by the tracer particles (the right panel of Fig.~\ref{fig:fracvn-particles}), to the mean unprocessed (NL99) fractional abundances reported by the tracer particles (the left panel of Fig.~\ref{fig:fracvn-particles}), vs. $n_\rmn{H,tot}$. The post-processing changes the hydrogen abundances very little, but alleviates the problem of overproduced atomic carbon in NL99, as described in Section~\ref{subsection-ppvalidation}. In the right panel, we show the ratio of the mean post-processed fractional abundances after regridding (Fig.~\ref{fig:fracvn-postgrid}), to the mean post-processed fractional abundance reported by the tracer particles, that is, before regridding (the right panel of Fig.~\ref{fig:fracvn-particles}).

\section{Column density maps of fundamental species}
\label{appendix:Nmaps}

\begin{figure*}
    \centering
    \includegraphics[width=\linewidth]{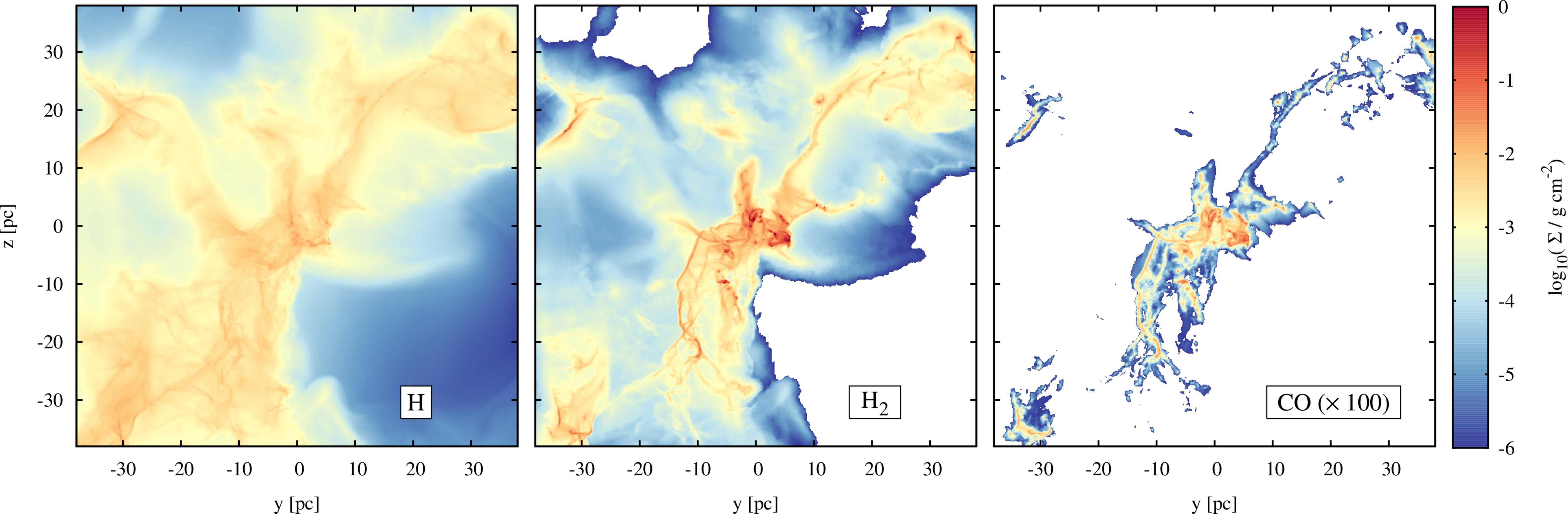}
    \caption{Projections along the $x$-direction of the column mass density of H, H\textsubscript{2}, and CO (upscaled by 100), for \mbox{MC1-HD} at $t_\rmn{evol}=4$~Myr, generated from the interpolated grids of post-processed tracer abundances. The maps are roughly centred on the densest region in the cloud. The atomic hydrogen (left) dominates in the more diffuse, outlying areas. Meanwhile, H\textsubscript{2} (centre) reaches its highest density in the heart of the cloud, but is also present in a diffuse envelope surrounding the core. CO (right) is only present in the densest parts of the cloud. These maps are very similar to the analogous AMR-derived maps in fig.~3 of \citet{seifried_silcc-zoom_2017}, supporting the accuracy of the regridding algorithm.}
    \label{fig:maps_mc1}
\end{figure*}

In Section~\ref{section-regridding}, we discuss the algorithm by which we allocate and interpolate post-processed tracer abundances into a uniformly-resolved density grid. Line integration of these grids along a chosen line of sight produces column density maps. In the text we provide such maps for HCO\textsuperscript{+}; here we present and discuss column density maps for atomic and molecular hydrogen, as well as CO, to validate our methods.

We present in Fig.~\ref{fig:maps_mc1} column mass density ($\Sigma$) maps for H, H\textsubscript{2}, and CO (upscaled by 100) for the $y$-$z$ projection of cloud \mbox{MC1-HD} at $t_\rmn{evol}=4$~Myr. These maps were presented at approximately the same $t_\rmn{evol}$ in fig.~3 of \cite{seifried_silcc-zoom_2017}, the work from which this SILCC-Zoom simulation originated. Although our results here are post-processed using the more extensive chemical network which originated in \cite{grassi_detailed_2017}, the distribution of different chemical species in the gas in Fig.~\ref{fig:maps_mc1} is approximately congruent with the original on-the-fly results, as predicted in \cite{seifried_silcc-zoom_2017}. The atomic hydrogen covers the entire domain and reaches a peak density around $10^{-2}$~g~cm$^{-2}$. An envelope of H\textsubscript{2} with column densities between $10^{-6}$ and $10^{-3}$~g~cm$^{-2}$ surrounds a dense core region where maximal values of $\Sigma \rmn{(H_2)} \simeq 1$~g~cm$^{-2}$ are reached. CO is only abundant in these core regions, with column densities around $10^{-3}$~g~cm$^{-2}$. The relative distributions of of H\textsubscript{2} and CO reflect the extent of the CO-dark molecular gas. The close correspondence of these maps with the results from \cite{seifried_silcc-zoom_2017} strongly supports the regridding algorithm by which we construct density grids from limited tracer particle data.

\bsp
\label{lastpage}
\end{document}